\documentclass[11pt,english]{article}
\usepackage[latin9]{inputenc}
\usepackage{color}
\usepackage{textcomp}
\usepackage{amsmath}
\usepackage{amssymb}
\usepackage{graphicx}
\usepackage{setspace}
\usepackage{subscript}
\makeatletter

\DeclareFontEncoding{LGR}{}{}
\DeclareRobustCommand{\greektext}{%
  \fontencoding{LGR}\selectfont\def\encodingdefault{LGR}}
\DeclareRobustCommand{\textgreek}[1]{\leavevmode{\greektext #1}}
\ProvideTextCommand{\~}{LGR}[1]{\char126#1}

\providecommand{\tabularnewline}{\\}

\@ifundefined{date}{}{\date{}}
\usepackage{esint}
\usepackage{hyperref}
\usepackage{latexsym}
\usepackage{graphicx}\usepackage{bm}\usepackage{longtable}

\usepackage{xcolor}

\@addtoreset{equation}{section}

\makeatother

\usepackage{babel}
\begin{document}
\title{The D4/D8 model and holographic QCD}
\maketitle
\begin{center}
\emph{$^{*}$}Si-wen Li\footnote{Email: siwenli@dlmu.edu.cn}, \emph{$^{\dagger}$}Xiao-tong
Zhang\footnote{Email: zxt@dlmu.edu.cn},
\par\end{center}

\begin{center}
\emph{$^{*}$Department of Physics, College of Science,}\\
\emph{Dalian Maritime University, }\\
\emph{Dalian 116026, China}\\
\vspace{4mm}
\par\end{center}

\begin{center}
\emph{$^{\dagger}$College of Innovation and Entrepreneurship ,}\\
\emph{Dalian Maritime University, }\\
\emph{Dalian 116026, China}
\par\end{center}

\vspace{12mm}

\begin{abstract}
As a top-down holographic approach, the D4/D8 model is expected to
be the holographic version of QCD since it almost includes all the
elementary features of QCD based on string theory. In this manuscript,
we review the fundamental properties of the D4/D8 model with respect
to the D4-brane background, embedding of flavor branes and holographic
quark, gluon, meson, baryon and glueball with various symmetries,
then we also take a look at some interesting applications and developments
based on this model. \vspace{8mm}

\noindent \textbf{Keywords:} Gauge-gravity duality; AdS/CFT correspondence;
Holographic QCD
\end{abstract}
\newpage{}

\tableofcontents{}

\section{Introduction}

Although it has been about 25 years since the proposal of AdS/CFT
and gauge-gravity duality with holography \cite{key-1,key-2,key-3},
it remains to attract great interests today. The most significant
part of AdS/CFT and gauge-gravity duality is that people can evaluate
strongly coupled quantum field theory (QFT) by analyzing quantitatively
its associated gravity theory in the weak coupling region. So it provides
a holographic way to study the strongly coupled QFT in which traditional
QFT based on perturbation method is out of reach. Accordingly, a large
number of publications about strongly coupled QFT through AdS/CFT
and gauge-gravity duality have appeared, for example\textcolor{blue}{{}
}on Wilson loop and quark potential \cite{key-4,key-5,key-6}, transport
coefficient \cite{key-7,key-8,key-9}, fermionic correlation function
\cite{key-10,key-11}, the Schwinger effect \cite{key-12}, quantum
entanglement entropy \cite{key-13,key-14} and quantum information
on black hole \cite{key-15} which have become most remarkable works
in this field.

On the other hand, QCD (quantum chromodynamics) as the fundamental
theory describing the property of strong interaction is extremely
complex in the strong coupling region, especially at finite temperature
with dense matter due to its asymptotic freedom \cite{key-16,key-17},
hence the holographic version of QCD is natural to become an interesting
topic. While there are several models and theories attempting to give
a holographic version of QCD (e.g. bottom-up approaches \cite{key-18,key-19,key-1+1},
D3/D7 approach \cite{key-20}, D4/D6 approach \cite{key-21}), one
of the most successful achievements in holography is the D4/D8 model
(also named as Witten-Sakai-Sugimoto model) \cite{key-22,key-23,key-24}
which includes almost all the elementary features of QCD in a very
simple way, e.g. meson, baryon \cite{key-25,key-26,key-27}, glueball
\cite{key-28,key-29,key-30,key-31}, deconfinement transition \cite{key-32,key-33,key-34},
chiral phase \cite{key-35,key-36}, heavy flavor \cite{key-37,key-38,key-39},
$\theta$ term and QCD axion \cite{key-40,key-41,key-42,key-43,key-44},
nucleon interaction \cite{key-45,key-46,key-47,key-48,key-49,key-50,key-51}.
The D4/D8 model is based on the holographic duality between the 11-dimensional
(11d) M-theory on $\mathrm{AdS_{7}}\times S^{4}$ and the $\mathcal{N}=\left(2,0\right)$
super conformal field theory (SCFT) on $N_{c}$ M5-branes \cite{key-52}.
By using the dimensional reduction in \cite{key-22,key-53}, it reduces
to the correspondence of the pure Yang-Mills theory on $N_{c}$ D4-branes
compactified on a circle and 10d IIA supergravity (SUGRA). Flavors
as $N_{f}$ pairs of D8- and anti D8-branes ($\mathrm{D8/\overline{D8}}$)
can be further introduced into the geometric background produced by
the $N_{c}$ D4-branes, so the dual theory includes flavored fundamental
quarks which would be more close to the realistic QCD. Beside, as
the D4/D8 model is a T-dualized version of D3/D9 system \cite{key-54},
the fundamental quark and meson in the D4/D8 model can therefore be
identified to the $4-8$\footnote{The $4-8$ string refers to the open string connects the $N_{c}$
D4-brane and $N_{f}$ D8-branes. And it is similar for e.g. the $8-8$
or $4-\bar{8}$ string.} and $8-8$ strings respectively by following the same discussion
in D3/D9 system \cite{key-54}. Moreover, the baryon vertex is identified
as a D4-brane wrapped on $S^{4}$ \cite{key-25} and the glueball
is recognized to be the bulk gravitational polarization \cite{key-28,key-29}.
The chiral phase is determined by the embedding configuration of the
$\mathrm{D8/\overline{D8}}$-branes due to the gauge symmetry on their
worldvolume \cite{key-55} while the deconfinement transition is suggested
to be the Hawking-Page transition in this model \cite{key-32,key-33,key-34,key-35}.
Altogether, this model includes all the fundamental elements of QCD
thus can be treated as a holographic version of QCD.

In this review, we will revisit the above properties of the D4/D8
model, then take a brief look at some recently relevant developments
and holographic approaches with this model. The outline of this review
is as follows. In Section 2, we will review the relation of 11d M-theory
and IIA SUGRA with respect to the case of M5-brane and D4-brane. Afterwards,
it is the embedding configuration of the $\mathrm{D8/\overline{D8}}$-branes
to the D4-brane background, the holographic quark, gluon, meson, baryon
and glueball with various symmetries which are all the relevant objects
in hadron physics. In Section 3, we review several topics about the
developments and holographic approaches of this model which includes
deconfinement transition, chiral transition, Higgs mechanism and heavy-light
meson or baryon, interaction involving glueball and QCD $\theta$
term in holography. In the appendix, we give the general form of the
black brane solution in type II SUGRA, the relevant dimensional reduction
for spinor and discussion about supersymmetric meson which are useful
to the main content of this paper.

\section{The D4/D8 model}

In this section, we will revisit the D4-brane background and the embedding
of the probe $\mathrm{D8/\overline{D8}}$-branes. Then we will review
how to identify quarks, gluon, meson, baryon and glueball with various
symmetries in this model.

\subsection{11d supergravity and D4-brane background}

The D4-brane background of the D4/D8 model is based on the holographic
duality between the type $\mathcal{N}=\left(2,0\right)$ super conformal
field theory (SCFT) on coincident $N_{c}$ M5-branes and 11-dimensional
(11d) M-theory on $\mathrm{AdS}_{7}\times S^{4}$ \cite{key-52}.
In order to obtain a geometric solution, the effective action of the
M-theory is necessary which is known as the 11d supergravity action.
In the large-$N_{c}$ limit, the geometric background can be obtained
by solving its bosonic part which is consisted of metric (elfbein)
and a three-from $C_{3}$ as \cite{key-56},

\begin{equation}
S_{\mathrm{SUGRA}}^{11\mathrm{d}}=\frac{1}{2\kappa_{11}^{2}}\int d^{11}x\sqrt{-g}\left[\mathcal{R}^{(11)}-\frac{1}{2}\left|F_{4}\right|^{2}\right]-\frac{1}{12\kappa_{11}^{2}}\int C_{3}\wedge F_{4}\wedge F_{4},\label{eq:1}
\end{equation}
where $F_{4}=dC_{3}$. The convention in (\ref{eq:1}) is as follows.
$\mathcal{R}^{(11)}$ refers to the 11d scalar curvature, $\kappa_{11}$
is the 11d gravity coupling constant given by,

\begin{equation}
2\kappa_{11}^{2}=16\pi G_{11}=\frac{1}{2\pi}\left(2\pi l_{p}\right)^{9},
\end{equation}
where $G_{11}$ is 11d Newton's constant, $l_{p}$ is the Planck length.
The quantity $\left|F_{4}\right|^{2}$ can be obtained by a general
notation of an $n$-form $F_{n}$ as,

\begin{equation}
\left|F_{n}\right|^{2}=\frac{1}{n!}g^{A_{1}B_{1}}g^{A_{2}B_{2}}...g^{A_{n}B_{n}}F_{A_{1}A_{2}...A_{n}}F_{B_{1}B_{2}...B_{n}},
\end{equation}
where $g_{AB}$ refers to the metric on the manifold. We note that
in (\ref{eq:1}) the last term is a Chern-Simons structure which is
independent on the metric or elfbein while the first term depends
on the metric or elfbein through the metric combination

\begin{equation}
g_{AB}=e_{A}^{a}\eta_{ab}e_{B}^{b}.
\end{equation}

The solution for extremal coincident $N_{c}$ M5-branes is obtained
as,

\begin{align}
ds_{\mathrm{11d}}^{2}= & H_{5}\left(r\right)^{-\frac{1}{3}}\eta_{MN}dx^{M}dx^{N}+H_{5}\left(r\right)^{\frac{2}{3}}\left(dr^{2}+r^{2}d\Omega_{4}^{2}\right),\nonumber \\
H_{5}\left(r\right)= & 1+\frac{r_{5}^{3}}{r^{3}},\ \left(^{*}F_{4}\right)_{r012...5}=-\frac{3r_{5}^{3}}{r^{4}H_{5}^{2}},\label{eq:5}
\end{align}
where $r$ denotes the radial coordinate to the M5-branes and $M,N$
run over the M5-branes. Using the BPS condition for M5-brane, 

\begin{equation}
2\kappa_{11}^{2}T_{\mathrm{M5}}N_{c}=\int_{S^{4}}F_{4},
\end{equation}
it leads to

\begin{equation}
r_{5}^{3}=\pi N_{c}l_{p}^{3},\label{eq:7}
\end{equation}
where $T_{\mathrm{M5}}=\frac{1}{\left(2\pi\right)^{5}l_{p}^{6}}$
refers to the tension of the M5-brane. Taking the near horizon limit
$H_{5}\rightarrow\frac{r_{5}^{3}}{r^{3}}$ and replacing the variables
as

\begin{equation}
\left\{ r,r_{5},x^{i},\Omega_{4}\right\} \rightarrow\left\{ r^{2},\frac{L}{2},\frac{1}{\sqrt{L}}x^{i},\frac{1}{\sqrt{L}}\Omega_{4}\right\} ,\label{eq:8}
\end{equation}
the metric presented in (\ref{eq:5}) reduces to,

\begin{equation}
ds_{\mathrm{11d}}^{2}=\frac{r^{2}}{L^{2}}\eta_{MN}dx^{M}dx^{N}+\frac{L^{2}}{r^{2}}\left(dr^{2}+\frac{r^{2}}{4}d\Omega_{4}^{2}\right),\label{eq:9}
\end{equation}
describing the standard form of $\mathrm{AdS}_{7}\times S^{4}$ where
the radius of $S^{4}$ is $L/2$. In addition, the action (\ref{eq:1})
also allows the near-extremal M5-brane solution which, after taking
near horizon limit and replacement (\ref{eq:8}), is

\begin{align}
ds_{\mathrm{11d}}^{2} & =\frac{r^{2}}{L^{2}}\left[-f\left(r\right)\left(dx^{0}\right)^{2}+\sum_{i=1}^{5}dx^{i}dx^{i}\right]+\frac{L^{2}}{r^{2}}\left[\frac{dr^{2}}{f\left(r\right)}+\frac{r^{2}}{4}d\Omega_{4}^{2}\right],\nonumber \\
f\left(r\right) & =1-\frac{r_{H}^{6}}{r^{6}},\label{eq:10}
\end{align}
The constant $r_{H}$ refers to the location of the horizon which
can be determined by omitting the conical singularity as,

\begin{equation}
\beta_{T}=\frac{2\pi L^{2}}{3r_{H}},\label{eq:11}
\end{equation}
where $\beta_{T}$ is the size of the compactified direction $x^{0}$.

In order to obtain a QCD-like low-energy theory in holography, Witten
proposed a scheme in \cite{key-22} based on the above M5-brane solution.
Specifically, the first step is to campactify one spacial direction
(denoted by $x^{5}$) of M5-branes on a circle with periodic condition
for fermions, which means the supersymmetry remains. Accordingly the
resultant theory is a supersymmetric gauge theory above the size of
the circle. Then, recall the relation between M-theory and IIA string
theory, the 11d metric presented in (\ref{eq:9}) or (\ref{eq:10})
reduces to a 10d metric as,

\begin{equation}
ds_{\mathrm{11d}}^{2}=e^{-\frac{2}{3}\phi}ds_{\mathrm{10d}}^{2}+e^{\frac{4}{3}\phi}\left(dx^{5}\right)^{2},\label{eq:12}
\end{equation}
with the non-trivial dilaton $e^{\phi}=\left(r/L\right)^{3/2}$. For
the later use, let us introduce another radial coordinates $U\in\left[U_{H},\infty\right)$
by

\begin{equation}
U=\frac{r^{2}}{2L},L=2R.
\end{equation}
So in the large $N_{c}$ limit, the 11th direction $x^{5}$ presented
in (\ref{eq:12}) vanishes due to (\ref{eq:7}) (\ref{eq:8}), which
means the coincident $N_{c}$ M5-branes corresponds to coincident
$N_{c}$ D4-branes for $N_{c}\rightarrow\infty$. And the remained
10d metric in (\ref{eq:12}) becomes the near-extremal black D4-brane
solution as\footnote{The extremal D4-brane solution can be obtained by setting $U_{H}\rightarrow0$.},

\begin{align}
ds_{\mathrm{10d}}^{2} & =\left(\frac{U}{R}\right)^{3/2}\left[-f_{T}\left(U\right)\left(dx^{0}\right)^{2}+\sum_{i=1}^{4}dx^{i}dx^{i}\right]+\left(\frac{R}{U}\right)^{3/2}\left[\frac{dU^{2}}{f_{T}\left(U\right)}+U^{2}d\Omega_{4}^{2}\right],\nonumber \\
f_{T}\left(U\right) & =1-\frac{U_{H}^{3}}{U^{3}},F_{4}=3R^{3}g_{s}^{-1}\epsilon_{4},e^{\phi}=\left(\frac{U}{R}\right)^{3/4},R^{3}=\pi g_{s}N_{c}l_{s}^{3}\label{eq:14}
\end{align}
where $\epsilon_{4}$ refers to the volume form of a unit $S^{4}$.
Once the formula (\ref{eq:12}) is imposed to action (\ref{eq:1}),
the 11d SUGRA action reduces to the 10d type IIA SUGRA action exactly
which is given as,\footnote{There could be a Chern-Simons term to the IIA SUGRA action (\ref{eq:15})
as $S_{\mathrm{CS}}^{\mathrm{10d}}=-\frac{1}{4\kappa_{10}^{2}}\int B_{2}\wedge F_{4}\wedge F_{4}$
with $B_{mn}=A_{mn5}$. For the purely black brane solution, the $B_{2}$
field can be gauged away by setting $B_{2}=0$ which implies this
CS term could be absent in 10d action \cite{key-52,key-56}.},

\begin{equation}
S_{\mathrm{IIA}}^{10\mathrm{d}}=\frac{1}{2\kappa_{10}^{2}}\int d^{10}x\sqrt{-G}e^{-2\phi}\left[\mathcal{R}^{(10)}+4\partial_{\mu}\phi\partial\phi\right]-\frac{1}{4\kappa_{10}^{2}}\int d^{10}x\sqrt{-G}\left|F_{4}\right|^{2},\label{eq:15}
\end{equation}
where $2\kappa_{10}^{2}=16\pi G_{10}=\left(2\pi\right)^{7}l_{s}^{8}g_{s}^{2}$
is the 10d gravity coupling constant. And it would be straightforward
to verify that the solution (\ref{eq:14}) satisfies the equation
of motion obtained by varying action (\ref{eq:15}).

\begin{figure}
\begin{centering}
\includegraphics[scale=0.3]{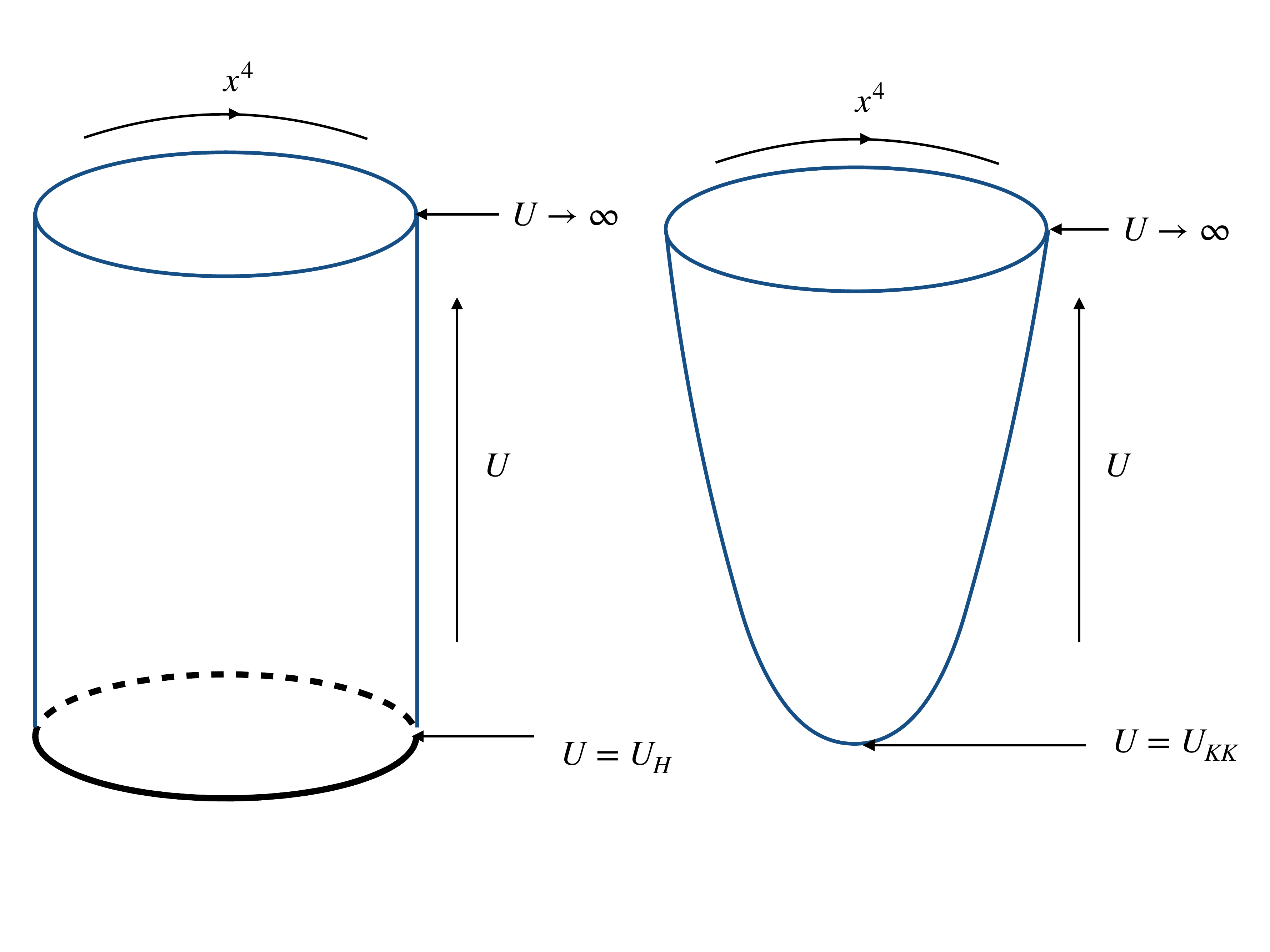}
\par\end{centering}
\caption{\label{fig:1} The compactified D4-brane geometry on $U-x^{4}$ plane.
\textbf{Left:} The black D4-brane background geometry. \textbf{Right:
}The D4 bubble geometry. }

\end{figure}

The next step is to perform the double Wick rotation $\left\{ x^{0}\rightarrow-ix^{4},x^{4}\rightarrow-ix^{0}\right\} $
on the metric (\ref{eq:14}) leading to a bubble D4-brane solution
as,

\begin{align}
ds_{\mathrm{10d}}^{2} & =\left(\frac{U}{R}\right)^{3/2}\left[\eta_{\mu\nu}dx^{\mu}dx^{\nu}+f\left(U\right)\left(dx^{4}\right)^{2}\right]+\left(\frac{R}{U}\right)^{3/2}\left[\frac{dU^{2}}{f\left(U\right)}+U^{2}d\Omega_{4}^{2}\right],\nonumber \\
f\left(U\right) & =1-\frac{U_{KK}^{3}}{U^{3}},\label{eq:16}
\end{align}
which is defined only for $U\in\left[U_{KK},\infty\right)$. We have
renamed $U_{H}$ as $U_{KK}$ in (\ref{eq:16}) since there is not
a horizon in the bubble solution as it is illustrated in Figure \ref{fig:1}
.
 Now the direction of $x^{4}$ is periodic as,

\begin{equation}
x^{4}\sim x^{4}+\delta x^{4},\delta x^{4}=\frac{2\pi}{M_{KK}}=\frac{4\pi R^{3/2}}{3U_{KK}^{1/2}},\label{eq:17}
\end{equation}
because it is identified as the time direction in the black brane
solution (\ref{eq:14}). $M_{KK}$ refers to the Klein-Kaluza (KK)
energy scale and the supersymmetry on the D4-branes breaks down below
$M_{KK}$ by imposing the anti-periodic condition to the supersymmetric
fermions along $x^{4}$. Accordingly, the low-energy zero modes on
the D4-branes are the massless gauge field on the D4-branes and the
scalar fields as the transverse modes of the D4-branes. While the
scalar fields acquire mass via one-loop corrections, the trace part
of the scalars $\phi^{i}$ and gauge field along $x^{4}$ direction
$a_{4}$ remain to be massless. As they give irrelevant coupling terms
in the low-energy effective theory on the D4-branes, it means the
dual theory below $M_{KK}$ only contains 4d pure Yang-Mills gauge
field as it is expected. Note that the three-from $C_{3}$ in 11d
SUGRA (\ref{eq:1}) corresponds to the Ramond-Ramond (R-R) three-form
in the type IIA string theory. 

Besides, as the wrap factor $\left(U/R\right)^{3/2}$ in (\ref{eq:16})
never goes to zero, the dual theory would be able to exhibit confinement
according to the behavior of the Wilson loop in this geometry. Since
the solution (\ref{eq:16}) allows an arbitrarily large period for
$x^{0}$, it implies the dual theory on the D4-brane could be defined
at zero (or very low) temperature. Furthermore, in order to obtain
a deconfined version of holographic QCD based on (\ref{eq:16}) at
finite temperature, it is also possible to compactify one spacial
direction (denoted by $x^{4}$) of the D4-branes in the background
(\ref{eq:14}) with the anti-periodic condition for the supersymmetric
fermions\footnote{There might be an issue if we identify the black brane background
(\ref{eq:14}) to the deconfinement phase exactly since Wilson loop
on this background may mot match to the deconfinement QCD \cite{key-57,key-58}.
Nevertheless, we can identify the black brane background (\ref{eq:14})
to QCD phase at high temperature in which the deconfinement will occur.} as it is displayed in (\ref{eq:17}) and Figure \ref{fig:1}. In
this case, the Hawking temperature $T$ in compactified background
(\ref{eq:14}) is given by (\ref{eq:11}) as,
\begin{equation}
\beta_{T}=\frac{1}{T}=\frac{4\pi R^{3/2}}{3U_{H}^{1/2}},
\end{equation}
which can therefore be identified as the temperature in the dual theory.
And the variables in terms of the dual theory is summarized as,

\begin{equation}
R^{3}=\frac{1}{2}\frac{g_{\mathrm{YM}}^{2}N_{c}l_{s}^{2}}{M_{KK}},\ U_{KK}=\frac{2}{3}g_{\mathrm{YM}}^{2}N_{c}M_{KK}l_{s}^{2},\ g_{s}=\frac{1}{2\pi}\frac{g_{\mathrm{YM}}^{2}}{M_{KK}l_{s}},
\end{equation}
where $g_{\mathrm{YM}}$ to the Yang-Mills (YM) coupling constant.

\subsection{Embedding the probe $\mathrm{D8/\overline{D8}}$-branes }

\begin{figure}
\begin{centering}
\includegraphics[scale=0.3]{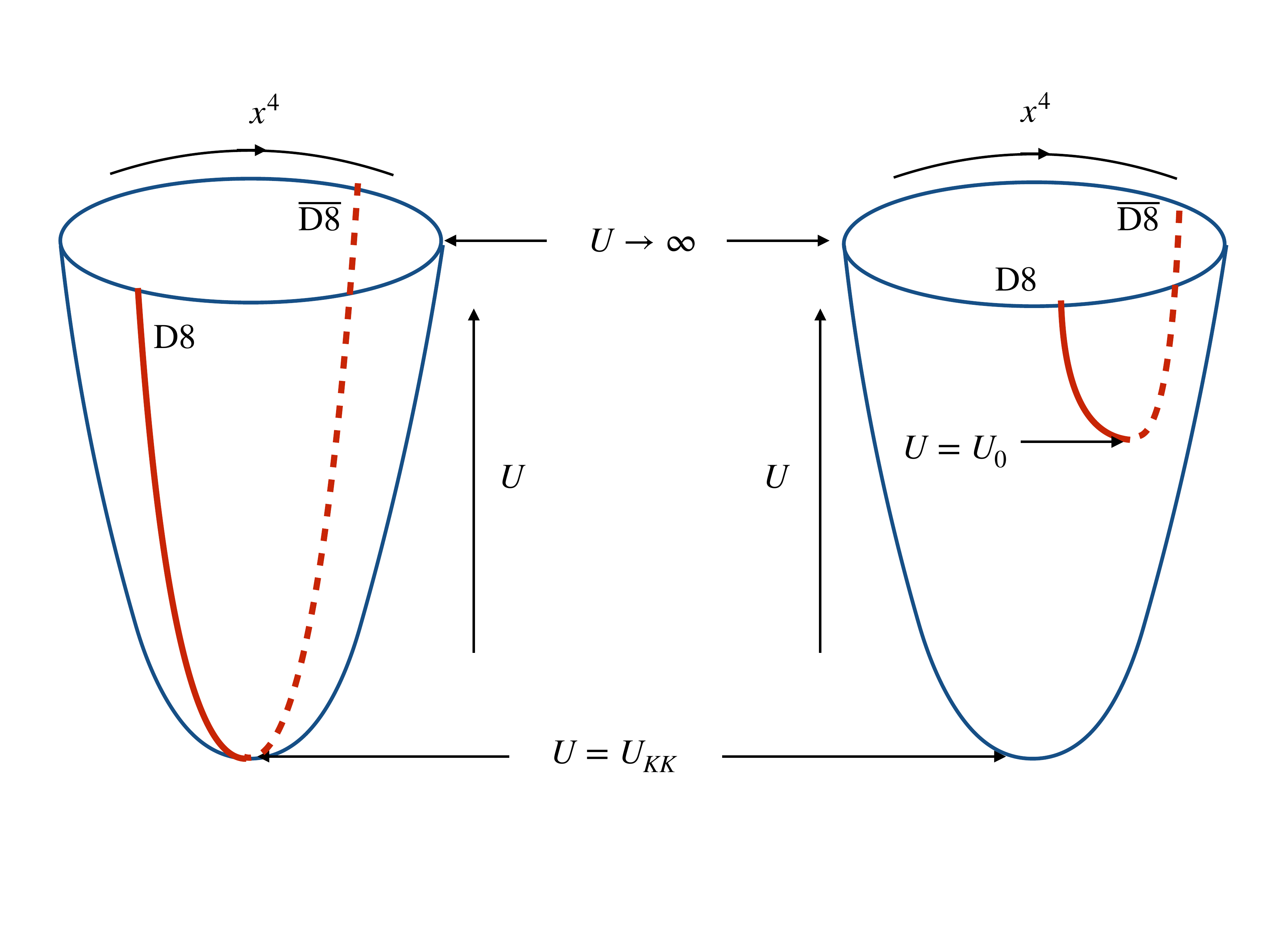}
\par\end{centering}
\caption{\label{fig:2} The D8-brane configuration in the bubble D4-brane background.
\textbf{Left:} D8- and $\overline{\mathrm{D8}}$-branes are located
at the antipodal points of $x^{4}$. \textbf{Right:} D8- and $\overline{\mathrm{D8}}$-branes
are located at the non-antipodal points of $x^{4}$.}
\end{figure}

\begin{figure}
\begin{centering}
\includegraphics[scale=0.3]{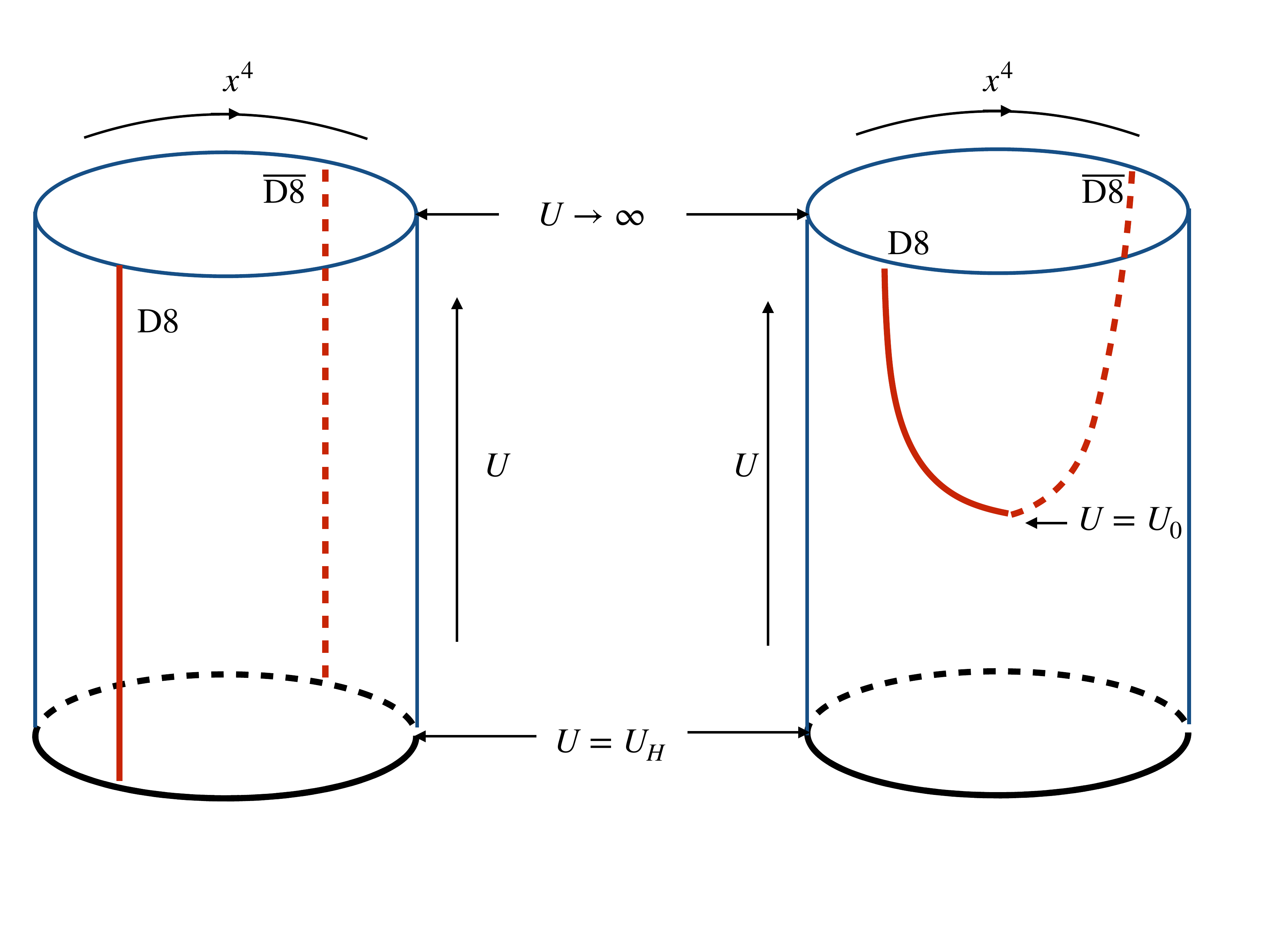}
\par\end{centering}
\caption{\label{fig:3} The D8-brane configuration in the black D4-brane background.
\textbf{Left:} D8- and $\overline{\mathrm{D8}}$-branes are parallel
located.\textbf{ Right:} D8- and $\overline{\mathrm{D8}}$-branes
are connected at $U=U_{0}$.}
\end{figure}

In the D4/D8 model, there is a stack of coincident $N_{f}$ pairs
of D8- and (anti-D8) $\overline{\mathrm{D8}}$-branes as probes embedded
into the bulk geometry illustrated in Figure \ref{fig:1}. The relevant
D-brane configuration is given in Table \ref{tab:1}. 
\begin{table}
\begin{centering}
\begin{tabular}{|c|c|c|c|c|c|c|c|c|c|c|}
\hline 
 & 0 & 1 & 2 & 3 & 4 & 5($U$) & 6 & 7 & 8 & 9\tabularnewline
\hline 
\hline 
$N_{c}$ D4-branes & - & - & - & - & - &  &  &  &  & \tabularnewline
\hline 
$N_{f}$ $\mathrm{D8/\overline{D8}}$-branes & - & - & - & - &  & - & - & - & - & -\tabularnewline
\hline 
\end{tabular}
\par\end{centering}
\caption{\label{tab:1} The D-brane configuration in the D4/D8 model. ``-''
denotes the D-brane extends along this direction.}

\end{table}
 The embedding configuration of $\mathrm{D8/\overline{D8}}$-branes
is determined by solving the bosonic action for a $\mathrm{D}_{p}$-branes,
which consists of Dirac-Born-Infeld (DBI) and Wess-Zumino (WZ) terms.
The action reads \cite{key-59},

\begin{align}
S_{\mathrm{D_{p}}} & =S_{\mathrm{DBI}}+S_{\mathrm{WZ}},\nonumber \\
S_{\mathrm{DBI}} & =-T_{\mathrm{D}_{p}}\int_{\mathrm{D}_{p}}d^{p+1}xe^{-\phi}\mathrm{STr}\left[\sqrt{-\det\left[E_{ab}+E_{aI}\left(Q^{IJ}-\delta^{IJ}\right)E_{Ib}+2\pi\alpha^{\prime}F_{ab}\right]}\sqrt{\det\left(Q_{\ J}^{I}\right)}\right],\nonumber \\
S_{\mathrm{WZ}} & =g_{s}T_{\mathrm{D}_{p}}\int_{\mathrm{D}_{p}}\sum_{n=0,1...}C_{p-2n+1}\wedge\frac{1}{n!}\left(2\pi\alpha^{\prime}\right)^{n}\mathrm{Tr}F^{n},\label{eq:20}
\end{align}
with the D-brane tension $T_{\mathrm{D}_{p}}=g_{s}^{-1}\left(2\pi\right)^{-p}l_{s}^{-p-1}$
and $e^{\Phi}=g_{s}e^{\phi}$,

\begin{align}
E_{ab}= & \left(G_{MN}+B_{MN}\right)\partial_{a}X^{M}\partial_{b}X^{N},\ E_{aI}=\left(G_{MI}+B_{MI}\right)\partial_{a}X^{M},\nonumber \\
Q_{\ J}^{I}= & \delta_{\ J}^{I}+i2\pi\alpha^{\prime}\left[\Psi^{I},\Psi^{J}\right],\ F_{ab}=\partial_{a}A_{b}-\partial_{b}A_{a}+i\left[A_{a},A_{b}\right],\nonumber \\
a,b= & 0,1...p,\ M,N=0,1...d,\ I,J=p+1,p+2...d.
\end{align}
Here $G_{MN},B_{MN}$ and $\Phi$ refers respectively to metric, the
antisymmetric tensor and dilaton field in the background spacetime.
$\Psi^{I}$ refers to the transverse mode of the $\mathrm{D}_{p}$-brane
under the T-duality. By choosing $p=8$, the action (\ref{eq:20})
leads to the action for D8-brane on the $N_{c}$ D4-brane background
as\footnote{In the D4/D8 approach, the antisymmetric tensor $B_{MN}$ has been
gauged away. },

\begin{equation}
S_{\mathrm{D8}}=-T_{\mathrm{D8}}\int_{\mathrm{D8}}d^{9}xe^{-\phi}\mathrm{STr}\sqrt{-\det\left[g_{ab}+\left(2\pi\alpha^{\prime}\right)F_{ab}\right]}+\frac{g_{s}}{3!}\left(2\pi\alpha^{\prime}\right)^{3}T_{\mathrm{D8}}\int_{\mathrm{D8}}C_{3}\wedge\mathrm{Tr}\left(F\wedge F\wedge F\right).\label{eq:22}
\end{equation}
Using the induced metric on $\mathrm{D8/\overline{D8}}$-branes with
respect to the bubble D4 background (\ref{eq:16}),

\begin{equation}
ds_{\mathrm{D8}}^{2}=\left(\frac{U}{R}\right)^{3/2}\eta_{\mu\nu}dx^{\mu}dx^{\nu}+\left(\frac{U}{R}\right)^{3/2}\left[f\left(U\right)\left(x^{4\prime}\right)^{2}+\left(\frac{R}{U}\right)^{3}\frac{1}{f\left(U\right)}\right]dU^{2}+R^{3/2}U^{1/2}d\Omega_{4}^{2},
\end{equation}
and the black D4-brane background (\ref{eq:14}) as,

\begin{align}
ds_{\mathrm{D8}}^{2}= & \left(\frac{U}{R}\right)^{3/2}\left[-f_{T}\left(U\right)\left(dx^{0}\right)^{2}+\delta_{ij}dx^{i}dx^{j}\right]\nonumber \\
 & +\left(\frac{U}{R}\right)^{3/2}\left[\left(x^{4\prime}\right)^{2}+\left(\frac{R}{U}\right)^{3}\frac{1}{f_{T}\left(U\right)}\right]dU^{2}+R^{3/2}U^{1/2}d\Omega_{4}^{2},
\end{align}
the DBI action for D8-branes becomes respectively,

\begin{equation}
S_{\mathrm{DBI}}=-g_{s}T_{\mathrm{D8}}V_{3}\beta_{4}\Omega_{4}\int_{U_{KK}}^{\infty}dUU^{4}\left[f\left(U\right)\left(x^{4\prime}\right)^{2}+\left(\frac{R}{U}\right)^{3}\frac{1}{f\left(U\right)}\right]^{1/2},\label{eq:25}
\end{equation}
and

\begin{equation}
S_{\mathrm{DBI}}=-g_{s}T_{\mathrm{D8}}V_{3}\beta_{T}\Omega_{4}\int_{U_{H}}^{\infty}dUU^{4}\left[\left(x^{4\prime}\right)^{2}f_{T}\left(U\right)+\left(\frac{R}{U}\right)^{3}\right]^{1/2}.\label{eq:26}
\end{equation}
Here we use $\Omega_{4}=8\pi^{2}/3$ to refer to the volume of a unit
$S^{4}$. Note that the WZ action is independent on the metric or
elfbein. Vary the action (\ref{eq:25}) and (\ref{eq:26}) with respect
to $x^{4}$, the associated equation of motion is respectively obtained
as,

\begin{equation}
\frac{d}{dU}\left[\frac{U^{4}f\left(U\right)x^{4\prime}}{\sqrt{f\left(U\right)\left(x^{4\prime}\right)^{2}+\left(\frac{R}{U}\right)^{3}\frac{1}{f\left(U\right)}}}\right]=0,\label{eq:27}
\end{equation}
and,

\begin{equation}
\frac{d}{dU}\left[\frac{U^{4}f_{T}\left(U\right)x^{4\prime}}{\sqrt{\left(x^{4\prime}\right)^{2}f_{T}\left(U\right)+\left(\frac{R}{U}\right)^{3}}}\right]=0.\label{eq:28}
\end{equation}
As the D8- and $\overline{\mathrm{D8}}$-branes are the only probe
branes, they could be connected smoothly at the location $U=U_{0}$
which means $x^{4}|_{U=U_{0}}\rightarrow\infty$. With this boundary
condition, (\ref{eq:27}) and (\ref{eq:28}) reduce respectively to
the following solutions as,

\begin{equation}
\left(x^{4\prime}\right)^{2}=\frac{U_{0}^{8}f\left(U_{0}\right)}{U^{3}f\left(U\right)^{2}}\frac{R^{3}}{U^{8}f\left(U\right)-U_{0}^{8}f\left(U_{0}\right)},\label{eq:29}
\end{equation}
and

\begin{equation}
\left(x^{4\prime}\right)^{2}=\frac{U_{0}^{8}f_{T}\left(U_{0}\right)}{U^{3}f_{T}\left(U\right)}\frac{R^{3}}{U^{8}f_{T}\left(U\right)-U_{0}^{8}f_{T}\left(U_{0}\right)}.\label{eq:30}
\end{equation}
In particular, in the bubble D4-brane background, solution (\ref{eq:29})
implies $x^{4}|_{U\rightarrow\infty,U_{0}\rightarrow U_{KK}}=\beta_{4}/4$
and $x^{4\prime}|_{U_{0}=U_{KK}}=0$. Thus $x^{4}=\beta_{4}/4$ is
a solution to (\ref{eq:29}) representing D8- and $\overline{\mathrm{D8}}$-branes
are located at the antipodal points of $S^{1}$ while they are connected
at $U=U_{KK}$, because the size of $x^{4}$ shrinks to zero at $U=U_{KK}$.
On the other hand, in the black D4-brane background, if $U_{0}=U_{H}$,
(\ref{eq:30}) also implies a constant solution for $x^{4}$ while
the separation of the D8- and $\overline{\mathrm{D8}}$-branes could
be arbitrary but no more than $\beta_{4}/2$. For $U_{0}>U_{KK},U_{H}$,
the solutions (\ref{eq:28}) (\ref{eq:29}) represent D8- and $\overline{\mathrm{D8}}$-branes
are joined into a single brane at $U=U_{0}$. The configuration of
the D8- and $\overline{\mathrm{D8}}$-branes in bubble and black D4-brane
background is illustrated in Figure \ref{fig:2} and \ref{fig:3}.
 
\subsection{Gluon, quark and symmetries}

As the dual theory in the D4/D8 model is expected to be QCD in the
large $N_{c}$ limit, it is natural to identify the effective theory
on $N_{c}$ D4-branes below $M_{KK}$ to the color sector in QCD which
implies the gauge field $A_{\mu}^{\left(\mathrm{D4}\right)}$ on the
D4-branes can be interpreted as gluon in holography. The reason is
that the low-energy theory on $N_{c}$ D4-branes is $U\left(N_{c}\right)$
pure Yang-Mills theory and it has a SUGRA duality in the strong coupling
region in the large $N_{c}$ limit as it is discussed in Section 2.1.
We note that the Lorentz symmetry of the 10d spacetime breaks down
to $SO\left(1+4\right)\times SO\left(5\right)$ when a stack of D4-branes
is introduced. However the worldvolume symmetry of the D4-branes becomes
$SO\left(1+3\right)$ since the D4-branes are compactified on a circle
in the D4/D8 model. Furthermore, when the flavors as D8- and $\overline{\mathrm{D8}}$-branes
are introduced, it is possible to create chiral fermions in the low-energy
theory which can be obtained by analyzing the spectrum of $4-8$ or
$4-\bar{8}$ string in R-sector (Ramond-sector). Both the spectra
of $4-8$ and $4-\bar{8}$ string in R-sector contain spinors with
positive and negative chirality as the representations of the Lorentz
group $SO\left(1,3\right)$. Since the GSO (Gliozzi-Scherk-Olive)
projection will remove the spinor with one of the chiralities in string
theory, we can choose the spinor with positive and negative chirality
as the massless fermionic modes (denoted by $q_{L,R}$) of $4-8$
and $4-\bar{8}$ string respectively which accordingly can be identified
as the fundamental chiral quarks in the dual theory. We note that
these chirally fermionic fields are complex spinors since the $4-8$
and $4-\bar{8}$ strings have two orientations. And they are also
the fundamental representation of $U\left(N_{c}\right)$ and $U\left(N_{f}\right)$.
The massless modes and symmetries in the D4/D8 system are collected
in Table \ref{tab:2}. 
\begin{table}
\begin{centering}
\begin{tabular}{|c|c|c|c|c|}
\hline 
Fields & $U\left(N_{c}\right)$ & $SO\left(1,3\right)$ & $SO\left(5\right)$ & $U\left(N_{f}\right)_{L}\times U\left(N_{f}\right)_{R}$\tabularnewline
\hline 
\hline 
$A_{\mu}^{\left(\mathrm{D4}\right)}$ & \textbf{adj.} & \textbf{4} & \textbf{1} & \textbf{(1,1)}\tabularnewline
\hline 
$q_{L}$ & \textbf{fund.} & \textbf{2}\textbf{\scriptsize{}}\textsubscript{\textbf{\scriptsize{}+}} & \textbf{1} & \textbf{(fund.,1)}\tabularnewline
\hline 
$q_{R}$ & \textbf{fund.} & \textbf{2}\textbf{\scriptsize{}}\textsubscript{\textbf{\scriptsize{}-}} & \textbf{1} & \textbf{(1,fund.)}\tabularnewline
\hline 
$a_{4}$ & \textbf{1} & \textbf{1} & \textbf{1} & \textbf{(1,1)}\tabularnewline
\hline 
$\phi^{i}$ & \textbf{1} & \textbf{1} & \textbf{5} & \textbf{(1,1)}\tabularnewline
\hline 
\end{tabular}
\par\end{centering}
\caption{\label{tab:2} The fields in the D4/D8 model. Here \textbf{2}\textbf{\scriptsize{}}\protect\textsubscript{\textbf{\scriptsize{}+}},\textbf{
2}\textbf{\scriptsize{}}\protect\textsubscript{-} denote the positive
and negative chirality spinor representations of $SO\left(1,3\right)$.
$a_{4}$ and $\phi^{i}$ are the trace parts of the gauge field along
$x^{4}$ direction on the $N_{c}$ D4-branes and the transverse modes
of the $N_{c}$ D4-branes which are decouple to the gluon and fundamental
quarks in the low-energy theory.}
\end{table}

Due to the above holographic correspondence, the chirally symmetric
and broken phase in the dual theory can be identified respectively
to the disconnected and connected configuration of the $\mathrm{D8/\overline{D8}}$-branes.
It would be clear if we employ the configuration presented in Figure
\ref{fig:2} for example. The effective action for the gauge fields
$A_{\mu}^{\left(\mathrm{D4}\right)}$ and fundamental fermions $q_{L,R}$
on the $N_{c}$ D4-branes with $N_{f}$ $\mathrm{D8/\overline{D8}}$-branes
can be evaluated by expanding the DBI action which leads to,

\begin{align}
S= & \int_{\mathrm{D4}}d^{5}x\sqrt{-g}\left[\delta\left(x^{4}-X_{L}\right)q_{L}^{\dagger}\bar{\sigma}^{\mu}\left(i\nabla_{\mu}+A_{\mu}\right)q_{L}+\delta\left(x^{4}-X_{R}\right)q_{R}^{\dagger}\bar{\sigma}^{\mu}\left(i\nabla_{\mu}+A_{\mu}\right)q_{R}\right]\nonumber \\
 & -\frac{1}{4g_{\mathrm{YM}}^{2}}\int_{\mathrm{D4}}d^{5}x\sqrt{-g}\mathrm{Tr}F_{MN},\ M,N=0,1...4,
\end{align}
where $X_{L,R}$ denotes the intersection of the D4- and D8-branes,
D4- and $\overline{\mathrm{D8}}$-branes and we have omitted the notation
``D4'' in $A_{\mu}$. As all the fields depend on $\left\{ x^{\mu},x^{4}\right\} $,
$q_{L}$ is identified to be $q_{R}$ if $X_{L}=X_{R}$ which leads
to an action with single flavor symmetry $U\left(N_{f}\right)$. For
the connected configurations, we can therefore see the D8- and $\overline{\mathrm{D8}}$-branes
are separated at high energy ($U\rightarrow\infty$, $X_{L}\neq X_{R}$)
according to the solutions (\ref{eq:29}) and (\ref{eq:30}), which
leads to an approximated $U\left(N_{f}\right)_{L}\times U\left(N_{f}\right)_{R}$
chiral symmetry. However, at low energy $(U\rightarrow U_{0},X_{L}\rightarrow X_{R})$,
D8- and $\overline{\mathrm{D8}}$-branes are joined into a single
pair of D8-branes at $U=U_{0}$ ($X_{L}=X_{R}$) which implies the
$U\left(N_{f}\right)_{L}\times U\left(N_{f}\right)_{R}$ symmetry
breaks down to a single $U\left(N_{f}\right)$. This configuration
of $\mathrm{D8/\overline{D8}}$-branes provides nicely a geometric
interpretation of chiral symmetry in this model \cite{key-55}.

\subsection{Mesons on the flavor brane}

As meson is the bound state in the adjoint representation of the chiral
symmetry group, it is identified as the gauge field on the flavor
branes which is the massless mode excited by $8-8$ string\footnote{Massless mode excited by $\bar{8}-\bar{8}$ string is therefore identified
as anti-meson. }. The reason is that the gauge field excited by $8-8$ (and $\bar{8}-\bar{8}$)
is the generator of $U\left(N_{f}\right)_{L}$ (and $U\left(N_{f}\right)_{R}$).
Hence we consider the gauge field on the flavor branes with non-zero
components as $A_{M}=\left\{ A_{\mu}\left(x,z\right),A_{z}\left(x,z\right)\right\} ,\mu=0,1...3$
in the bubble D4-brane background (\ref{eq:16}). We note that, while
the supersymmetry on $N_{c}$ D4-branes breaks down by compactifying
$x^{4}$ on a circle, there is not any mechanism to break down the
supersymmetry on $\mathrm{D8/\overline{D8}}$-branes since $\mathrm{D8/\overline{D8}}$-branes
are vertical to $x^{4}$. Therefore the $8-8$ string is supersymmetric
leading to a super partner fermion $\Psi$ of the gauge field $A_{M}$
in the low-energy theory. And we will see in Appendix C, this supersymmetric
fermion is Majorana spinor which leads to the fermionic meson (mesino)
while they are absent in the realistic QCD. 

Nonetheless let us assume the supersymmetry on the flavor branes somehow
breaks down and the supersymmetric meson can be turned off in order
to continue the discussion about the QCD sector of this model. Since
the D8-branes are probes, the worldvolume gauge field is fluctuation.
Thus the effective action for $A_{M}$ can be obtained by expanding
(\ref{eq:22}) which, for Abelian case $N_{f}=1$, is

\begin{align}
S_{\mathrm{D8}}= & -T_{\mathrm{D8}}\int_{\mathrm{D8}}d^{9}xe^{-\phi}\sqrt{-\det\left[g_{ab}+\left(2\pi\alpha^{\prime}\right)F_{ab}\right]}\nonumber \\
= & -T_{\mathrm{D8}}\int_{\mathrm{D8}}d^{9}x\sqrt{-g}e^{-\phi}\left[1+\frac{1}{4}\left(2\pi\alpha^{\prime}\right)^{2}F_{MN}F^{MN}+\mathcal{O}\left(F^{4}\right)\right],\label{eq:32}
\end{align}
leading to the Yang-Mills (YM) action as,

\begin{align}
S_{\mathrm{YM}}= & -T_{\mathrm{D8}}\int_{\mathrm{D8}}d^{9}x\sqrt{-g}e^{-\phi}\frac{1}{4}\left(2\pi\alpha^{\prime}\right)^{2}F_{MN}F^{MN}\nonumber \\
= & -\frac{2}{3}R^{3/2}U_{KK}^{1/2}\left(2\pi\alpha^{\prime}\right)^{2}T_{\mathrm{D8}}\Omega_{4}\int d^{4}xdz\left(\frac{R^{3}}{4U}\eta^{\mu\rho}\eta^{\nu\sigma}F_{\mu\nu}F_{\rho\sigma}+\frac{9U^{3}}{8U_{KK}}\eta^{\mu\nu}F_{\mu z}F_{\nu z}\right)\nonumber \\
= & -\kappa\int d^{4}xdZ\left(\frac{1}{2}K^{-1/3}\eta^{\mu\rho}\eta^{\nu\sigma}F_{\mu\nu}F_{\rho\sigma}+KM_{KK}^{2}\eta^{\mu\nu}F_{\mu Z}F_{\nu Z}\right),\label{eq:33}
\end{align}
where we have used the Cartesian coordinates $z$ and dimensionless
$Z$ defined as,

\begin{equation}
U^{3}=U_{KK}^{3}+U_{KK}z^{2},Z=\frac{z}{U_{KK}},K\left(Z\right)=1+Z^{2}=\frac{U^{3}}{U_{KK}^{3}},\label{eq:34}
\end{equation}
and 
\begin{equation}
\kappa=\frac{1}{3}R^{9/2}U_{KK}^{1/2}\left(2\pi\alpha^{\prime}\right)^{2}T_{\mathrm{D8}}\Omega_{4}=\frac{\lambda N_{c}}{216\pi^{3}},\lambda=g_{\mathrm{YM}}^{2}N_{c},
\end{equation}
with the induced metric on the D8-branes as,

\begin{equation}
ds_{\mathrm{D8}}^{2}=\left(\frac{U}{R}\right)^{3/2}\eta_{\mu\nu}dx^{\mu}dx^{\nu}+\frac{4R^{3/2}U_{KK}}{9U^{5/2}}dz^{2}+R^{3/2}U^{1/2}d\Omega_{4}^{2}.
\end{equation}
We have employed the configuration that $\mathrm{D8/\overline{D8}}$-branes
are located at the antipodal points of $S^{1}$. Then, in order to
obtain a 4d mesonic action, let us assume that $A_{\mu}\left(x,z\right),A_{z}\left(x,z\right)$
can be expanded in terms of complete sets $\left\{ \psi_{n}\left(z\right),\phi_{n}\left(z\right)\right\} $
as,

\begin{equation}
A_{\mu}\left(x,z\right)=\sum_{n}B_{\mu}^{\left(n\right)}\left(x\right)\psi_{n}\left(z\right),A_{z}\left(x,z\right)=\sum_{n}\varphi^{\left(n\right)}\left(x\right)\phi_{n}\left(z\right),\label{eq:37}
\end{equation}
where $B_{\mu}^{\left(n\right)}\left(x\right),\varphi^{\left(n\right)}\left(x\right)$
refers to the 4d meson field. To obtain a finite action, the normalization
condition for $\psi_{n}\left(z\right)$ is chosen as,

\begin{equation}
2\kappa\int dZK^{-1/3}\psi_{n}\psi_{m}=\delta_{mn},\label{eq:38}
\end{equation}
with the eigen equation ($n\geq1$),

\begin{equation}
-K^{1/3}\partial_{Z}\left(K\partial_{Z}\psi_{n}\right)=\lambda_{n}\psi_{n},
\end{equation}
where $\lambda_{n}$ is the associated eigen value. In this sense,
the basic function $\phi_{n}\left(z\right)$ can be chosen as ($n\geq1$),

\begin{equation}
\phi_{0}=\frac{1}{\sqrt{2\pi\kappa}}\frac{1}{U_{KK}M_{KK}}\frac{1}{K},\phi_{n}=m_{n}^{-1}\partial_{Z}\psi_{n},m_{n}=\lambda_{n}M_{KK}.\label{eq:40}
\end{equation}
Keeping these in hand, impose (\ref{eq:38}) - (\ref{eq:40}) into
(\ref{eq:33}) then define the vector field $V_{\mu}^{\left(n\right)}\left(x\right)$
by a gauge transformation,

\begin{equation}
V_{\mu}^{\left(n\right)}=B_{\mu}^{\left(n\right)}-m_{n}^{-1}\partial_{\mu}\varphi^{\left(n\right)},
\end{equation}
the Yang-Mills action (\ref{eq:33}) reduces to a 4d effective action
for mesons as,

\begin{equation}
S_{\mathrm{YM}}=-\kappa\int d^{4}x\left(\frac{1}{2}\partial_{\mu}\varphi^{\left(0\right)}\partial^{\mu}\varphi^{\left(0\right)}+\sum_{n=1}^{\infty}\left[\frac{1}{4}F_{\mu\nu}^{\left(n\right)}F^{\left(n\right)\mu\nu}+\frac{1}{2}m_{n}^{2}V_{\mu}^{\left(n\right)}V^{\left(n\right)\mu}\right]\right),
\end{equation}
where $F_{\mu\nu}^{\left(n\right)}=\partial_{\mu}V_{\nu}^{\left(n\right)}-\partial_{\nu}V_{\mu}^{\left(n\right)}$.
Accordingly, $\varphi^{\left(0\right)}$ can be interpreted as pion
meson which is the Nambu-Goldstone boson associated to the chiral
symmetry breaking. By analyzing the parity, it turns out that $\varphi^{\left(0\right)}$
is pseudo-scalar field as it is expected.

The above discussion implicitly assumes that the gauge field $A_{M}$
and its field strength $F_{MN}$ should vanish at $\left|z\right|\rightarrow\infty$
in order to obtain a finite 4d mesonic action. However there is an
alternative gauge choice $A_{z}=0$ which is recognized as a gauge
transformation

\begin{equation}
A_{M}\rightarrow A_{M}-\partial_{M}\Lambda,
\end{equation}
to (\ref{eq:37}). Here $\Lambda$ is solved as,

\begin{equation}
\Lambda\left(x,z\right)=\varphi^{\left(0\right)}\left(x\right)\psi_{0}\left(z\right)+\sum_{n=1}^{\infty}m_{n}^{-1}\varphi^{\left(n\right)}\left(x\right)\psi_{n}\left(z\right),
\end{equation}
where

\begin{equation}
\psi_{0}\left(z\right)=\int dz\phi_{0}\left(z\right)=\frac{1}{\sqrt{2\pi\kappa}}\frac{1}{M_{KK}}\arctan\left(\frac{z}{U_{KK}}\right).
\end{equation}
Thus the components of $A_{M}$ under gauge condition $A_{z}=0$ becomes,

\begin{align}
A_{\mu}\left(x,z\right)= & -\partial_{\mu}\varphi^{\left(0\right)}\left(x\right)\psi_{0}\left(z\right)+\sum_{n=1}^{\infty}\left[B_{\mu}^{\left(n\right)}\left(x\right)-m_{n}^{-1}\partial_{\mu}\varphi^{\left(n\right)}\right]\psi_{n}\left(z\right)\nonumber \\
= & -\partial_{\mu}\varphi^{\left(0\right)}\left(x\right)\psi_{0}\left(z\right)+\sum_{n=1}^{\infty}V_{\mu}^{\left(n\right)}\left(x\right)\psi_{n}\left(z\right),\nonumber \\
A_{z}\left(x,z\right)= & 0.\label{eq:46}
\end{align}
In the region $\left|z\right|\rightarrow\infty$, the gauge potential
$A_{\mu}\left(x,z\right)\rightarrow\pm\sqrt{\frac{\pi}{8\kappa}}\frac{1}{M_{KK}}$
which implies the gauge field strength remains to be vanished and
the effective 4d action remains to be finite.

The above setup for mesons can be generalized into multi-flavor case
by taking into account the non-Abelian version of (\ref{eq:33}),

\begin{equation}
S_{\mathrm{YM}}^{\left(N_{f}\right)}=-\kappa N_{f}\int d^{4}xdZ\left[\frac{1}{2}K^{-1/3}\eta^{\mu\rho}\eta^{\nu\sigma}\mathrm{Tr}\left(F_{\mu\nu}F_{\rho\sigma}\right)+KM_{KK}^{2}\eta^{\mu\nu}\mathrm{Tr}\left(F_{\mu Z}F_{\nu Z}\right)\right],\label{eq:47}
\end{equation}
where $F_{MN}=\partial_{M}A_{N}-\partial_{N}A_{M}+\left[A_{M},A_{N}\right]$,
$M,N=0,1,2,3,z$ is the gauge field strength of $U\left(N_{f}\right)$.
As it has been discussed, in order to obtain a finite 4d action, the
gauge field strength must vanish in the limit $\left|z\right|\rightarrow\infty$.
Under the gauge condition $A_{z}=0$, $A_{\mu}$ must take asymptotically
a pure gauge configuration for $\left|z\right|\rightarrow\infty$
as,

\begin{equation}
A_{\mu}\left(x,z\right)|_{z\rightarrow\pm\infty}\rightarrow\xi_{\pm}\left(x\right)\partial_{\mu}\xi_{\pm}^{-1}\left(x\right).\label{eq:48}
\end{equation}
Compare this with (\ref{eq:46}), the gauge potential can be expanded
with boundary condition (\ref{eq:48}) as,

\begin{align}
A_{\mu}\left(x,z\right)= & \xi_{+}\left(x\right)\partial_{\mu}\xi_{+}^{-1}\left(x\right)\psi_{+}\left(z\right)+\xi_{-}\left(x\right)\partial_{\mu}\xi_{-}^{-1}\left(x\right)\psi_{-}\left(z\right)+\sum_{n=1}^{\infty}V_{\mu}^{\left(n\right)}\left(x\right)\psi_{n}\left(z\right)\nonumber \\
\equiv & \alpha_{\mu}\left(x\right)\hat{\psi}_{0}\left(z\right)+\beta_{\mu}\left(x\right)+\sum_{n=1}^{\infty}V_{\mu}^{\left(n\right)}\left(x\right)\psi_{n}\left(z\right),\label{eq:49}
\end{align}
with

\begin{align}
\psi_{\pm}= & \frac{1}{2}\pm\hat{\psi}_{0},\hat{\psi}_{0}=\frac{1}{\pi}\arctan\left(\frac{z}{U_{KK}}\right).\nonumber \\
\alpha_{\mu}\left(x\right)= & \xi_{+}\left(x\right)\partial_{\mu}\xi_{+}^{-1}\left(x\right)-\xi_{-}\left(x\right)\partial_{\mu}\xi_{-}^{-1}\left(x\right)\nonumber \\
\beta_{\mu}\left(x\right)= & \frac{1}{2}\left[\xi_{+}\left(x\right)\partial_{\mu}\xi_{+}^{-1}\left(x\right)+\xi_{-}\left(x\right)\partial_{\mu}\xi_{-}^{-1}\left(x\right)\right].
\end{align}
To obtain the chiral Lagrangian for mesons from the Yang-Mills action
(\ref{eq:47}), we identify the lowest vector meson field $V_{\mu}^{\left(n\right)}$
as the $\rho$ meson $V_{\mu}^{\left(1\right)}=\rho_{\mu}$ and choose
the following gauge conditions 

\begin{equation}
\xi_{+}^{-1}\left(x\right)=U\left(x\right),\xi_{-}\left(x\right)=1,\label{eq:51}
\end{equation}
or

\begin{equation}
\xi_{+}^{-1}\left(x\right)=\xi_{-}\left(x\right)=\exp\left[i\pi\left(x\right)/f_{\pi}\right].\label{eq:52}
\end{equation}
Inserting (\ref{eq:49}) into action (\ref{eq:47}) with the gauge
condition (\ref{eq:51}), the 4d Yang-Mills action (\ref{eq:47})
includes a part of Skyrme model \cite{key-60} as,

\begin{equation}
S_{\mathrm{YM}}^{\left(N_{f}\right)}=\int d^{4}x\left[\frac{f_{\pi}^{2}}{4}\mathrm{Tr}\left(U^{-1}\partial_{\mu}U\right)^{2}+\frac{1}{32e^{2}}\mathrm{Tr}\left(U^{-1}\partial_{\mu}U,U^{-1}\partial_{\mu}U\right)^{2}+...\right],\label{eq:53}
\end{equation}
where the coupling constants $f_{\pi},e$ are given as,

\begin{align}
f_{\pi}^{2} & =6R^{3/2}U_{KK}^{1/2}T_{\mathrm{D8}}\Omega_{4}\left(2\pi\alpha^{\prime}\right)^{2}\int dz\frac{U\left(z\right)^{3}}{U_{KK}}\left(\partial_{z}\psi_{+}\right)^{2}=\frac{\lambda M_{KK}N_{c}}{54\pi^{2}},\nonumber \\
e^{2} & =\left[\frac{32}{3}R^{3/2}U_{KK}^{1/2}T_{\mathrm{D8}}\Omega_{4}\left(2\pi\alpha^{\prime}\right)^{2}\int dz\frac{R^{3}}{U\left(z\right)}\psi_{+}^{2}\left(\psi_{+}-1\right)^{2}\right]^{-1}=\frac{27\pi^{7}}{2b\lambda N_{c}},\nonumber \\
b & =\int\frac{dZ}{\left(1+Z^{2}\right)^{1/3}}\left(\arctan Z+\frac{\pi}{2}\right)^{2}\left(\arctan Z-\frac{\pi}{2}\right)^{2}\simeq15.25.
\end{align}
And the interaction terms of $\pi,\rho$ mesons would be determined
by the Yang-Mills action (\ref{eq:47}) with the gauge condition (\ref{eq:52})
as,

\begin{align}
S_{\mathrm{YM}}^{\left(N_{f}\right)}= & \int d^{4}x\mathrm{Tr}\left[-\partial_{\mu}\pi\partial^{\mu}\pi+\frac{1}{2}W_{\mu\nu}^{2}+m_{v}^{2}\rho_{\mu}^{2}+a_{\rho^{3}}\left[\rho_{\mu},\rho_{\nu}\right]W^{\mu\nu}+a_{\rho\pi^{2}}\left[\partial_{\mu}\pi,\partial_{\nu}\pi\right]W^{\mu\nu}\right]\nonumber \\
 & +\mathcal{O}\left(\pi^{4},\rho_{\mu}^{4}\right),
\end{align}
where $W_{\mu\nu}$ is the gauge field strength of $\rho_{\mu}$ and
the associated coupling constants are given by,

\begin{align}
m_{\rho}^{2} & =\Lambda_{1}M_{KK}^{2},a_{\rho^{3}}=\frac{\left(6\pi\right)^{3/2}}{\sqrt{\lambda N_{c}}}b_{\rho^{3}},a_{\rho\pi^{2}}=\frac{\pi\left(3\pi\right)^{3/2}}{M_{KK}^{2}\sqrt{2\lambda N_{c}}}b_{\rho\pi^{2}},\nonumber \\
b_{\rho^{3}} & \simeq0.45,b_{\rho\pi^{2}}\simeq1.6,\Lambda_{1}\simeq0.67.
\end{align}
Therefore, we can reach to the meson tower or chiral Lagrangian starting
with the D8-brane action which provides description of meson in holography. 

To close this section, let us finally take a look at the WZ term presented
in action (\ref{eq:22}), which can be integrated as,

\begin{align}
S_{\mathrm{WZ}}^{\mathrm{D8}} & =\frac{g_{s}}{3!}\left(2\pi\alpha^{\prime}\right)^{3}T_{\mathrm{D8}}\int_{\mathrm{D8}}C_{3}\wedge\mathrm{Tr}\left(F\wedge F\wedge F\right)\nonumber \\
 & =\frac{g_{s}}{3!}\left(2\pi\alpha^{\prime}\right)^{3}T_{\mathrm{D8}}\int_{\mathrm{D8}}F_{4}\wedge\omega_{5}\nonumber \\
 & =\frac{N_{c}}{24\pi^{2}}\mathrm{Tr}\int_{M_{4}\times\mathbb{R}}\omega_{5}\left(A\right).\label{eq:57}
\end{align}
Here $F_{4}=dC_{3}$ is the Ramond-Ramond field given in (\ref{eq:14})
and $\omega_{5}\left(A\right)$ is the Chern-Simons (CS) 5-form given
as,

\begin{equation}
\omega_{5}\left(A\right)=AF^{2}-\frac{1}{2}A^{3}F+\frac{1}{10}A^{5},
\end{equation}
where $F=dA+\frac{1}{2}\left[A,A\right]$ is the gauge field strength.
Under the gauge transformation on the D8-brane,

\begin{equation}
\delta_{\Lambda}A=d\Lambda+\left[A,\Lambda\right],\delta_{\Lambda}F=\left[F,\Lambda\right],
\end{equation}
we can compute

\begin{align}
\mathrm{Tr}\left(\delta_{\Lambda}\omega_{5}\right) & =\mathrm{Tr}\left[d\Lambda d\left(AdA+\frac{1}{2}A^{3}\right)\right]\nonumber \\
 & =d\left(\mathrm{Tr}\left[\Lambda d\left(AdA+\frac{1}{2}A^{3}\right)\right]\right)\nonumber \\
 & \equiv d\chi_{4}\left(\Lambda,A\right).
\end{align}
Hence by defining the boundary value of the gauge potential as, 

\begin{equation}
A_{\mu}^{L}\left(x\right)=\lim_{z\rightarrow+\infty}A_{\mu}\left(x,z\right),A_{\mu}^{R}\left(x\right)=\lim_{z\rightarrow-\infty}A_{\mu}^{R}\left(x,z\right),
\end{equation}
the WZ term reduces to the chiral anomaly of $U\left(N_{f}\right)_{L}\times U\left(N_{f}\right)_{R}$
in QCD as,

\begin{align}
\delta_{\Lambda}S_{\mathrm{WZ}}^{\mathrm{D8}} & =\frac{N_{c}}{24\pi^{2}}\mathrm{Tr}\int_{M_{4}\times\mathbb{R}}\delta_{\Lambda}\omega_{5}\left(A\right)=\frac{N_{c}}{24\pi^{2}}\int_{M_{4}}\chi_{4}\left(\Lambda,A\right)\big|_{z\rightarrow-\infty}^{z\rightarrow+\infty}\nonumber \\
 & =\frac{N_{c}}{24\pi^{2}}\int_{M_{4}}\left[\chi_{4}\left(\Lambda_{L},A_{L}\right)-\chi_{4}\left(\Lambda_{R},A_{R}\right)\right].
\end{align}
And the formula for the chiral anomaly can also be expressed in the
gauge condition $A_{z}=0$ which is to perform the gauge transformation 

\begin{equation}
A^{g}=gAg^{-1}+gdg^{-1}.
\end{equation}
Then the CS 5-form reduces to

\begin{equation}
\omega_{5}\left(A^{g}\right)=\omega_{5}\left(A\right)+\frac{1}{10}\left(gdg^{-1}\right)^{5}+d\alpha_{4},
\end{equation}
where

\begin{align}
\alpha_{4} & =-\frac{1}{2}W_{1}\left(AdA+dAA+A^{3}\right)+\frac{1}{4}W_{1}AW_{1}A-W_{1}^{3}A,\nonumber \\
W_{1} & =dg^{-1}g.
\end{align}
Recall the formulas (\ref{eq:49}) in the gauge $A_{z}=0$ and choose
the gauge condition (\ref{eq:51}), the WZ term (\ref{eq:57}) can
be rewritten, after somewhat lengthy but straightforward calculations,
as

\begin{equation}
S_{\mathrm{WZ}}^{\mathrm{D8}}=-\frac{N_{c}}{48\pi^{2}}\mathrm{Tr}\int_{M_{4}}L_{\mathrm{WZW}}-\frac{N_{c}}{240\pi^{2}}\mathrm{Tr}\int_{M_{4}\times\mathbb{R}}\left(gdg^{-1}\right)^{5},
\end{equation}
where $L_{\mathrm{WZW}}$ is the Wess-Zumino-Witten (WZW) term in
\cite{key-61,key-62} given as,

\begin{align}
L_{\mathrm{WZW}}= & \left[\left(A_{R}dA_{R}+dA_{R}A_{R}+A_{R}^{3}\right)\left(U^{-1}A_{L}U+U^{-1}dU\right)-\mathrm{p.c.}\right]\nonumber \\
 & +\left(dA_{R}dU^{-1}A_{L}U-\mathrm{p.c.}\right)+\left[A_{R}\left(dU^{-1}U\right)^{3}-\mathrm{p.c.}\right]\nonumber \\
 & +\frac{1}{2}\left[\left(A_{R}dU^{-1}U\right)^{2}-\mathrm{p.c.}\right]+\left[UA_{R}U^{-1}A_{L}dUdU^{-1}-\mathrm{p.c.}\right]\nonumber \\
 & -\left[A_{R}dU^{-1}UA_{R}U^{-1}A_{L}U-\mathrm{p.c.}\right]+\frac{1}{2}\left(A_{R}U^{-1}A_{L}U\right)^{2}.
\end{align}
Note that ``$\mathrm{p.c.}$'' refers to the terms by exchanging
$A_{L}\leftrightarrow A_{R},U\leftrightarrow U^{-1}$. And one can
further work out the couplings to the vector mesons by using (\ref{eq:49})
with $A_{z}=0$.

\subsection{The wrapped D4-brane and baryon vertex}

In the $SU\left(N_{c}\right)$ gauge theory, a baryon vertex connects
to $N_{c}$ external fundamental quarks with the color wave function
combined together by an $N_{c}$-th antisymmetric tensor of $SU\left(N_{c}\right)$
group. Accordingly the baryon vertex in gauge-gravity duality is recognized
as a D-brane wrapped on the internal sphere \cite{key-25,key-63}.
To clarify this briefly, let us first recall the baryon vertex in
the holographic duality between $\mathcal{N}=4$ super Yang-Mills
theory and IIB string theory on $\mathrm{AdS}_{5}\times S^{5}$. As
the fundamental quarks in the super Yang-Mills theory is created by
the $N_{c}$ elementary superstrings in $\mathrm{AdS}_{5}\times S^{5}$,
it is represented by the endpoints of elementary superstrings at the
boundary of $\mathrm{AdS}_{5}$. Hence we need $N_{c}$ elementary
superstrings with same orientation somehow to terminate in the $\mathrm{AdS}_{5}\times S^{5}$.
On the other hand, since the baryon current must be conserved, one
needs to find a source to cancel the $N_{c}$ charges (baryon charge)
contributed by the $N_{c}$ elementary superstrings. To figure out
these problems and work out a baryon vertex, a probe D5-brane wrapped
on $S^{5}$ provides us a good answer. The $N_{c}$ elementary superstrings
end on the D5-brane contributes $N_{c}$ to the D5-brane, however
the WZ action for such a wrapped D5-brane, 

\begin{equation}
S_{\mathrm{WZ}}^{\left(\mathrm{D5}\right)}\sim\int_{S^{5}\times\mathbb{R}}C_{4}F\sim\int_{S^{5}\times\mathbb{R}}F_{5}A\sim N_{c}\int_{\mathbb{R}}A,
\end{equation}
(where $F_{5}=dC_{4}$ is the Ramond-Ramond field strength) nicely
provides $N_{c}$ charges to cancel the charges given by $N_{c}$
elementary superstrings\footnote{The sign of the $N_{c}$ charge depends on the orientation of the
elementary superstrings and D5-branes.}. Therefore the $A$ current would be conserved, which implies the
D5-brane is a baryon vertex.

The construction of the baryon vertex can also been employed in the
D4/D8 model which is identified as a D4-brane\footnote{To distinguish the D4-branes as the baryon vertex from the $N_{c}$
D4-branes, we denote the baryon vertex as $\mathrm{D4}^{\prime}$-brane
in the rest of this paper.} wrapped on $S^{4}$ with $N_{c}$ elementary superstrings ending
on it. A remarkable point here is that the $\mathrm{D4}^{\prime}$-branes
can be described equivalently by the instanton configuration of the
gauge field on the D8-branes \cite{key-64,key-65}. To see this clearly,
let us consider a $\mathrm{D}_{p}$-brane with its worldvolume gauge
field strength $F$. According to (\ref{eq:20}), The WZ action for
such a $\mathrm{D}_{p}$-brane includes a term as a source,

\begin{equation}
S_{\mathrm{WZ}}\sim g_{s}T_{D_{p}}\left(2\pi\alpha^{\prime}\right)^{2}\int C_{p-3}\mathrm{Tr}F^{2}.\label{eq:69}
\end{equation}
For the single instanton configuration, the gauge field strength can
be integrated as,

\begin{equation}
\mathrm{Tr}\int F^{2}=8\pi^{2}.\label{eq:70}
\end{equation}
Hence (\ref{eq:69}) can be written as,

\[
g_{s}T_{D_{p}}\left(2\pi\alpha^{\prime}\right)^{2}\int C_{p-3}\mathrm{Tr}F^{2}=g_{s}T_{D_{p-4}}\int C_{p-3},
\]
giving rise to a same source included by a $\mathrm{D}_{p-4}$-brane.
Accordingly, we obtain a simple and interesting conclusion here, that
is the instanton in the $\mathrm{D}_{p}$-brane is the same object
as a $\mathrm{D}_{p-4}$-brane inside it. 

Let us return to the D4/D8 model, so it implies the $\mathrm{D4}^{\prime}$-branes
as the baryon vertex are equivalent to the instanton in the $\mathrm{D8/\overline{D8}}$-branes.
For multiple instantons, (\ref{eq:70}) is replaced by,

\begin{equation}
\mathrm{Tr}\int F^{2}=8\pi^{2}n,\label{eq:71}
\end{equation}
where $n$ refers to the instanton number. Inserting the instanton
configuration of the gauge field denoted as $A_{\mathrm{cl}}$ with
a $U\left(N_{f}\right)_{V}$ fluctuation $\tilde{A}$ into the WZ
action (\ref{eq:57}) of D8-brane, it reduces to

\begin{equation}
\frac{N_{c}}{8\pi^{2}}\int\tilde{A}\mathrm{Tr}F^{2}=nN_{c}\int\tilde{A},
\end{equation}
which implies the instantons take $U\left(1\right)_{V}$ charge $nN_{c}$.
Since the baryon number is defined as $1/N_{c}$ times the charge
of the diagonal $U\left(1\right)_{V}$ subgroup of the $U\left(N_{f}\right)_{V}$
symmetry, it is obvious that the instanton number is equivalent to
the baryon number in this holographic system. 

Moreover, when (\ref{eq:71}) is integrated out to be a Chern-Simons
3-form $\omega_{3}$ as,

\begin{equation}
\int_{M_{4}}\mathrm{Tr}F^{2}=\int_{\partial M_{4}}\mathrm{Tr}\omega_{3}=\int\mathrm{Tr}\left(AF-\frac{1}{3}A^{3}\right),
\end{equation}
the baryon number can be obtained as,

\begin{equation}
n=\frac{1}{8\pi^{2}}\mathrm{Tr}\int_{\mathbb{R}^{4}}F^{2}=\mathrm{Tr}\int_{\mathbb{R}^{3}}\omega_{3}|_{z\rightarrow-\infty}^{z\rightarrow+\infty}=-\frac{1}{24\pi^{2}}\mathrm{Tr}\int_{\mathbb{R}^{3}}\left(U^{-1}dU\right)^{3},\label{eq:74}
\end{equation}
where we have imposed the similar boundary condition as it is given
in (\ref{eq:51}),

\begin{equation}
\lim_{z\rightarrow+\infty}A_{\mu}\left(x^{\mu},z\right)=U^{-1}\left(x^{i}\right)\partial_{i}U\left(x^{i}\right),\lim_{z\rightarrow-\infty}A_{\mu}\left(x^{\mu},z\right)=0.
\end{equation}
The (\ref{eq:74}) gives the winding number of $U$ which means the
homotopy is $\pi_{3}\left[U\left(N_{f}\right)\right]\simeq\mathbb{Z}$.
And it agrees with the baryon number charge in the Skyrme model \cite{key-61,key-66}. 

\begin{figure}
\begin{centering}
\includegraphics[scale=0.3]{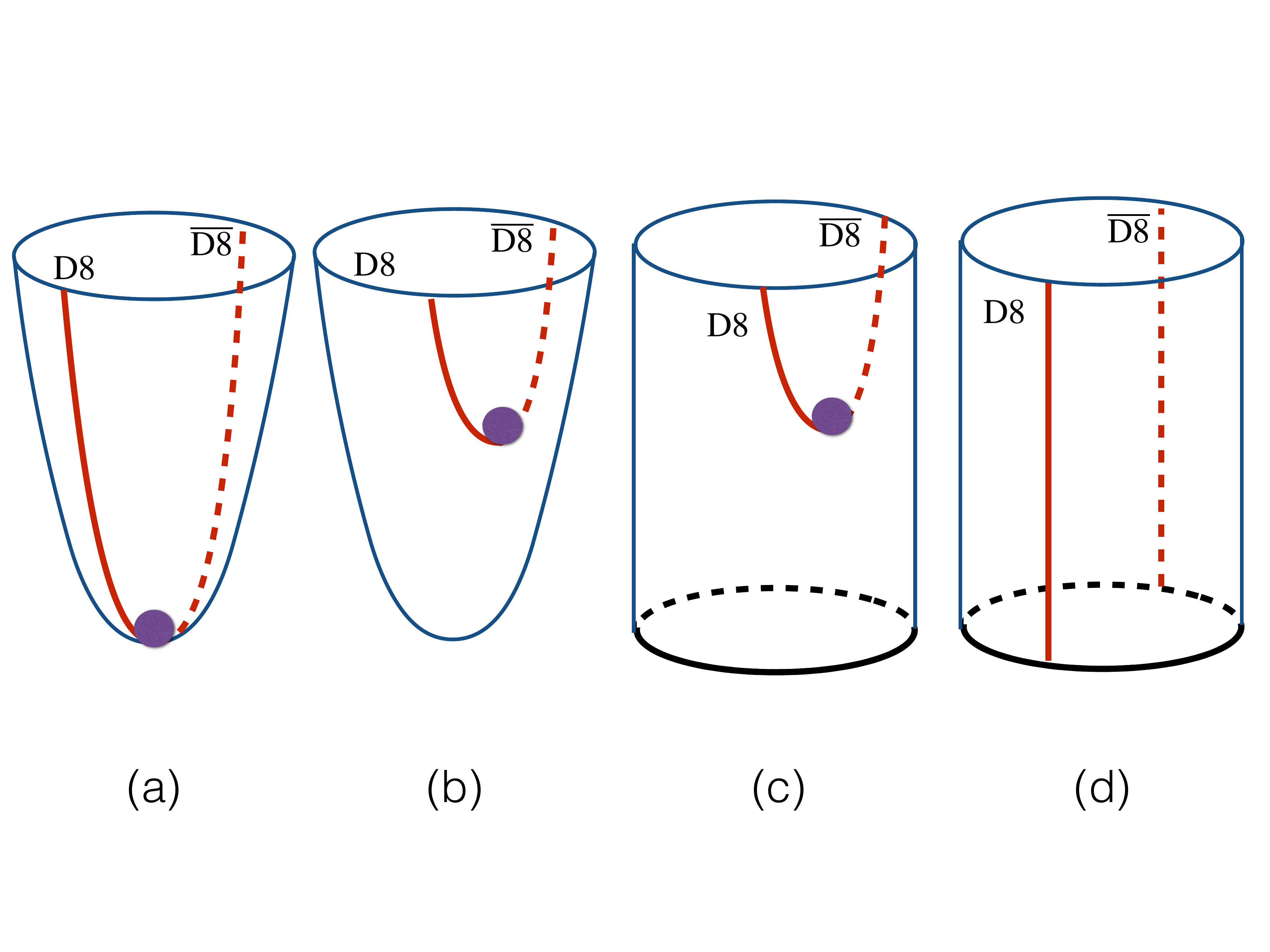}
\par\end{centering}
\caption{\label{fig:4} Various D-brane configurations in the D4/D8 model.
Purple refers to the baryon vertex ($\mathrm{D4}^{\prime}$-brane)
locates at the stable position. When the $\mathrm{D8/\overline{D8}}$-branes
are connected as (a), (b), (c), the baryon vertex can locate stably
at the connected position inside the $\mathrm{D8/\overline{D8}}$-branes.
There is not a stable position for baryon vertex if $\mathrm{D8/\overline{D8}}$-branes
are parallel as (d).}
\end{figure}

The baryon mass $m_{B}$ can be roughly obtained by evaluating the
energy carried by the $\mathrm{D4}^{\prime}$-branes which can be
read from its DBI action as,

\begin{align}
S_{\mathrm{D4}^{\prime}}= & -T_{\mathrm{D4}}\int dx^{0}d\Omega_{4}e^{-\phi}\sqrt{-g_{00}g_{S^{4}}}\nonumber \\
= & -\frac{1}{27}M_{KK}\lambda N_{c}^{2}\int_{\mathbb{R}}dx^{0},\nonumber \\
= & -m_{B}\int_{\mathbb{R}}dx^{0}\label{eq:76}
\end{align}
where the bubble D4-brane background has been chosen for the confined
property of baryon and $g_{S^{4}}$ refers to the metric on $S^{4}$
presented in (\ref{eq:16}). This formula illustrates a stable position
of the baryonic $\mathrm{D4}^{\prime}$-brane by minimizing its energy,
which is $U=U_{KK}$ since the bubble geometry shrinks at $U=U_{KK}$.
In the black D4-brane, one can follow the same formula (\ref{eq:76})
to evaluate the baryon mass. However, if the baryonic $\mathrm{D4}^{\prime}$-brane
is the only probe brane, it can not stay at $U=U_{H}$ stably in the
black D4-brane background since gravity will pull it into the horizon.
In this sense, the baryon vertex exists in the bubble D4-brane background
only and it is consistent with its property of confinement. When the
probe $\mathrm{D8/\overline{D8}}$-branes are embedded into the bulk
geometry, due to the balance condition, the baryonic $\mathrm{D4}^{\prime}$-brane
can be restricted inside the D8-branes if $\mathrm{D8/\overline{D8}}$-branes
are connected, as it is displayed in Figure \ref{fig:4}\footnote{Authors of \cite{key-67} claim that, according to the numerical calculation,
there is not a wrapped configuration for the baryonic $\mathrm{D4}^{\prime}$-brane
in the black D4-brane background thus this background may correspond
to the deconfinement phase of QCD. We note that this issue is not
figured out even if the baryon vertex is introduced into the black
D4-brane background.}. 
 Therefore it can be described equivalently by the instanton configuration
on the D8-branes.

To obtain the baryon mass or baryon spectrum in this model, it is
worth searching for an exact instanton solution for the gauge field
on the D8-branes. As baryon lives in the low-energy region of QCD,
we may find an approximated solution for the instanton configuration
in the strong coupling limit i.e. $\lambda\rightarrow\infty$. To
this goal, let us take a look at the gauge field on the D8-branes
whose dynamic is described by the Yang-Mills action (\ref{eq:33})
plus the Chern-Simons action presented in (\ref{eq:57}) \footnote{As the size of instanton takes order of $\lambda^{-1/2}$, it may
lead to a puzzle if the Yang-Mills action is taken into account only
because the high order derivatives in the DBI action contributes more
importantly. However, according to the holographic duality, taking
the near-horizon limit requires $\alpha^{\prime}\rightarrow0$ which
implies the Yang-Mills action dominates the dynamics in the DBI action.
This puzzle is not figured out completely in \cite{key-26,key-27}
and we may additionally set $\alpha^{\prime}\lambda\rightarrow0$
when Yang-Mills action is taken into account only in this setup.}. Since the size of instanton is of order $\lambda^{-1/2}$, it would
be convenient to rescale the coordinate $\left\{ x^{0},x^{i},z\right\} $
and the gauge potential $A$ as,

\begin{align}
x^{M}\rightarrow\lambda^{-1/2}x^{M},\  & x^{0}\rightarrow x^{0}\nonumber \\
A_{M}\rightarrow\lambda^{1/2}A_{M},\  & A_{0}\rightarrow A_{0}\nonumber \\
F_{MN}\rightarrow\lambda F_{MN},\  & F_{0M}\rightarrow\lambda^{1/2}F_{0M},\label{eq:77}
\end{align}
where $M,N=1,2,3,z$. In the large $\lambda$ limit, the Yang-Mills
action (\ref{eq:33}) can be expanded as,

\begin{align}
S_{\mathrm{YM}}= & -\frac{\kappa}{\lambda}\int d^{4}xdz\mathrm{Tr}\left[\frac{\lambda}{2}F_{MN}^{2}+\left(-\frac{z^{2}}{6}F_{ij}^{2}+z^{2}F_{iz}^{2}-F_{0M}^{2}\right)+\mathcal{O}\left(\lambda^{-1}\right)\right]\nonumber \\
 & -\frac{\kappa}{2\lambda}\int d^{4}xdz\left[\frac{\lambda}{2}\hat{F}_{MN}^{2}+\left(-\frac{z^{2}}{6}\hat{F}_{ij}^{2}+z^{2}\hat{F}_{iz}^{2}-\hat{F}_{0M}^{2}\right)+\mathcal{O}\left(\lambda^{-1}\right)\right],\label{eq:78}
\end{align}
while the Chern-Simons action (\ref{eq:57}) remains under the rescaling
(\ref{eq:77}). We have employed the D-brane configuration presented
in (a) of Figure \ref{fig:4}. The $U\left(N_{f}\right)$ group is
decomposed as $U\left(N_{f}\right)\simeq U\left(1\right)\times SU\left(N_{f}\right)$
and correspondingly its generator is decomposed as,

\begin{equation}
A+\frac{1}{\sqrt{2N_{f}}}\hat{A}=A^{a}t^{a}+\frac{1}{\sqrt{2N_{f}}}\hat{A},
\end{equation}
where $\hat{A},A$ refers to the generators of $U\left(1\right)$,
$SU\left(N_{f}\right)$ respectively and $t^{a}$ ($a=1,2...N_{f}^{2}-1$)
are the normalized bases satisfying

\begin{equation}
\mathrm{Tr}\left(t^{a}t^{b}\right)=\frac{1}{2}\delta^{ab}.
\end{equation}
So the Chern-Simons action (\ref{eq:57}) can be derived as, 

\begin{align}
S_{\mathrm{CS}}= & \frac{N_{c}}{24\pi^{2}}\int\omega_{5}^{SU\left(N_{f}\right)}\left(A\right)+\frac{N_{c}}{24\pi^{2}}\sqrt{\frac{2}{N_{f}}}\epsilon_{MNPQ}\int d^{4}xdz\bigg[\frac{3}{8}\hat{A}_{0}\mathrm{Tr}\left(F_{MN}F_{PQ}\right)\nonumber \\
 & -\frac{3}{2}\hat{A}_{M}\mathrm{Tr}\left(\partial_{0}A_{N}F_{PQ}\right)+\frac{3}{4}\hat{F}_{MN}\mathrm{Tr}\left(A_{0}F_{PQ}\right)+\frac{1}{16}\hat{A}_{0}\hat{F}_{MN}\hat{F}_{PQ}-\frac{1}{4}\hat{A}_{M}\hat{F}_{0N}\hat{F}_{PQ}\nonumber \\
 & +\left(\mathrm{total\ derivatives}\right)\bigg].\label{eq:81}
\end{align}
And the equation of motion for $\hat{A},A$ can be derived by varying
action (\ref{eq:78}) plus (\ref{eq:81}) which allows an instanton
solution as,

\begin{equation}
A_{M}^{\mathrm{cl}}=-if\left(\xi\right)g\left(x\right)\partial_{M}g^{-1},\label{eq:82}
\end{equation}
where

\begin{align}
f\left(\xi\right) & =\frac{\xi^{2}}{\xi^{2}+\rho^{2}},\xi=\sqrt{\left(x^{M}-X^{M}\right)^{2}},\nonumber \\
g\left(x\right) & =\left(\begin{array}{cc}
g^{SU\left(2\right)}\left(x\right) & 0\\
0 & \mathbf{1}_{N_{f}-2}
\end{array}\right),g^{SU\left(2\right)}\left(x\right)=\frac{1}{\xi}\left[\left(z-Z\right)\mathbf{1}_{2}-i\left(x^{i}-X^{i}\right)\tau^{i}\right].\label{eq:83}
\end{align}
Here $\mathbf{1}_{N}$ is an $N\times N$ identity matrix and $\tau^{i}$'s
are the Pauli matrices. The position and the size of the instanton
are denoted by the constants $X^{M}=\left\{ X^{i},Z\right\} $ and
$\rho$ respectively to which have been rescaled as (\ref{eq:77}).
The configuration (\ref{eq:82}) (\ref{eq:83}) is the Belavin--Polyakov--Schwarz--Tyupkin
(BPST) solution embedding into $SU\left(N_{f}\right)$ which represents
the $SU\left(2\right)$ Euclidean instanton, and one may verify this
solution satisfies (\ref{eq:70}). Then the $U\left(1\right)$ part
of the gauge field is solved as,

\begin{equation}
\hat{A}_{0}^{\mathrm{cl}}=\sqrt{\frac{2}{N_{f}}}\frac{1}{8\pi^{2}a}\frac{1}{\xi^{2}}\left[1-\frac{\rho^{4}}{\left(\xi^{2}+\rho^{2}\right)^{2}}\right],\ \hat{A}_{M}^{\mathrm{cl}}=0.
\end{equation}
which leads to a non-zero $A_{0}$ as,

\begin{equation}
A_{0}^{\mathrm{cl}}=\frac{1}{16\pi^{2}a}\frac{1}{\xi^{2}}\left[1-\frac{\rho^{4}}{\left(\xi^{2}+\rho^{2}\right)^{2}}\right]\left(\mathcal{P}_{2}-\frac{2}{N_{f}}\mathbf{1}_{N_{f}}\right),\label{eq:85}
\end{equation}
where $\mathcal{P}_{2}$ is an $N_{f}\times N_{f}$ matrix $\mathcal{P}_{2}=\mathrm{diag}\left(1,1,0,...0\right)$.

Keeping these in hand, it is possible to evaluate the classical baryon
mass through the soliton mass $M$ with respect to the D8-brane action
as $S\left[A^{\mathrm{cl}}\right]=-\int Mdt$ which is obtained as,

\begin{equation}
M=8\pi^{2}\kappa+\frac{8\pi^{2}\kappa}{\lambda}\left(\frac{\rho^{2}}{6}+\frac{Z^{2}}{3}+\frac{1}{320\pi^{4}a^{2}\rho^{2}}\right),\label{eq:86}
\end{equation}
by inserting (\ref{eq:82}) - (\ref{eq:85}) into action (\ref{eq:78})
plus (\ref{eq:81}). On the other hand, since the low-energy effective
theory on the D8-branes can reduce to Skyrme model, we can further
employ the idea in the Skyrme model of baryon, which is identified
as the excitation of the collective modes, in order to search for
the baryon spectrum. The classically effective Lagrangian for baryon
describing the dynamics of the collective coordinates $\mathcal{X}^{\alpha}$
in the moduli space by the one instanton solution. which refers to
the world line element with a baryonic potential $U\left(\mathcal{X}^{\alpha}\right)$
in the moduli space 

\begin{equation}
L\left(\mathcal{X}^{\alpha}\right)=\frac{m_{X}}{2}\mathcal{G}_{\alpha\beta}\dot{\mathcal{X}}^{\alpha}\dot{\mathcal{X}}^{\beta}-U\left(\mathcal{X}^{\alpha}\right)+\mathcal{O}\left(\lambda^{-1}\right),\label{eq:87}
\end{equation}
where ``$\cdot$'' refers to the derivative respected to time, the
collective coordinates $\mathcal{X}^{\alpha}$ denotes to $\left\{ X^{M},\rho,y^{a}\right\} $
and $W=y^{a}t^{a}$ is the $SU\left(N_{f}\right)$ orientation of
the instanton. The potential $U\left(\mathcal{X}^{\alpha}\right)$
is the classical soliton mass given by $S\left[A^{\mathrm{cl}}\right]=-\int dtU\left(\mathcal{X}^{\alpha}\right)$.
The basic idea to quantize the classical Lagrangian (\ref{eq:87})
is to move slowly the classical soliton so that the collective coordinates
$\mathcal{X}^{\alpha}$ are promoted to be time-dependent \cite{key-68}.
Approximately, the $SU\left(N_{f}\right)$ gauge field potential is
becomes time-dependent by a gauge transformation,

\begin{align}
A_{M}\left(t,x\right) & =W\left(t\right)A_{M}^{\mathrm{cl}}\left(x,\mathcal{X}^{\alpha}\right)W\left(t\right)^{-1}-iW\left(t\right)\partial_{M}W\left(t\right)^{-1},\nonumber \\
A_{0}\left(t,x\right) & =W\left(t\right)A_{0}^{\mathrm{cl}}\left(x,\mathcal{X}^{\alpha}\right)W\left(t\right)^{-1}+\Delta A_{0},\nonumber \\
\hat{A}_{M}\left(t,x\right) & =0,\ \hat{A}_{0}\left(t,x\right)=\hat{A}_{0}^{\mathrm{cl}}\left(t,x\right),
\end{align}
and the associated field strength becomes,

\begin{align}
F_{MN}= & W\left(t\right)F_{MN}^{\mathrm{cl}}W\left(t\right)^{-1},\nonumber \\
F_{0M}= & W\left(t\right)\left(\dot{\mathcal{X}}^{\alpha}\frac{\partial}{\partial\mathcal{X}^{\alpha}}A_{M}^{\mathrm{cl}}-D_{M}^{\mathrm{cl}}\Sigma-D_{M}^{\mathrm{cl}}A_{0}^{\mathrm{cl}}\right)W\left(t\right)^{-1},\nonumber \\
\hat{F}_{0M}= & \hat{F}_{0M}^{\mathrm{cl}},\ \hat{F}_{MN}=\hat{F}_{MN}^{\mathrm{cl}},
\end{align}
where 

\begin{align}
D_{M}^{\mathrm{cl}}A_{0} & =\partial_{M}A_{0}+i\left[A_{M}^{\mathrm{cl}},A_{0}\right],\nonumber \\
\Sigma & =W\left(t\right)^{-1}\Delta A_{0}W\left(t\right)-i\dot{W}\left(t\right)^{-1}W\left(t\right).
\end{align}
 $\Delta A_{0}$ must be determined by its equation of motion as,

\begin{equation}
D_{M}^{\mathrm{cl}}\left(\dot{X}^{N}\frac{\partial}{\partial X^{N}}A_{M}^{\mathrm{cl}}+\dot{\rho}\frac{\partial}{\partial\rho}A_{M}^{\mathrm{cl}}-D_{M}^{\mathrm{cl}}\Sigma\right)=0.
\end{equation}
While for generic $N_{f}$, the exact solution for $\Sigma$ may be
out of reach, the solution with $N_{f}=2,3$ is collected respectively
in \cite{key-26,key-27}. Accordingly the Lagrangian of the collective
modes is given by

\begin{align}
S\left[\mathcal{A}\right]-S\left[\mathcal{A}^{\mathrm{cl}}\right] & =\int dt\left[L_{\mathrm{YM}}\left(\mathcal{X}^{\alpha}\right)+L_{\mathrm{CS}}\left(\mathcal{X}^{\alpha}\right)\right]=\int dtL\left(\mathcal{X}^{\alpha}\right)\nonumber \\
S_{\mathrm{YM}}\left[\mathcal{A}\right]-S_{\mathrm{YM}}\left[\mathcal{A}^{\mathrm{cl}}\right] & =\int dtL_{\mathrm{YM}}\left(\mathcal{X}^{\alpha}\right),\nonumber \\
S_{\mathrm{CS}}\left[\mathcal{A}\right]-S_{\mathrm{CS}}\left[\mathcal{A}^{\mathrm{cl}}\right] & =\int dtL_{\mathrm{CS}}\left(\mathcal{X}^{\alpha}\right),
\end{align}
which leads to,

\begin{align}
L\left(\mathcal{X}^{\alpha}\right) & =-M+aN_{c}\mathrm{Tr}\int d^{3}xdz\left(\dot{X}^{N}F_{MN}^{\mathrm{cl}}+\dot{\rho}\frac{\partial}{\partial\rho}A_{M}-\dot{X}^{N}D_{M}^{\mathrm{cl}}A_{N}^{\mathrm{cl}}-D_{M}^{\mathrm{cl}}\Sigma\right)^{2}+\mathcal{O}\left(\lambda^{-1}\right)\nonumber \\
 & =-M_{0}+\frac{m_{X}}{2}\delta_{ij}\dot{X}^{i}\dot{X}^{j}+L_{Z}+L_{\rho}+L_{\rho W}+\mathcal{O}\left(\lambda^{-1}\right),\label{eq:93}
\end{align}
where

\begin{align}
L_{Z}= & \frac{m_{Z}}{2}\left(\dot{Z}^{2}-\omega_{Z}^{2}Z^{2}\right),L_{\rho}=\frac{m_{\rho}}{2}\left(\dot{\rho}-\omega_{\rho}^{2}\rho^{2}\right)-\frac{K}{m_{\rho}\rho^{2}},\nonumber \\
L_{\rho W}= & \frac{m_{\rho}\rho^{2}}{2}\sum_{a}C_{a}\left[\mathrm{Tr}\left(-iW^{-1}\dot{W}t^{a}\right)\right]^{2},a=1,2...N_{f}^{2}-1
\end{align}
and

\begin{equation}
M_{0}=8\pi^{2}\kappa,m_{X}=m_{Z}=\frac{m_{\rho}}{2}=8\pi^{2}\kappa\lambda^{-1},K=\frac{2}{5}N_{c}^{2},\omega_{Z}^{2}=4\omega_{\rho}^{2}=\frac{2}{3}.
\end{equation}
Here we note that the formulas are in the unit of $M_{KK}=1$, $C_{a}$'s
are constants dependent on the $SU\left(N_{f}\right)$ instanton solution
and the metric of the moduli space can be further obtained by comparing
(\ref{eq:93}) with (\ref{eq:87}). For example, we have $C_{1,2,3}=1$
for $N_{f}=2$, and $C_{1,2,3}=1,C_{4,5,6,7}=1/2,C_{8}=0$ for $N_{f}=3$.
Afterwards, the baryon states can be obtained by quantizing the Lagrangian
(\ref{eq:93}) that is to replace the derivative term by $\dot{X}^{\alpha}\rightarrow-\frac{i}{m_{X}}\partial_{\alpha}$
straightforwardly. Hence the quantized Hamiltonian associated to (\ref{eq:93})
is collected as\footnote{We note that for generic $N_{f}$, the baryonic Hamiltonian must be
supported by additional constraint according to \cite{key-69} although
it may not change the baryon spectrum. },

\begin{align}
H= & M_{0}+H_{Z}+H_{\rho}+H_{\rho W},\nonumber \\
H_{Z}= & -\frac{1}{2m_{Z}}\partial_{Z}^{2}+\frac{1}{2}m_{Z}\omega_{Z}^{2}Z^{2},\nonumber \\
H_{\rho}= & -\frac{1}{2m_{\rho}}\frac{1}{\rho^{\eta}}\partial_{\rho}\left(\rho^{\eta}\partial_{\rho}\right)+\frac{1}{2}m_{\rho}\omega_{\rho}^{2}\rho^{2}+\frac{K}{m_{\rho}\rho^{2}},\nonumber \\
H_{\rho W}= & \frac{m_{\rho}\rho^{2}}{2}\sum_{a}C_{a}\left[\mathrm{Tr}\left(-iW^{-1}\dot{W}t^{a}\right)\right]^{2}=\frac{2}{m_{\rho}\rho^{2}}\sum_{a}C_{a}\left(J^{a}\right)^{2},\label{eq:96}
\end{align}
where $\eta=N_{f}^{2}-1$ and $J^{a}$'s are the operators of the
angular momentum of $SU\left(N_{f}\right)$. The baryon spectrum can
be finally obtained by evaluating the eigen values of the Hamiltonian
(\ref{eq:96}) which takes a fortunately analytical formula \cite{key-26,key-27}.

\subsection{Gravitational wave as glueball}

According to AdS/CFT and gauge-gravity duality \cite{key-28,key-29,key-30,key-31},
the glueball operator $\hat{O}$ can be identified as the source of
gravitational fluctuation in bulk since it is included in energy-momentum
tensor of Yang-Mills theory in the dual theory as $\hat{O}\sim F_{\mu\nu}F^{\mu\nu}\sim T_{\mu\nu}$
coupling to metric. Thus due to the confined property of glueball,
we can choose the bubble D4-brane background (\ref{eq:16}) compactified
on a circle with gravitational fluctuation in order to investigate
glueball in holography. 

The dual field to the glueball operator is the gravitational fluctuation
coupling the energy-momentum which therefore refers to the gravitational
polarization. By employing the relation of 11d M-theory and 10d type
IIA string theory in Section 2.1, it would be convenient to find the
gravitational polarization in 11d theory. For example, the lowest
exotic scalar glueball with quantum number $J^{CP}=0^{++}$ corresponds
to the exotic polarizations of the bulk gravitational polarization
and its 11d components are given as ($\mu,\nu=0,1,2,3$),

\begin{eqnarray}
\delta g_{44} & = & -\frac{r^{2}}{L^{2}}f\left(r\right)H_{E}\left(r\right)G_{E}\left(x\right),\nonumber \\
\delta g_{\mu\nu} & = & \frac{r^{2}}{L^{2}}H_{E}\left(r\right)\left[\frac{1}{4}\eta_{\mu\nu}-\left(\frac{1}{4}+\frac{3r_{KK}^{6}}{5r^{6}-2r_{KK}^{6}}\right)\frac{\partial_{\mu}\partial_{\nu}}{M_{E}^{2}}\right]G_{E}\left(x\right),\nonumber \\
\delta g_{55} & = & \frac{r^{2}}{4L^{2}}H_{E}\left(r\right)G_{E}\left(x\right),\nonumber \\
\delta g_{rr} & = & -\frac{L^{2}}{r^{2}}\frac{1}{f\left(r\right)}\frac{3r_{KK}^{6}}{5r^{6}-2r_{KK}^{6}}H_{E}\left(r\right)G_{E}\left(x\right),\nonumber \\
\delta g_{r\mu} & = & \frac{90r^{7}r_{KK}^{6}}{M_{E}^{2}L^{2}\left(5r^{6}-2r_{KK}^{6}\right)^{2}}H_{E}\left(r\right)\partial_{\mu}G_{E}\left(x\right),\label{eq:97}
\end{eqnarray}
where $H_{E}\left(r\right)$ must be determined by its eigen equation
given by,

\begin{equation}
\frac{1}{r^{3}}\frac{d}{dr}\left[r\left(r^{6}-r_{KK}^{6}\right)\frac{d}{dr}H_{E}\left(r\right)\right]+\left[\frac{432r^{2}r_{KK}^{12}}{\left(5r^{6}-2r_{KK}^{6}\right)^{2}}+L^{4}M_{E}^{2}\right]H_{E}\left(r\right)=0,\label{eq:98}
\end{equation}
and $G_{E}\left(x\right)$ refers to the 4d glueball field. By imposing
metric with gravitational fluctuation (\ref{eq:97}) as $g_{MN}=g_{MN}^{\left(0\right)}+\delta g_{MN}$
and solution of $C_{3}$ into 11d SUGRA action (\ref{eq:1}), we can
obtain

\begin{equation}
S_{\mathrm{SUGRA}}^{11\mathrm{d}}=-\frac{1}{2}\mathcal{C}_{E}\beta_{4}\beta_{5}\int d^{4}x\left[\left(\partial_{\mu}G_{E}\right)^{2}+M_{E}^{2}G_{E}^{2}\right],
\end{equation}
representing the standard kinetic action for scalar glueball field
$G_{E}$, where $\beta_{5}$, $g_{MN}^{\left(0\right)}$ refers respectively
to the size of $x^{5}$ and the bubble version of (\ref{eq:10}).
$\mathcal{C}_{E}$ is a numerical constant given by,

\begin{align}
\mathcal{C}_{E} & =0.057396\left[H_{E}\left(r_{KK}\right)\right]^{2}\frac{r_{KK}^{4}}{L^{3}},\nonumber \\
\left[H_{E}\left(r_{KK}\right)\right]^{-1} & =0.0069183\lambda^{1/2}N_{c}M_{KK}.
\end{align}
Hence the eigen value of (\ref{eq:98}) determines the mass spectrum
of exotic scalar glueball. The mass spectrum for various glueballs
can be obtained by taking into account different gravitational polarizations
e.g. dilatonic scalar glueball with $J^{CP}=0^{++}$,

\begin{align}
\delta G_{11,11} & =-3\frac{r^{2}}{L^{2}}H_{D}\left(r\right)G_{D}\left(x\right),\nonumber \\
\delta G_{\mu\nu} & =\frac{r^{2}}{L^{2}}H_{D}\left(r\right)\left[\eta^{\mu\nu}-\frac{\partial^{\mu}\partial^{\nu}}{M_{D}^{2}}\right]G_{D}\left(x\right),
\end{align}
and tensor glueball with $J^{CP}=2^{++}$,

\begin{equation}
\delta G_{\mu\nu}=-\frac{r^{2}}{L^{2}}H_{T}\left(r\right)T_{\mu\nu}\left(x\right).
\end{equation}
The eigen equations for $H_{D}\left(r\right)$ and $H_{T}\left(r\right)$
are given as,

\begin{equation}
\frac{1}{r^{3}}\frac{d}{dr}\left[r\left(r^{6}-r_{KK}^{6}\right)\frac{d}{dr}H_{D,T}\left(r\right)\right]+L^{4}M_{D,T}^{2}H_{D,T}\left(r\right)=0,
\end{equation}
which determines the mass spectrum of dilatonic scalar and tensor
glueball. The mass spectrum of various glueballs are collected in
Table \ref{tab:3} for the reader's convenience. The labels $\mathrm{S,T,V,N,M,L}$
refer to the solutions for six independent wave equations for various
scalar, vector tensor modes of glueballs. 
\begin{table}
\begin{centering}
\begin{tabular}{|c|c|c|c|c|c|c|}
\hline 
Mode & $\mathrm{S}_{4}$ & $\mathrm{T}_{4}$ & $\mathrm{V}_{4}$ & $\mathrm{N}_{4}$ & $\mathrm{M}_{4}$ & $\mathrm{L}_{4}$\tabularnewline
$J^{PC}$ & $0^{++}$ & $0^{++}/2^{++}$ & $0^{-+}$ & $1^{+-}$ & $1^{--}$ & $0^{++}$\tabularnewline
\hline 
$n=0$ & 7.30835 & 22.0966 & 31.9853 & 53.3758 & 83.0449 & 115.002\tabularnewline
\hline 
$n=1$ & 46.9855 & 55.5833 & 72.4793 & 109.446 & 143.581 & 189.632\tabularnewline
\hline 
$n=2$ & 94.4816 & 102.452 & 126.144 & 177.231 & 217.397 & 227.283\tabularnewline
\hline 
$n=3$ & 154.963 & 162.699 & 193.133 & 257.959 & 304.531 & 378.099\tabularnewline
\hline 
$n=4$ & 228.709 & 236.328 & 273.482 & 351.895 & 405.011 & 492.171\tabularnewline
\hline 
\end{tabular}
\par\end{centering}
\caption{\label{tab:3} Glueball mass spectrum $m_{n}^{2}$ of $\mathrm{AdS}_{7}$
in the units of $r_{KK}^{2}/L^{4}=M_{KK}^{2}/9$ in \cite{key-31}.}
\end{table}
 And this model predicts the properties of glueball in a very simple
and powerful way.

\section{Developments and holographic approaches to QCD}

In this section, we will review some holographic approaches to QCD
by using the D4/D8 model and some developments of this model in recent
years which includes basically the topics about phase transition,
heavy flavor, hadron interaction and theta angle in QCD. 

\subsection{QCD deconfinement transition}

While the confinement phase of QCD corresponds to the bubble D4-brane
geometry given in (\ref{eq:16}), it is less clear whether the black
D4-brane background (\ref{eq:14}) corresponds to the deconfinement
phase in holography exactly \cite{key-57,key-58}. This issue is recognized
by investigating the associated Wilson loop in the bubble (\ref{eq:16})
and black brane background (\ref{eq:14}) respectively. Nonetheless,
it would be interesting to compare the deconfinement transition in
QCD with the Hawking-Page transition in D4-brane system through the
gauge-gravity duality to find a holographic description of the deconfinement
transition exactly. To this goal, let us first recall the holographic
relation between the partition functions $Z$ of the bulk gravity
and its dual field theory,

\begin{equation}
Z=e^{-F}=e^{-S_{\mathrm{SUGRA}}^{ren}},\label{eq:104}
\end{equation}
which implies the classical (onshell) renormalized SUGRA action $S_{\mathrm{SUGRA}}^{ren}$
is equivalent to the free energy $F$ of the dual theory (in the Euclidean
version). The  classical SUGRA action $S_{\mathrm{SUGRA}}^{ren}$
can be collected by

\begin{equation}
S_{\mathrm{SUGRA}}^{ren}=S_{\mathrm{IIA}}^{E}+S_{\mathrm{GH}}+S_{\mathrm{CT}}^{\mathrm{bulk}},\label{eq:105}
\end{equation}
where $S_{\mathrm{IIA}}^{E}$ refers to the Euclidean version of IIA
SUGRA action given in (\ref{eq:15}). And $S_{\mathrm{GH}}$ refers
to the standard Gibbons-Hawking term given by \cite{key-70},

\begin{equation}
S_{\mathrm{GH}}=\frac{1}{\kappa_{10}^{2}}\int_{\partial\mathcal{M}}e^{-2\Phi}\sqrt{h}K,\label{eq:106}
\end{equation}
where $h_{MN}$ refers to the metric on the holographic boundary $\partial\mathcal{M}$
with $\partial\mathcal{M}=\left\{ r=\varepsilon\right\} $ for $\varepsilon\rightarrow0$,
and 
\begin{equation}
K=h^{MN}\nabla_{M}n_{N}=-\frac{1}{\sqrt{\left|g\right|}}\partial_{r}\left(\frac{\sqrt{\left|g\right|}}{\sqrt{g_{rr}}}\right)\big|_{r=\varepsilon},
\end{equation}
is trace of the extrinsic curvature. $S_{\mathrm{CT}}^{\mathrm{bulk}}$
refers to the counterterm of the bulk fields presented in action (\ref{eq:15})
given as \cite{key-71},

\begin{equation}
S_{\mathrm{CT}}^{\mathrm{bulk}}=-\frac{5}{2\kappa_{10}^{2}}\frac{g_{s}^{1/3}}{R}\int d^{9}xe^{-7\Phi/3}\sqrt{h}.\label{eq:108}
\end{equation}
Using (\ref{eq:105}) - (\ref{eq:108}) by picking up the bubble (\ref{eq:16})
and black brane background (\ref{eq:14}) solution respectively, we
can obtain the free energy of the dual theory by a simple formula
as,

\begin{align}
F_{\mathrm{conf.}} & =S_{\mathrm{SUGRA}}^{ren,\mathrm{bub}}=-\frac{2N_{c}^{2}M_{KK}^{4}\lambda V_{4}}{3^{7}\pi^{2}},\nonumber \\
F_{\mathrm{deconf}.} & =S_{\mathrm{SUGRA}}^{ren,\mathrm{black}}=-\frac{2^{7}N_{c}^{2}T^{6}\pi^{4}\lambda V_{4}}{3^{7}M_{KK}^{2}},\label{eq:109}
\end{align}
where $V_{4}$ refers to the volume of $\mathbb{R}^{4}$, $T$ is
the Hawking temperature in the black D4-brane solution (\ref{eq:16}).
Comparing the free energies given in (\ref{eq:109}), we obtain the
critical temperature by $F_{\mathrm{conf.}}=F_{\mathrm{deconf}.}\left(T=T_{c}\right)$
for the Hawking-Page transition as,

\begin{equation}
T_{c}=\frac{M_{KK}}{2\pi},\label{eq:110}
\end{equation}
which is expected to be the deconfinement transition in QCD in the
large $N_{c}$ limit. While this may be a trivial result for QCD,
it is theoretically expected in gravity side since the bubble solution
(\ref{eq:14}) is obtained by a double Wick rotation to the black
brane solution (\ref{eq:16}) i.e. (\ref{eq:110}) means exactly $\beta_{4}=\beta_{T}$.
However, this does not mean that Hawking-Page transition has nothing
to do with the QCD deconfinement transition because the fundamental
flavored matter has not been taken into account.

In order to obtain a critical temperature close to the realistic QCD
with the D4/D8 model, the flavored matter on the D8-branes must somehow
contribute to the free energy. It means in the gravity side, flavor
branes have to back react to the bulk geometry thus they would not
be probes. For such a holographic setup, we have to require $N_{f}/N_{c}$
is fixed in the large $N_{c}$ limit in order to go beyond the probe
approximation for the flavor branes. Besides we further need $N_{f}/N_{c}\ll1$
otherwise the dynamics of the dual theory is determined by flavors
instead of colors, and in the gravity side $N_{f}/N_{c}\ll1$ is also
necessary since $N_{c}$ D4-branes must dominate the bulk geometry
otherwise the holographic duality given in the previous sections would
not be valid\footnote{See similar setups in \cite{key-72,key-73,key-74} for the D3/D7 system.}.
Then the next step is to confirm the embedding configuration of the
$\mathrm{D8/\overline{D8}}$-branes. Since the configuration of the
$\mathrm{D8/\overline{D8}}$-branes relates to the chiral symmetry
discussed in Section 2.2 - 2.3, we can identify respectively the bubble
D4-brane background where $\mathrm{D8/\overline{D8}}$-branes are
located at the antipodal points of $x^{4}$ (the left one in Figure
\ref{fig:2}) as the confined phase with broken chiral symmetry and
the black D4-brane background where $\mathrm{D8/\overline{D8}}$-branes
are parallel (the left one in Figure \ref{fig:3}) as the deconfined
phase with the restored chiral symmetry. While this identification
does not distinguish exactly the chiral transition from deconfinement
transition and is not unique, it is the most simple setup to include
the elementary features in the QCD deconfinement transition. However,
keep the above requirements in hand, it is not enough to give a holographic
setup quantitatively, because when the flavored backreaction is considered,
it would be extremely challenging to search for a SUGRA solution technologically
with respect to the D-brane configuration in the D4/D8 model as in
Table \ref{tab:1}. To simplify the calculation and keep the fundamental
features of QCD, authors of \cite{key-33,key-34} suggest to consider
the case that the $\mathrm{D8/\overline{D8}}$-branes are smeared
on the $x^{4}$ direction homogeneously so that the harmonic function
for the D8-branes are identified uniquely thus it is possible to search
for a geometric solution in this setup. 

Altogether, let us write down the IIA SUGRA action plus the dynamics
of $N_{f}$ D8-branes smeared on $x^{4}$ as the total action $S$,

\begin{align}
S= & \frac{1}{2\kappa_{10}^{2}}\int d^{10}x\sqrt{-G}e^{-2\phi}\left[\mathcal{R}^{(10)}+4\partial_{\mu}\phi\partial\phi\right]-\frac{1}{4\kappa_{10}^{2}}\int d^{10}x\sqrt{-G}\left|F_{4}\right|^{2}\nonumber \\
 & -\frac{g_{s}N_{f}T_{\mathrm{D8}}M_{KK}}{\pi}\int d^{10}x\frac{\sqrt{-\det\left[g+\left(2\pi\alpha^{\prime}\right)F\right]}}{\sqrt{g_{44}}}e^{-\phi}.\label{eq:111}
\end{align}
To search for an approximate solution under the condition $N_{f}/N_{c}\ll1$,
the solution with $N_{f}=0$ is therefore the zero-th order solution
to the equations of motion from (\ref{eq:111}) which is nothing but
the bubble and black D4-brane solution given in (\ref{eq:16}) and
(\ref{eq:14}). Then let us attempt to find a solution of $\mathcal{O}\left(N_{f}/N_{c}\right)$
to the bubble D4-brane (\ref{eq:16}) first. For a homogeneous solution,
the ansatz of the metric to solve the action (\ref{eq:111}) can be
chosen as \cite{key-32,key-33,key-34},

\begin{align}
ds^{2} & =e^{2\hat{\lambda}}\left(-dt^{2}+dx_{i}dx^{i}\right)+e^{2\tilde{\lambda}}\left(dx^{4}\right)^{2}+l_{s}^{2}e^{-2\varphi}d\rho^{2}+l_{s}^{2}e^{2\nu}d\Omega_{4}^{2},\nonumber \\
\varphi & =2\Phi-4\lambda-\tilde{\lambda}-4\nu,\label{eq:112}
\end{align}
where $\Phi$ refers to the dilaton field, $\hat{\lambda},\tilde{\lambda},\varphi,\nu$
are unknown functions depending on the holographic coordinate $\rho$
only. And $\rho$ is the logarithmic coordinate defined as,

\begin{equation}
e^{-3a\rho}=1-\frac{U_{KK}^{3}}{U^{3}},a=\frac{\sqrt{2}Q_{c}U_{KK}^{3}}{3R^{3}g_{s}}=\frac{U_{KK}^{3}}{l_{s}^{3}g_{s}^{2}}.
\end{equation}
Impose the ansatz (\ref{eq:112}) into (\ref{eq:111}), it reduces
to a 1d effective action,

\begin{align}
S & =\mathcal{V}\int d\rho\left[-4\dot{\hat{\lambda}}^{2}-\dot{\tilde{\lambda}}^{2}-4\dot{\nu}^{2}+\dot{\varphi}^{2}+V\right],\nonumber \\
V & =12e^{-2\nu-2\varphi}-Q_{c}^{2}e^{4\hat{\lambda}+\tilde{\lambda}-4\nu-4\varphi}-Q_{f}e^{2\hat{\lambda}-\frac{\tilde{\lambda}}{2}+2\nu-\frac{3}{2}\varphi},\label{eq:114}
\end{align}
which has to be supported by the zero-energy constraint (dot refers
to the derivative respected to $\rho$)

\begin{equation}
-4\dot{\hat{\lambda}}^{2}-\dot{\tilde{\lambda}}^{2}-4\dot{\nu}^{2}+\dot{\varphi}^{2}-V=0,\label{eq:115}
\end{equation}
with

\begin{equation}
Q_{c}=\frac{3\pi N_{c}}{\sqrt{2}}=\frac{3}{\sqrt{2g_{s}}}\frac{R^{3}}{l_{s}^{3}},Q_{f}=\frac{2\kappa_{10}^{2}N_{f}g_{s}T_{\mathrm{D8}}M_{KK}l_{s}^{2}}{\pi},\mathcal{V}=\frac{1}{2\kappa_{10}^{2}}V_{3}\Omega_{4}\beta_{T}\frac{2\pi}{M_{KK}}l_{s}^{3}.
\end{equation}
Here $\beta_{T}$ is the size of $x^{0}$ representing the temperature
in the dual theory $\beta_{T}=1/T$ and the only nonzero component
of the gauge field potential on the D8-branes is a constant $A_{0}$
representing the chemical potential in the dual theory. Next, we expand
the all the relevant functions $\hat{\lambda},\tilde{\lambda},\varphi,\nu$
up to $N_{f}/N_{c}$ as,

\begin{equation}
\Psi=\Psi_{0}+\epsilon_{f}\Psi_{1}+\mathcal{O}\left(\epsilon_{f}^{2}\right),\label{eq:117}
\end{equation}
where

\begin{equation}
\epsilon_{f}=\frac{R^{3/2}u_{0}^{1/2}g_{s}}{l_{s}}Q_{f}=\frac{1}{12\pi^{3}}\lambda^{2}\frac{N_{f}}{N_{c}}\ll1.
\end{equation}
So the zero-th order solution of $\hat{\lambda},\tilde{\lambda},\varphi,\nu$
reads by comparing the metric ansatz (\ref{eq:112}) with the bubble
D4-brane solution (\ref{eq:16}) as,

\begin{align}
\hat{\lambda}_{0} & =f_{0}+\frac{3}{4}\log\frac{U_{KK}}{R},\nonumber \\
\tilde{\lambda}_{0} & =f_{0}-\frac{3}{2}a\rho+\frac{3}{4}\log\frac{U_{KK}}{R},\nonumber \\
\Phi_{0} & =f_{0}+\frac{3}{4}\log\frac{U_{KK}}{R}+\log g_{s},\nonumber \\
\nu_{0} & =\frac{1}{3}f_{0}+\frac{1}{4}\log\frac{U_{KK}}{R}+\log\frac{R}{l_{s}},\nonumber \\
f_{0} & =-\frac{1}{4}\log\left(1-e^{-3a\rho}\right),\label{eq:119}
\end{align}
Put (\ref{eq:117}) - (\ref{eq:119}) back into the equation of motion
varied from action (\ref{eq:114}), we can obtain a series of equations
for $\hat{\lambda}_{1},\tilde{\lambda}_{1},\varphi_{1},\nu_{1}$ as

\begin{align}
\frac{\ddot{\hat{\lambda}}_{1}}{a^{2}}-\frac{9}{2}\frac{e^{-3a\rho}}{\left(1-e^{-3a\rho}\right)^{2}}\left(4\hat{\lambda}_{1}+\tilde{\lambda}_{1}-\Phi_{1}\right) & =\frac{1}{4}\frac{e^{-\frac{3}{2}a\rho}}{\left(1-e^{-3a\rho}\right)^{13/6}},\nonumber \\
\frac{\ddot{\tilde{\lambda}}_{1}}{a^{2}}-\frac{9}{2}\frac{e^{-3a\rho}}{\left(1-e^{-3a\rho}\right)^{2}}\left(4\hat{\lambda}_{1}+\tilde{\lambda}_{1}-\Phi_{1}\right) & =-\frac{1}{4}\frac{e^{-\frac{3}{2}a\rho}}{\left(1-e^{-3a\rho}\right)^{13/6}},\nonumber \\
\frac{\ddot{\Phi}_{1}}{a^{2}}-\frac{9}{2}\frac{e^{-3a\rho}}{\left(1-e^{-3a\rho}\right)^{2}}\left(4\hat{\lambda}_{1}+\tilde{\lambda}_{1}-\Phi_{1}\right) & =\frac{5}{4}\frac{e^{-\frac{3}{2}a\rho}}{\left(1-e^{-3a\rho}\right)^{13/6}},\nonumber \\
\frac{\ddot{\nu}_{1}}{a^{2}}-\frac{3}{2}\frac{e^{-3a\rho}}{\left(1-e^{-3a\rho}\right)^{2}}\left(4\hat{\lambda}_{1}+\tilde{\lambda}_{1}-5\Phi_{1}+12\nu_{1}\right) & =\frac{1}{4}\frac{e^{-\frac{3}{2}a\rho}}{\left(1-e^{-3a\rho}\right)^{13/6}},
\end{align}
which can be solved analytically by

\begin{align}
\hat{\lambda}_{1} & =\frac{3}{8}f+y-\frac{1}{4}\left(A_{2}-A_{1}\right)-\frac{1}{4}\left(B_{2}-B_{1}\right)a\rho,\nonumber \\
\tilde{\lambda}_{1} & =-\frac{1}{8}f+y-\frac{1}{4}\left(A_{2}+B_{2}a\rho\right)-\frac{3}{4}\left(A_{1}+B_{1}a\rho\right),\nonumber \\
\Phi_{1} & =\frac{11}{8}f+y+\frac{1}{4}\left(A_{1}+B_{1}a\rho\right)-\frac{5}{4}\left(A_{2}+B_{2}a\rho\right),\nonumber \\
\nu_{1} & =\frac{11}{24}f+q,
\end{align}
with hyper geometrical functions 

\begin{align}
f= & \frac{4}{9}e^{-3a\rho/2}\ _{3}F_{2}\left(\frac{1}{2},\frac{1}{2},\frac{13}{6};\frac{3}{2},\frac{3}{2};e^{-3a\rho}\right),\nonumber \\
y= & C_{2}-\coth\left(\frac{3a\rho}{2}\right)\left[C_{1}+C_{2}\left(\frac{3}{2}a\rho+1\right)\right]+z,\nonumber \\
q= & \frac{1}{12}\left[A_{1}-5A_{2}+a\rho\left(B_{1}-5B_{2}\right)\right]+\frac{5}{3}z-\coth\left(\frac{3a\rho}{2}\right)\left[M_{1}+M_{2}\left(3a\rho+2\right)\right]+2M_{2},\nonumber \\
z= & -\frac{e^{-9a\rho/2}\left(e^{-3a\rho}+1\right)\left[9e^{3a\rho}\ _{3}F_{2}\left(\frac{1}{2},\frac{1}{2},\frac{19}{6};\frac{3}{2},\frac{3}{2};e^{-3a\rho}\right)+\ _{3}F_{2}\left(\frac{3}{2},\frac{3}{2},\frac{19}{6};\frac{5}{2},\frac{5}{2};e^{-3a\rho}\right)\right]}{162\left(1-e^{-3a\rho}\right)}\nonumber \\
 & -\frac{8e^{-3a\rho/2}\left(10e^{-3a\rho}+3\right)}{819\left(1-e^{-3a\rho}\right)}\ _{2}F_{1}\left(\frac{1}{6},\frac{1}{2};\frac{3}{2};e^{-3a\rho}\right)+-\frac{e^{-15a\rho/2}\left(38e^{3a\rho}+8e^{6a\rho}-40\right)}{273\left(1-e^{-3a\rho}\right)^{13/6}},
\end{align}
where $A_{1,2},B_{1,2},C_{1,2},M_{1,2}$ are integration constants.
The integration constants can be determined by analyzing the asymptotics
and using the zero-energy constraint (\ref{eq:115}) which leads to

\begin{align}
A_{2} & =-2A_{1},A_{1}=\frac{81\sqrt{3}\pi^{2}\left(-9+\sqrt{3}\pi-12\log2+9\log3\right)}{43120\left(2\right)^{2/3}\Gamma\left(-14/3\right)\Gamma\left(-2/3\right)^{2}},\nonumber \\
k & =C_{1}+C_{2}=\frac{\pi^{3/2}\left(3+\sqrt{3}\pi-12\log2+9\log3\right)}{78\Gamma\left(-2/3\right)\Gamma\left(1/6\right)},\nonumber \\
\frac{5k}{3} & =M_{1}+M_{2},\label{eq:123}
\end{align}
while the other constants must be confirmed by imposing additional
physical condition. Nevertheless, one may find the phase transition
depends only on the integration constants given in (\ref{eq:123}).

Follow the same step, it is also possible to obtain a solution of
order $N_{f}/N_{c}$ to the black D4-brane solution by using the metric
ansatz,

\begin{align}
ds^{2} & =-e^{2\tilde{\lambda}}dt^{2}+e^{2\hat{\lambda}}dx_{i}dx^{i}+e^{2\lambda_{s}}\left(dx^{4}\right)^{2}+l_{s}^{2}e^{-2\varphi}d\rho^{2}+l_{s}^{2}e^{2\nu}d\Omega_{4}^{2},\nonumber \\
\varphi & =2\Phi-3\hat{\lambda}-\tilde{\lambda}-\lambda_{s}-4\nu,\label{eq:124}
\end{align}
with a non-zero dynamical chemical potential 

\begin{equation}
2\pi\alpha^{\prime}A=A_{t}dt.\label{eq:125}
\end{equation}
We note that $U_{KK}$ is replaced by $U_{H}$ in the black brane
case. Put (\ref{eq:124}) and (\ref{eq:125}) into action (\ref{eq:111}),
it leads to a 1d action as,

\begin{align}
S & =\mathcal{V}\int d\rho\left[-3\dot{\hat{\lambda}}^{2}-\dot{\tilde{\lambda}}^{2}-\dot{\lambda}_{s}^{2}-4\dot{\nu}^{2}+\dot{\varphi}^{2}+V\right],\nonumber \\
V & =12e^{-2\nu-2\varphi}-Q_{c}^{2}e^{3\hat{\lambda}+\tilde{\lambda}+\lambda_{s}-4\nu-4\varphi}-Q_{f}e^{\frac{3}{2}\hat{\lambda}-\frac{1}{2}\lambda_{s}+\frac{\tilde{\lambda}}{2}+2\nu-\frac{3}{2}\varphi}\sqrt{1-\frac{1}{l_{s}^{2}}e^{-2\tilde{\lambda}+2\varphi}\dot{A}_{t}^{2}}.\label{eq:126}
\end{align}
Taking into account the near-horizon geometry, the DBI action presented
in (\ref{eq:126}) can be expanded with respect to small gauge field
potential. Then keep the quadratic action for $A_{t}$, we can obtain
an analytical leading order solution by the equations of motion derived
from (\ref{eq:126}) as, 

\begin{align}
A_{t} & =\frac{2}{3}qU_{H}\left(1-\sqrt{1-e^{-3a\rho}}\right),q=\frac{g_{s}l_{s}^{4}}{R^{3/2}U_{H}^{5/2}}n\nonumber \\
\hat{\lambda}_{1} & =\frac{f}{28}-\frac{3}{16}q^{2}g+y-\frac{1}{4}\left(a_{2}-a_{1}-a_{3}\right)-\frac{1}{4}\left(b_{2}-b_{1}-b_{3}\right)a\rho,\nonumber \\
\lambda_{s1} & =\hat{\lambda}_{1}-\frac{f}{21}+\frac{q}{4}g-a_{1}-b_{1}a\rho,\nonumber \\
\tilde{\lambda}_{1} & =\hat{\lambda}_{1}+\frac{q^{2}}{2}g-a_{3}-b_{3}a\rho,\nonumber \\
\Phi_{1} & =\hat{\lambda}_{1}+\frac{2}{21}f-a_{2}-b_{2}a\rho,\nonumber \\
\nu_{1} & =\frac{11}{252}f-\frac{q^{2}}{16}g+w,
\end{align}
with

\begin{align}
f= & \frac{6}{\left(1-e^{-3a\rho}\right)^{1/6}}+\sqrt{3}\tan^{-1}\left[\frac{2\left(1-e^{-3a\rho}\right)^{1/6}-1}{\sqrt{3}}\right]+\sqrt{3}\tan^{-1}\left[\frac{2\left(1-e^{-3a\rho}\right)^{1/6}+1}{\sqrt{3}}\right]\nonumber \\
 & -2\tanh^{-1}\left[\left(1-e^{-3a\rho}\right)^{1/6}\right]-\coth^{-1}\left[\left(1-e^{-3a\rho}\right)^{1/6}+\frac{1}{\left(1-e^{-3a\rho}\right)^{1/6}}\right],\nonumber \\
g= & \frac{4}{9}\log\left(e^{-3a\rho/2}\sqrt{e^{3a\rho}-1}+1\right)-\frac{4}{9}e^{-3a\rho/2}\sqrt{e^{3a\rho}-1},\nonumber \\
y= & c_{2}-\left[c_{1}+\left(1+\frac{3}{2}a\rho\right)c_{2}\right]\coth\left(\frac{3a\rho}{2}\right)+q^{2}j+z,\nonumber \\
w= & 2m_{2}-\left[m_{1}+\left(2+3r\right)m_{2}\right]\coth\left(\frac{3a\rho}{2}\right)+\frac{1}{12}\left(a_{1}-5a_{2}+a_{3}+b_{1}a\rho-5b_{2}a\rho+b_{3}a\rho\right)\nonumber \\
 & +\frac{5}{3}z-q^{2}j,\nonumber \\
j= & \frac{1}{72}\left[4\sqrt{1-e^{-3a\rho}}+\left(-9a\rho+6\sqrt{1-e^{-3a\rho}}-6\log\left(\sqrt{1-e^{-3a\rho}}+1\right)\right)\coth\left(\frac{3a\rho}{2}\right)\right],\nonumber \\
z= & \frac{3e^{3a\rho}\left(1-e^{-3a\rho}\right)^{5/6}-\frac{\sqrt{3}}{2}\left(e^{3a\rho}+1\right)\left[\tan^{-1}\left(\frac{2\sqrt[6]{1-e^{-3a\rho}}-1}{\sqrt{3}}\right)+\tan^{-1}\left(\frac{2\sqrt[6]{1-e^{-3a\rho}}+1}{\sqrt{3}}\right)\right]}{546\left(e^{3a\rho}-1\right)}\nonumber \\
 & +\frac{1}{2}\left(e^{3a\rho}+1\right)\frac{2\tan^{-1}\left(\frac{2\sqrt[6]{1-e^{-3a\rho}}-1}{\sqrt{3}}\right)+\coth^{-1}\left[\left(1-e^{-3a\rho}\right)^{1/6}+\frac{1}{\left(1-e^{-3a\rho}\right)^{1/6}}\right]}{546\left(e^{3a\rho}-1\right)},
\end{align}
where $a_{1,2,3},b_{1,2,3},c_{1,2},m_{1,2}$ are integration constants
and $n$ refers to the $U\left(1\right)$ charge density. And the
zero-energy constraint is given by

\begin{align}
-3\dot{\hat{\lambda}}^{2}-\dot{\lambda}_{s}^{2}-\dot{\tilde{\lambda}}^{2}-4\dot{\nu}^{2}+\dot{\varphi}^{2}+\frac{Q_{f}}{2l_{s}}e^{\frac{3}{2}\hat{\lambda}-\frac{1}{2}\lambda_{s}+\frac{3\tilde{\lambda}}{2}+2\nu+\frac{1}{2}\varphi}\dot{A}_{t}^{2}-\mathcal{P} & =0,
\end{align}
with

\begin{equation}
\mathcal{P}=12e^{-2\nu-2\varphi}-Q_{c}^{2}e^{3\hat{\lambda}+\tilde{\lambda}+\lambda_{s}-4\nu-4\varphi}-Q_{f}e^{\frac{3}{2}\hat{\lambda}-\frac{1}{2}\lambda_{s}+\frac{\tilde{\lambda}}{2}+2\nu-\frac{3}{2}\varphi}.
\end{equation}

Now it is possible to obtain the free energy of the dual theory involving
the flavored matters by imposing the above leading order solutions
into the action given in (\ref{eq:111}) after holographic renormalization.
Before this, we need to add an additional holographic counterterm
to (\ref{eq:105}) in order to cancel the divergence in the DBI action
presented in (\ref{eq:111}) which is turned out to be \cite{key-33,key-75,key-76,key-77}

\begin{equation}
S_{\mathrm{CT}}^{\mathrm{D8}}=\frac{Q_{f}}{\kappa_{10}^{2}l_{s}^{2}}\int_{\partial\mathcal{M}}d^{9}x\frac{\sqrt{h}}{\sqrt{h_{44}}}\left[\frac{R}{g_{s}^{1/3}}\chi_{1}e^{-2\Phi/3}-\frac{2R^{2}}{g_{s}^{2/3}}\chi_{2}e^{-\Phi/3}\left(K-\frac{8}{3}\boldsymbol{n}\cdot\nabla\Phi-\boldsymbol{n}\cdot\frac{\nabla\left(\sqrt{g_{44}}\right)}{\sqrt{g_{44}}}\right)\right].
\end{equation}
Here $\boldsymbol{n}$ refers to the normal vector of $\partial\mathcal{M}$
and $\chi_{1,2}$ are renormalized constants. For example, with the
back reaction from D8-branes to the bubble background, it leads to,

\begin{equation}
\chi_{1}=-\frac{631}{5005},\chi_{2}=-\frac{2}{2145}.
\end{equation}
For the black brane background, it leads to

\begin{equation}
\chi_{1}=-\frac{607}{5005},\chi_{2}=-\frac{4}{15015}.
\end{equation}
Hence we finally obtained the renormalized SUGRA action as,

\begin{equation}
F=S_{\mathrm{SUGRA}}^{ren}=S_{\mathrm{IIA}}^{E}+S_{\mathrm{GH}}+S_{\mathrm{CT}}^{\mathrm{bulk}}+S_{\mathrm{CT}}^{\mathrm{D8}},\label{eq:134}
\end{equation}
with suitable choice of $\chi_{1,2}$. Respectively, the confined
and deconfined free energy with flavors can be computed straightforwardly
by plugging the solutions of order $N_{f}/N_{c}$ into (\ref{eq:134}),
as 
\begin{figure}
\begin{centering}
\includegraphics[scale=0.25]{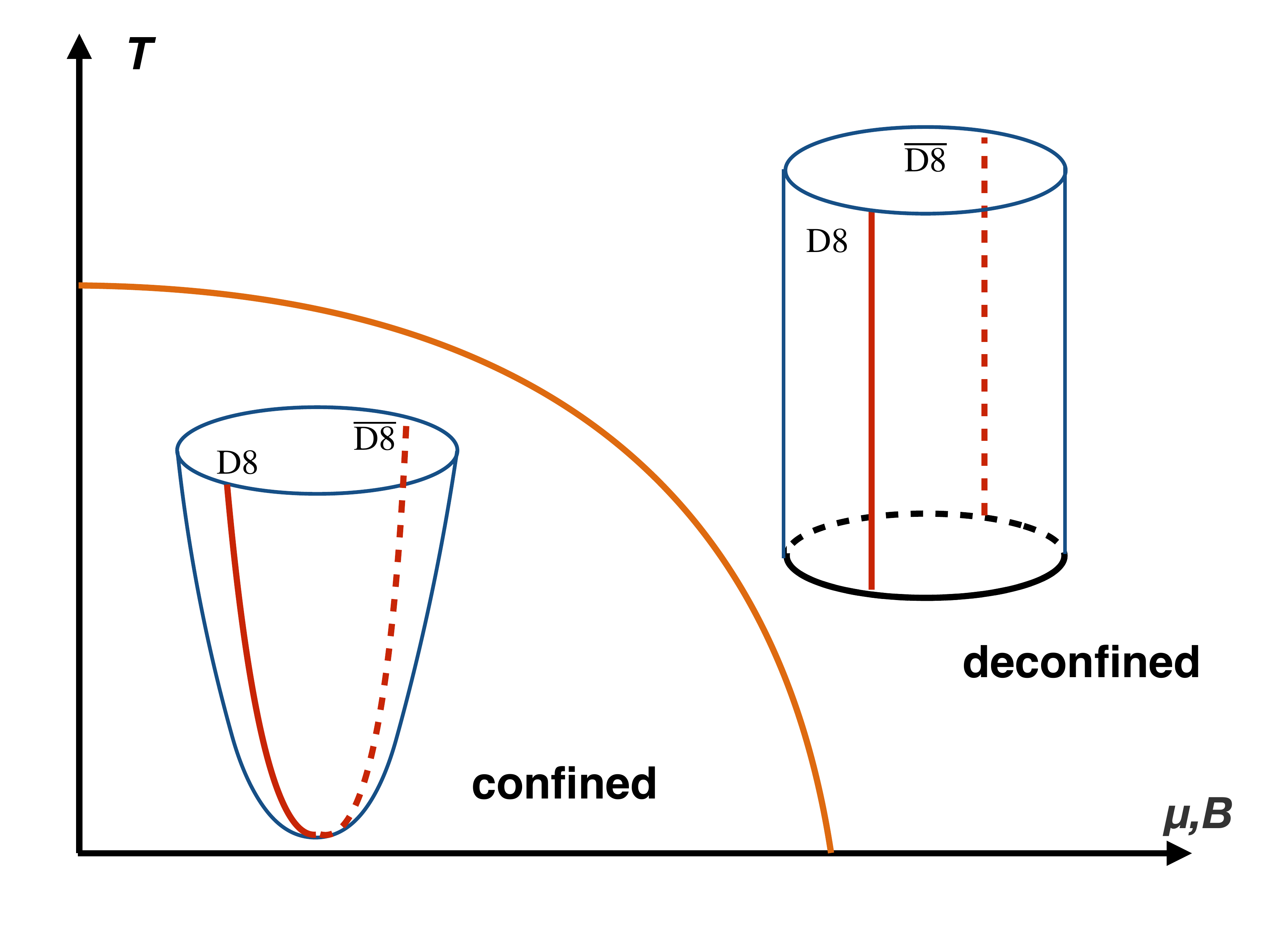}
\par\end{centering}
\caption{\label{fig:5} The critical temperature of Hawking-Page transition
as the temperature of QCD deconfinement transition in the D4/D8 model
with chemical potential $\mu$ or magnetic field $B$.}
\end{figure}

\begin{align}
F_{\mathrm{conf.}} & =-\frac{2N_{c}^{2}M_{KK}^{4}\lambda V_{4}}{3^{7}\pi^{2}}\left[1-\frac{48}{7}\epsilon_{f}\frac{\pi^{3/2}}{\Gamma\left(-2/3\right)\Gamma\left(1/6\right)}\right],\nonumber \\
F_{\mathrm{deconf.}} & =-\frac{2^{7}N_{c}^{2}T^{6}\pi^{4}\lambda V_{4}}{3^{7}M_{KK}^{2}}\left[1+\frac{4}{7}\epsilon_{fT}\left(1+\frac{7}{6}q^{2}\right)\right],\label{eq:135}
\end{align}
where

\begin{equation}
\epsilon_{fT}=\frac{R^{3/2}U_{H}^{1/2}g_{s}Q_{f}}{l_{s}^{2}}=\frac{\lambda^{2}}{12\pi^{3}}\frac{2\pi T}{M_{KK}}\frac{N_{f}}{N_{c}}\ll1,
\end{equation}
and we have used the choice of the relevant constants given in (\ref{eq:123}).
Therefore, compare the free energy given in (\ref{eq:135}), we can
obtain the critical temperature with flavors as,

\begin{equation}
\frac{2\pi T_{c}}{M_{KK}}=1-\frac{1}{126\pi^{3}}\lambda^{2}\frac{N_{f}}{N_{c}}\left[1+\frac{12\pi^{3/2}}{\Gamma\left(-2/3\right)\Gamma\left(1/6\right)}\right]-\frac{27}{16\pi}\frac{N_{f}}{N_{c}}\frac{\mu^{2}}{M_{KK}^{2}},\label{eq:137}
\end{equation}
where $\mu$ is the chemical potential in the dual theory given by
$A_{t}\big|_{U\rightarrow\infty}=\mu$. And the behavior of the Hawking-Page
transition given in (\ref{eq:137}) agrees qualitatively with the
QCD deconfinement transition \cite{key-78,key-79,key-80,key-81}. 

Moreover, when the backreaction to the bulk geometry of the flavor
branes is picked up, it is also possible to evaluate QCD deconfinement
transition under an external magnetic field. Because extremely strong
magnetic field may also give rise to deconfinement transition in QCD
\cite{key-82,key-83,key-84,key-85}. The setup follows mostly the
same discussion given above while we need to turn on a constant magnetic
field in the DBI action presented in (\ref{eq:111}), as the only
non-zero component of the gauge field strength. Then we can derive
the effective 1d action by using the metric ansatz (\ref{eq:112})
and (\ref{eq:124}) with respect to the bubble and black D4-brane
background. Fortunately, it is possible to find an analytical solution,
which leads to critical temperature $T_{c}$ as

\begin{equation}
\frac{2\pi T_{c}}{M_{KK}}=1+\frac{N_{f}}{N_{c}}\left(X\lambda^{2}+\frac{B^{2}}{M_{KK}^{4}}Y\right),\label{eq:138}
\end{equation}
by comparing free energy in the same way. Here $B$ refers to the
external magnetic field, and $X,Y$ are numerical numbers. For the
probe approximation limit of the D8-brane, $X,Y$ are calculated as,

\begin{align}
X & =-\frac{1}{126\pi^{3}}\left[1-\frac{8\pi^{3/2}}{3\Gamma\left(1/6\right)\Gamma\left(4/3\right)}\right]\simeq5\times10^{-4},\nonumber \\
Y & =-\frac{81}{16\pi}\left[1-\frac{2\pi^{3/2}}{3\Gamma\left(1/6\right)\Gamma\left(4/3\right)}\right]\simeq-0.408.
\end{align}
By considering the backreaction of the D8-branes, $X,Y$ are calculated
as,

\begin{equation}
X\simeq5\times10^{-4},Y\simeq-2.44.
\end{equation}
And the behavior of the critical temperature illustrated in (\ref{eq:138})
also coincides qualitatively with the QCD deconfinement transition
under external magnetic field predicted by lattice QCD \cite{key-82}.
We plot out the behavior of the critical temperature given in (\ref{eq:137})
and (\ref{eq:138}) in Figure \ref{fig:5}. In this sense, we could
conclude at least, investigating the Hawking-Page transition in the
D4/D8 model is very suggestive to study the QCD deconfinement transition
in holography which also covers partly the discussion in some bottom-up
approaches \cite{key-1+2,key-1+3}.

\subsection{Phase diagram with chiral transition}

As we have reviewed the deconfinement transition in the D4/D8 model
which can not be distinguished from the chiral transition, let us
focus on the chiral transition in the D4/D8 model since QCD has various
phases with chiral symmetry.

Recall the relation of $\mathrm{D8/\overline{D8}}$-brane configuration
and chiral symmetry, the chiral transition is identified as the transition
from connected to disconnected configuration of the $\mathrm{D8/\overline{D8}}$-branes.
Hence we need to choose the black D4-brane background in order to
include both the connected and disconnected $\mathrm{D8/\overline{D8}}$-brane
configuration. The main idea to evaluate the phase transition follows
the Section 3.1, which is to compute the free energy in holography.
As we will work with respect to the black D4-brane background only,
the contribution from the bulk geometry would be irrelevant, because
the difference of the free energy determines the phase transition.
Keeping this in mind, we can quickly write down the D8-brane action
for mesonic (broken chiral symmetry) and quark matter phase (restored
chiral symmetry) with a $U\left(1\right)$ chemical potential $\hat{A}_{0}$,
which corresponds to the connected and disconnected D8-brane configuration
respectively in Figure \ref{fig:3} as\footnote{We note that in this setup, the Chern-Simons action vanishes.},

\begin{equation}
S_{\mathrm{DBI}}^{\mathrm{D8}}=\mathcal{N}\frac{V_{3}}{T}\int_{u_{0}}^{\infty}duu^{5/2}\sqrt{1+u^{3}f_{T}\left(x_{4}^{\prime}\right)^{2}-\hat{a}_{0}^{\prime2}},\label{eq:141}
\end{equation}
where the variables in (\ref{eq:141}) are dimensionless as,

\begin{equation}
x_{4}=x^{4}M_{KK},u=\frac{U}{R}\frac{1}{\left(M_{KK}R\right)^{2}},\hat{a}_{0}=\hat{A}_{0}\frac{4\pi}{\lambda M_{KK}},\mathcal{N}=2T_{\mathrm{D8}}\Omega_{4}R^{5}\left(M_{KK}R\right)^{7}.
\end{equation}
The equation of motion can be obtained by varying (\ref{eq:141})
respected to $x_{4}$ and $\hat{a}_{0}$ which are,

\begin{align}
\frac{u^{5/2}\hat{a}_{0}^{\prime}}{\sqrt{1+u^{3}f_{T}x_{4}^{\prime2}-\hat{a}_{0}^{\prime2}}} & =n_{I},\nonumber \\
\frac{u^{11/2}f_{T}x_{4}^{\prime}}{\sqrt{1+u^{3}f_{T}x_{4}^{\prime2}-\hat{a}_{0}^{\prime2}}} & =f_{T}^{1/2}\left(u_{0}\right)u_{0}^{4},\label{eq:143}
\end{align}
where ``$\prime$'' refers to the derivative with respect to $u$.
The constant $n_{I}$ corresponds to the $U\left(1\right)$ charge
density which is therefore the baryon number in this setup. In the
mesonic phase, the baryon number is zero i.e. $n_{I}=0$, and the
equations of motion in (\ref{eq:143}) can be solved by the following
boundary condition according to the connected configuration in Figure
\ref{fig:3}, as

\begin{align}
\hat{a}_{0}^{\prime}\left(u_{0}\right) & =0,\ \hat{a}_{0}\left(\infty\right)=\mu,\nonumber \\
x_{4}^{\prime}\left(u_{0}\right) & =\infty,\ \frac{l}{2}=\int_{u_{0}}^{\infty}dux_{4}^{\prime},
\end{align}
where constant $\mu$ refers to the chemical potential in the dual
theory, constant $l$ refers to the separation of the $\mathrm{D8/\overline{D8}}$-branes
at boundary $u\rightarrow\infty$. Thus the solution is

\begin{equation}
\hat{a}_{0}=\mu,x_{4}^{\prime}=\frac{f_{T}^{1/2}\left(u_{0}\right)u_{0}^{4}}{f_{T}^{1/2}\left(u\right)u^{3/2}\sqrt{u^{8}f_{T}\left(u\right)-u_{0}^{8}f_{T}\left(u_{0}\right)}}.\label{eq:145}
\end{equation}
Put the solution (\ref{eq:145}) back into action (\ref{eq:141}),
we can obtain the free energy as,

\begin{equation}
F_{\mathrm{mesonic}}=\mathcal{N}\int_{u_{0}}^{\infty}duu^{5/2}\frac{u^{4}f_{T}^{1/2}\left(u\right)}{\sqrt{u^{8}f_{T}\left(u\right)-u_{0}^{8}f_{T}\left(u_{0}\right)}}.\label{eq:146}
\end{equation}

For the quark matter phase, the boundary condition reads from the
disconnected configuration in Figure \ref{fig:3} as,

\begin{equation}
\hat{a}_{0}\left(u_{H}\right)=0,\hat{a}_{0}\left(\infty\right)=\mu,x_{4}^{\prime}=0,
\end{equation}
which leads to a solution with hyper geometrical functions as,

\begin{equation}
\hat{a}_{0}\left(u\right)=\mu-\frac{n_{I}^{2/5}\Gamma\left(3/10\right)\Gamma\left(6/5\right)}{\sqrt{\pi}}+u\ _{2}F_{1}\left(\frac{1}{5},\frac{1}{2},\frac{6}{5},-\frac{u^{5}}{n_{I}^{2}}\right).
\end{equation}
Therefore the free energy is computed as,

\begin{equation}
F_{\mathrm{quarkmatter}}=\mathcal{N}\int_{u_{H}}^{\infty}du\frac{u^{5}}{\sqrt{u^{5}+n_{I}^{5}}}.\label{eq:149}
\end{equation}
We note that the condition $\hat{a}_{0}\left(u_{H}\right)=0$ implies
$n_{I}$ is a function of $\mu$ as,

\begin{equation}
0=\mu-\frac{n_{I}^{2/5}\Gamma\left(3/10\right)\Gamma\left(6/5\right)}{\sqrt{\pi}}+u_{H}\ _{2}F_{1}\left(\frac{1}{5},\frac{1}{2},\frac{6}{5},-\frac{u_{H}^{5}}{n_{I}^{2}}\right).\label{eq:150}
\end{equation}
Then the phase diagram can be obtained by compare the free energy
given in (\ref{eq:146}) and (\ref{eq:150}) with constraint (\ref{eq:149}).
Notice that while the free energy given in (\ref{eq:146}) and (\ref{eq:150})
is divergent, their difference which determines the phase diagram
remains to be finite. Thus it is not necessary to do the holographic
renormalization in this case.

\begin{figure}
\begin{centering}
\includegraphics[scale=0.2]{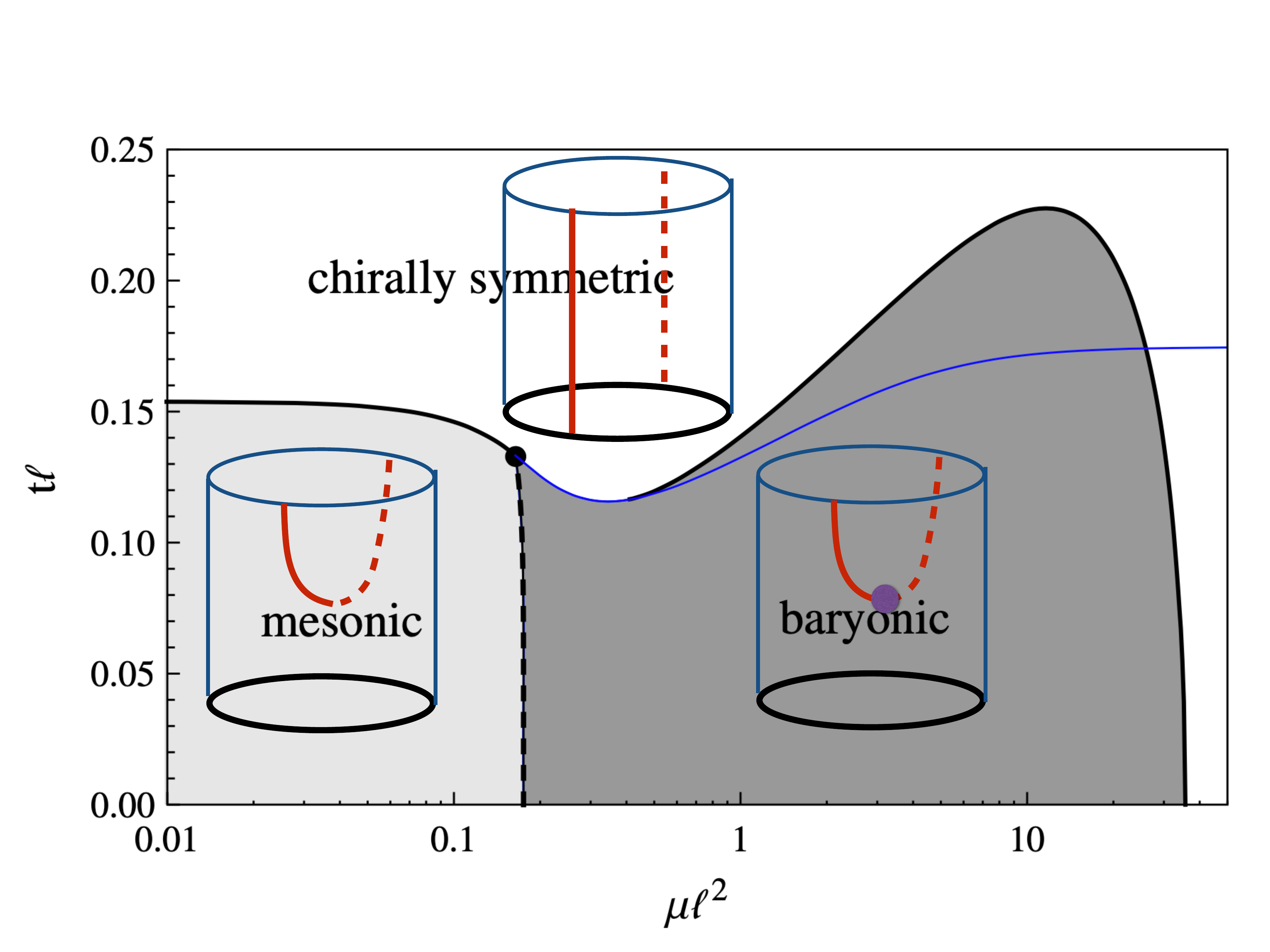}\includegraphics[scale=0.2]{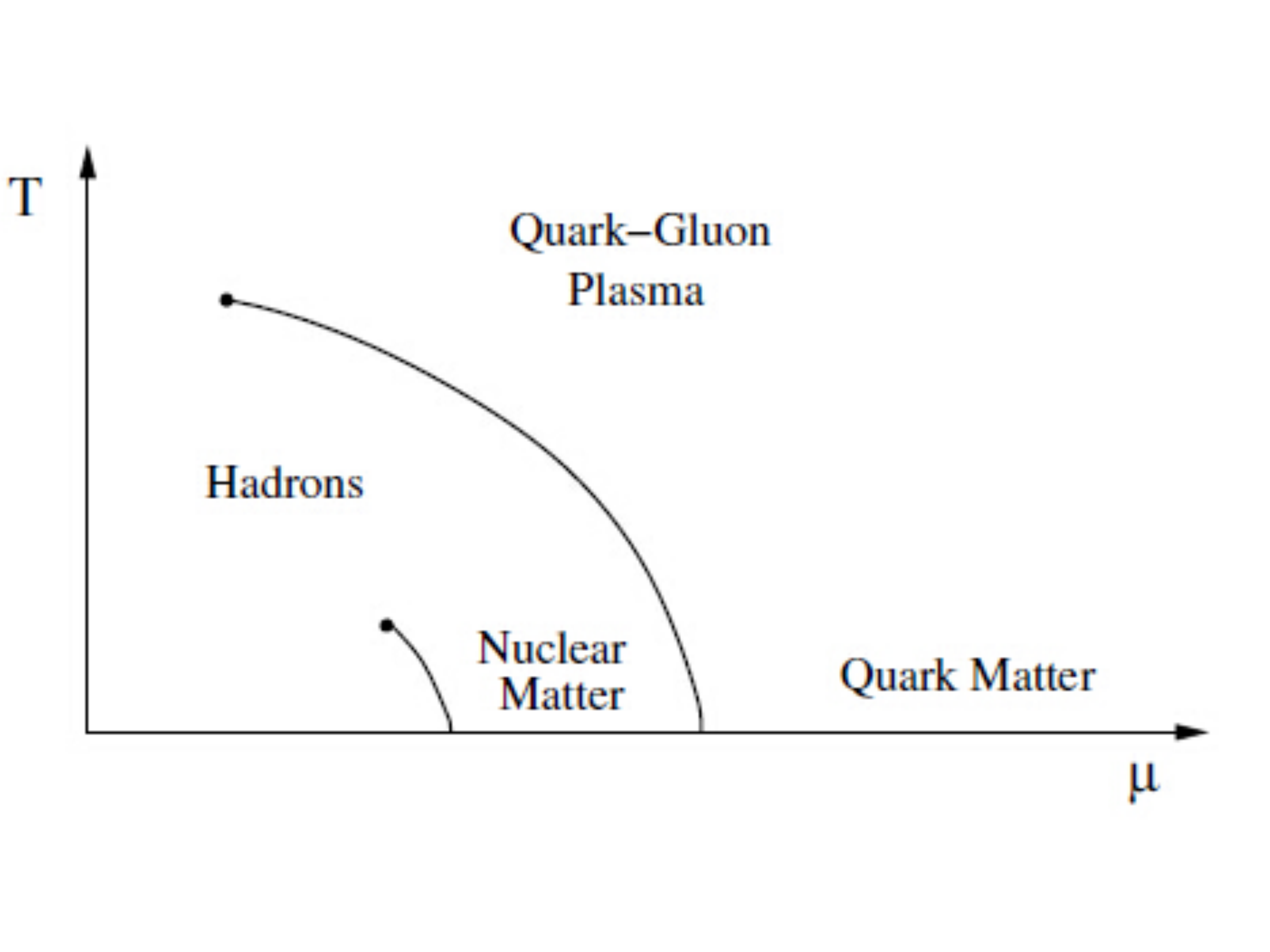}
\par\end{centering}
\caption{\label{fig:6} Holographic QCD phase diagram vs realistic QCD phase
diagram in the $T-\mu$ plane.}

\end{figure}

For a more ambitious approach, let us include the baryonic phase in
the black D4-brane background, that is to take into account the configuration
(c) in Figure \ref{fig:4} which has broken chiral symmetry with baryon
vertex. Since baryon vertex is identified as $\mathrm{D4}^{\prime}$-brane
described equivalently by instantons on the D8-branes, we employ the
BPST instanton solution given in (\ref{eq:83}) to represent baryon
on the flavor brane. For multiple baryons we can summarize the instanton
field strength as what we have discussed in Section 2.5, in this sense
baryons are treated as instanton gas on the flavor brane. On the other
hand, as the instanton size takes order of $\lambda^{-1/2}$ and the
DBI action (\ref{eq:20}) does not define how to treat it with non-Abelian
gauge field\footnote{The symmetrized trace in DBI action is usually used for all terms
of $\mathcal{O}\left(F^{4}\right)$ and higher, however it is known
to be incomplete starting from $\mathcal{O}\left(F^{6}\right)$ \cite{key-86}.}, we may generalize the DBI action by taking all order of gauge field
strength into non-Abelian case through the identity for Abelian gauge
field strength $\hat{F}$ as,

\begin{align}
\sqrt{\det\left(g+2\pi\alpha^{\prime}\hat{F}\right)}= & U^{4}\left(\frac{R}{U}\right)^{3/4}\big\{\left(2\pi\alpha^{\prime}\right)^{2}f_{T}\hat{F}_{iU}^{2}+\left(1+u^{3}f_{T}x_{4}^{\prime2}-\hat{a}_{0}^{\prime2}\right)\left[1+\left(\frac{R}{U}\right)^{3}\frac{\left(2\pi\alpha^{\prime}\right)^{2}}{2}\hat{F}_{ij}^{2}\right]\nonumber \\
 & +\left(\frac{R}{U}\right)^{3}\frac{\left(2\pi\alpha^{\prime}\right)^{4}f_{T}\left(\hat{F}_{ij}\hat{F}_{kU}\epsilon_{ijk}\right)^{2}}{4}\big\}^{1/2}.\label{eq:151}
\end{align}
For non-Abelian generalization, we follow \cite{key-87} to replace
the quadratic terms of $\hat{F}$ by its non-Abelian version $F$
then take trace of each term separately as,

\begin{equation}
\hat{F}_{iU}^{2}\rightarrow\mathrm{Tr}F_{iU}^{2},\hat{F}_{ij}^{2}\rightarrow\mathrm{Tr}F_{ij}^{2},\hat{F}_{ij}\hat{F}_{kU}\rightarrow\mathrm{Tr}F_{ij}F_{kU}.
\end{equation}
Afterwards we impose the BPST instanton solution (\ref{eq:83}) with
multiple number $n_{I}$ to $F$ in order to represent baryons \footnote{\cite{key-26,key-27} illustrate that in the case of $N_{f}=2$ the
non-Abelian part of $A_{0}$ presented in (\ref{eq:85}) vanishes.
And we do not attempt to consider baryon with $N_{f}>2$ in this section. }. Altogether, we reach to a generalized version of action for baryonic
phase as,

\begin{equation}
F_{\mathrm{baryonic}}=S^{\mathrm{D8}}=\mathcal{N}\frac{V_{3}}{T}\int_{u_{c}}^{\infty}\left[\sqrt{\left(1+g_{1}+u^{3}f_{T}x_{4}^{\prime2}-\hat{a}_{0}^{\prime2}\right)\left(1+g_{2}\right)}-n_{I}\hat{a}_{0}\left(u\right)q\left(u\right)\right],\label{eq:153}
\end{equation}
where 

\begin{align}
g_{1}\left(u\right) & =\frac{f_{T}\left(u\right)u^{1/2}}{3u_{c}^{2}\sqrt{f_{c}\left(u\right)}}n_{I}q\left(u\right),f_{c}\left(u\right)=1-\frac{u_{c}^{3}}{u^{3}},\nonumber \\
g_{2}\left(u\right) & =\frac{u_{c}^{2}\sqrt{f_{c}\left(u\right)}}{3u^{7/2}}n_{I}q\left(u\right),q\left(Z\right)=\frac{3\rho^{4}}{4\left(Z^{2}+\rho^{2}\right)^{5/2}}.
\end{align}
We note that the last term in (\ref{eq:153}) is the Chern-Simons
action and $q\left(Z\right)$ is the average instanton field strength
defined by the summary of the BPST instanton (\ref{eq:83}) as \cite{key-88}

\begin{equation}
\frac{1}{V_{3}}\sum_{n=1}^{N_{I}}\int d^{3}X\frac{4\left(\rho/\gamma\right)^{4}}{\left[\left(\vec{X}-\vec{X}_{0n}\right)^{2}+\left(Z/\gamma\right)^{2}+\left(\rho/\gamma\right)^{2}\right]^{4}}=\frac{2}{3}\pi^{2}\gamma q\left(Z\right)n_{I},
\end{equation}
with normalization condition

\begin{equation}
\int_{-\infty}^{+\infty}q\left(Z\right)dZ=1.
\end{equation}
$\vec{X}_{0n}$ refers to the center of the $n$-th instanton. $Z$
is the Cartesian coordinate, in the case of (c) in Figure \ref{fig:4},
it is defined as,

\begin{equation}
U^{3}=U_{c}^{3}+U_{c}Z^{2},
\end{equation}
where we use $U_{c}$ to denote the connected position of $\mathrm{D8/\overline{D8}}$-branes
with instantons to distinguish it from $U_{0}$ in which instantons
are absent. $N_{I}$ is the instanton number which relates to its
number density $n_{I}$ as $n_{I}=N_{I}96\pi^{4}/\left(\lambda^{2}M_{KK}^{3}\right)$.
With the boundary condition

\begin{equation}
\frac{l}{2}=\int_{u_{c}}^{\infty}dux_{4}^{\prime},\hat{a}_{0}\left(\infty\right)=\mu,
\end{equation}
the equations of motion obtained by varying (\ref{eq:153}),

\begin{align}
\frac{u^{5/2}\hat{a}_{0}^{\prime}\sqrt{1+g_{2}}}{\sqrt{1+g_{1}+u^{3}f_{T}x_{4}^{\prime2}-\hat{a}_{0}^{\prime2}}} & =n_{I}Q\nonumber \\
\frac{u^{11/2}f_{T}x_{4}^{\prime}\sqrt{1+g_{2}}}{\sqrt{1+g_{1}+u^{3}f_{T}x_{4}^{\prime2}-\hat{a}_{0}^{\prime2}}} & =k,
\end{align}
where $k$ is an integration constant to be determined and

\begin{equation}
Q\left(u\right)=\int_{u_{c}}^{u}q\left(v\right)dv=\frac{u^{3/2}\sqrt{f_{c}}}{2}\frac{3\rho^{2}u_{c}+2\left(u^{3}-u_{c}^{3}\right)}{\left(u^{3}-u_{c}^{3}+\rho^{2}u_{c}\right)^{3/2}},
\end{equation}
can be solved as,

\begin{align}
\hat{a}_{0}^{\prime2} & =\frac{\left(n_{I}Q\right)^{2}}{u^{5}}\frac{1+g_{1}}{1+g_{2}-\frac{k^{2}}{u^{8}f_{T}}+\frac{\left(n_{I}Q\right)^{2}}{u^{5}}},\nonumber \\
x_{4}^{\prime2} & =\frac{k^{2}}{u^{11}f_{T}^{2}}\frac{1+g_{1}}{1+g_{2}-\frac{k^{2}}{u^{8}f_{T}}+\frac{\left(n_{I}Q\right)^{2}}{u^{5}}}.\label{eq:161}
\end{align}
Plugging solution (\ref{eq:161}) back into (\ref{eq:153}), we can
obtain the free energy of the baryonic phase. In order to obtain the
phase transition, we need to further minimize the free energy given
in (\ref{eq:153}) with respect to $n_{I},\rho,u_{c}$ as the parameters
which leads to three constraints as,

\begin{align}
0= & \int_{u_{c}}^{\infty}\left[\frac{u^{5/2}}{2}\left(\frac{\partial g_{1}}{\partial n_{I}}\zeta^{-1}+\frac{\partial g_{2}}{\partial n_{I}}\zeta\right)+\hat{a}_{0}^{\prime}Q\right]-\mu,\nonumber \\
0= & \int_{u_{c}}^{\infty}\left[\frac{u^{5/2}}{2}\left(\frac{\partial g_{1}}{\partial\rho}\zeta^{-1}+\frac{\partial g_{2}}{\partial\rho}\zeta\right)+n_{I}\hat{a}_{0}^{\prime}\frac{\partial Q}{\partial\rho}\right],\nonumber \\
u_{c}^{5/2}\sqrt{\left[1+g_{1}\left(u_{c}\right)\right]\left[1+g_{2}\left(u_{c}\right)-\frac{k^{2}}{u_{c}^{8}f_{T}\left(u_{c}\right)}\right]}= & \int_{u_{c}}^{\infty}\left[\frac{u^{5/2}}{2}\left(\frac{\partial g_{1}}{\partial u_{c}}\zeta^{-1}+\frac{\partial g_{2}}{\partial u_{c}}\zeta\right)+n_{I}\hat{a}_{0}^{\prime}\frac{\partial Q}{\partial u_{c}}\right],
\end{align}
where

\begin{equation}
\zeta=\frac{\sqrt{1+g_{1}}}{\sqrt{1+g_{2}-\frac{k^{2}}{u^{8}f_{T}}+\frac{\left(n_{I}Q\right)^{2}}{u^{5}}}}.
\end{equation}
With all above in hand, we can obtain numerically a holographic diagram
included mesonic, baryonic and quark matter phase of QCD by comparing
the associated free energies given in (\ref{eq:146}) (\ref{eq:149})
and (\ref{eq:153}). The resultant phase diagram is given in Figure
\ref{fig:6}. 
 and we can see the holographic diagram includes all the elementary
phases in realistic QCD although the confined geometry is not included
in the current discussion.

We note that it is very difficult to work out a reasonable model describing
QCD matter over a very wide density regime with traditional models
or theories of QCD. For example, the quark-meson model (e.g. \cite{key-89,key-90})
and the Nambu-Jona-Lasinio (NJL) model \cite{key-91,key-92,key-93,key-94,key-95}
are very useful to get some insight into the chiral and deconfinement
phase transitions and quark matter phases, however the nuclear matter
is usually not included in these models. In addition, nucleon-meson
models e.g. \cite{key-96,key-97,key-98,key-99} are based on the properties
of nuclear matter and may be able to describe moderately dense nuclear
matter realistically, while they give a poor description of quark
matter with restored chiral symmetry. In this sense, this holographic
model provides a very powerful way to study QCD phase diagram in a
very wide density regime based on string theory.

\subsection{Higgs mechanism and heavy-light meson field}

One of the interesting developments of the D4/D8 model is to include
heavy flavor by using the Higgs mechanism in D-brane system. Recall
that the fundamental quarks in the D4/D8 model are identified to be
the $4-8$ and $4-\bar{8}$ strings. Since $N_{c}$ D4-branes and
$N_{f}$ D8-branes are coincident, we find that the $4-8$ and $4-\bar{8}$
strings have a vanishing vacuum expectation value (VEV). Therefore
the fundamental quarks created by $4-8$ and $4-\bar{8}$ strings
are massless which implies this model can describe the mesons with
light flavors only. Hence it is naturally motivated to include the
massive heavy flavor in this model. To this goal, in this section,
let us review the Higgs mechanism in D-brane system and see how to
use it to introduce heavy flavor.

\begin{figure}
\begin{centering}
\includegraphics[scale=0.3]{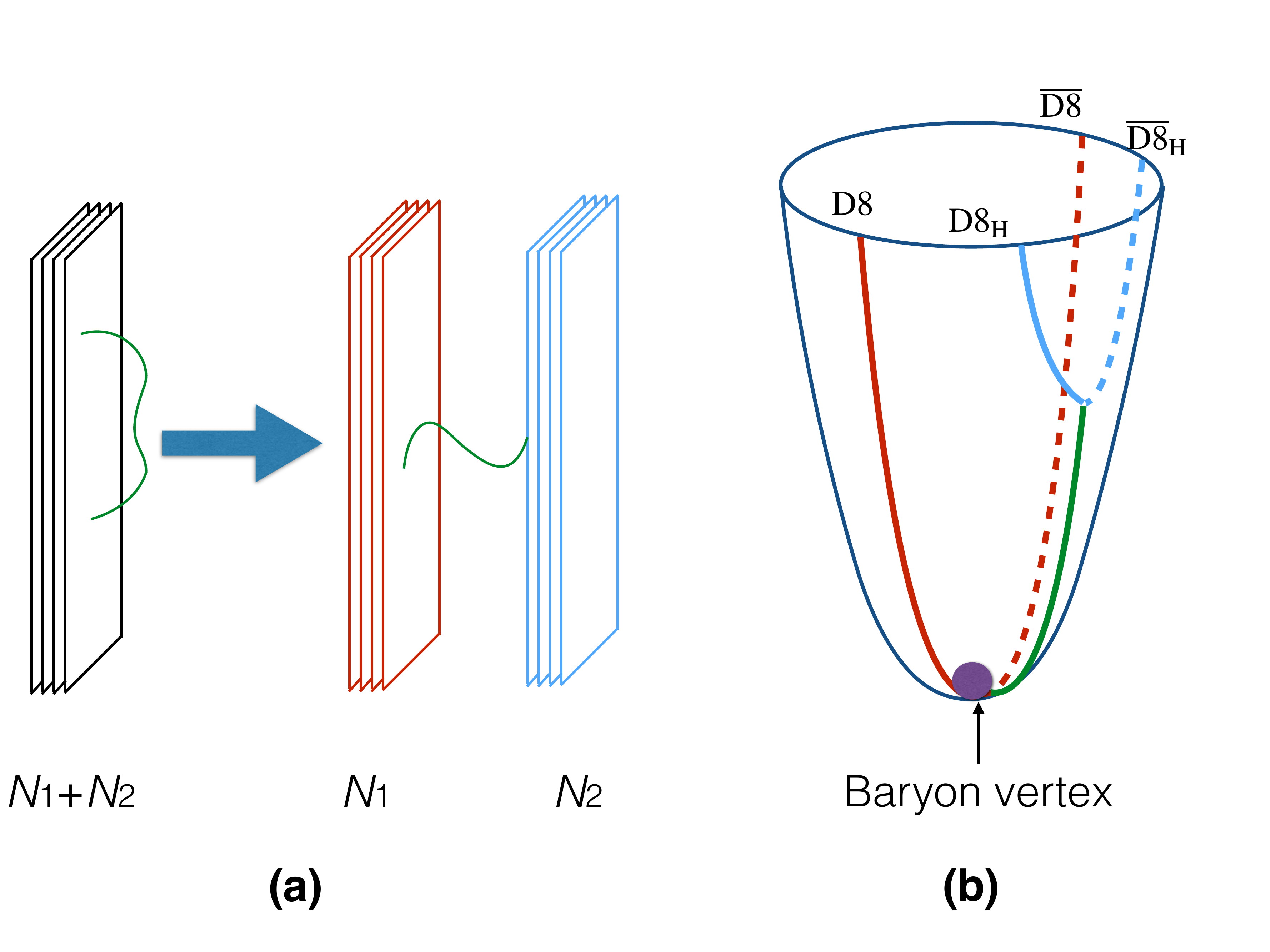}
\par\end{centering}
\caption{\label{fig:7} The Higgs mechanism in string theory (a) and the configuration
of heavy flavor in D4/D8 model (b). \textbf{(a)} Higgs mechanism in
string theory: The gauge symmetry on a stack of coincident $N_{1}+N_{2}$
D-branes could be $U\left(N_{1}+N_{2}\right)$, while it breaks down
to $U\left(N_{1}\right)\times U\left(N_{2}\right)$ if the D-branes
somehow move apart to become two stacks of coincident $N_{1}$ and
$N_{2}$ D-branes. The open string connecting the two stacks of the
D-branes has a non-zero VEV hence its ground states acquire non-zero
mass which corresponds to the separation of the D-branes. \textbf{(b)
}Configuration of heavy flavor in D4/D8 model: Red line refers to
the stack of $N_{f}$ $\mathrm{D8/\overline{D8}}$-branes in the original
model which now is identified as light flavor. Blue line refers to
another one pair of $\mathrm{D8/\overline{D8}}$-branes separated
from $N_{f}$ $\mathrm{D8/\overline{D8}}$-branes which is identified
as heavy flavor. The green line refers to the open string connecting
the light and heavy brane which is the heavy-light string, and it
acquires non-zero VEV to create massive ground states.}

\end{figure}

First, we take a look at the Higgs mechanism in D-brane system by
considering the configuration of an open string connecting two stacks
of the separated D-branes as it is illustrated in (a) of Figure \ref{fig:7}.
 In this D-brane configuration, the worldvolume symmetry $U\left(N_{1}+N_{2}\right)$
breaks down to $U\left(N_{1}\right)\times U\left(N_{2}\right)$ when
the D-branes move separately, where we use $N_{1},N_{2}$ to refer
to the D-branes number in each stack. Accordingly the transverse modes
of the D-brane acquire a non-zero VEV due to the separation of the
D-branes. Hence the multiplets created by the open string connecting
the separated D-branes become massive just like the Higgs mechanism
in the standard model of the particle physics \cite{key-52,key-59}.
Let us investigate this mechanism quantitatively by recalling the
D-brane action in (\ref{eq:20}). In the holographic approach, we
need the near-horizon limit i.e. $\alpha^{\prime}\rightarrow0$, thus
the D-brane action can be expanded to be Yang-Mills plus Wess-Zumino
action as it is in Section 2.4. Now we pick up the transverse modes
in the DBI action, it reduces to an additional quadratic action for
the transverse modes $\Psi^{I}$ with $\alpha^{\prime}\rightarrow0$
as,

\begin{align}
S\left[\Psi^{I}\right] & =-T_{\mathrm{D}_{p}}\frac{\left(2\pi\alpha^{\prime}\right)^{2}}{4}\int_{\mathrm{D}_{p}}d^{p+1}x\sqrt{-\det g}e^{-\phi}\mathrm{Tr}\left\{ 2D_{a}\Psi^{I}D_{a}\Psi^{I}+\left[\Psi^{I},\Psi^{J}\right]^{2}\right\} ,\nonumber \\
a & =0,1...p,\ I,J=p+1...d,\label{eq:164}
\end{align}
where the covariant derivative is

\begin{equation}
D_{a}\Psi^{I}=\partial_{a}\Psi^{I}+i\left[\mathbf{A}_{a},\Psi^{I}\right],\label{eq:165}
\end{equation}
and $\mathbf{A}_{a}$ is the gauge field potential on the D-brane.
Consider a stack of coincident $N_{1}+N_{2}$ D-branes, $\mathbf{A}_{a}$
could be the $U\left(N_{1}+N_{2}\right)$ generator as an $\left(N_{1}+N_{2}\right)\times\left(N_{1}+N_{2}\right)$
matrix. However, if the coincident $N_{1}+N_{2}$ D-branes move apart
to become two stacks of $N_{1}$ and $N_{2}$ D-branes, the gauge
potential $\mathbf{A}_{a}$ becomes,

\begin{equation}
\mathbf{A}_{a}=\left(\begin{array}{cc}
A_{a} & \Phi_{a}\\
\Phi_{a}^{\dagger} & 0
\end{array}\right),
\end{equation}
where $A_{a}$ is the $U\left(N_{1}\right)$ gauge potential as an
$N_{1}\times N_{1}$ matrix. $\Phi_{a}$ is the multiplet created
by the open string connecting to the two stacks of the D-brane which
is an $N_{1}\times N_{2}$ matrix-valued field and the last element
can be gauged away by the residual symmetry. On the other hand, when
the D-branes are separated, the transverse mode $\Psi^{I}$ will have
a non-zero VEV since the open string connecting the separated D-branes
can not shrink to zero. Therefore we can write down $\Psi^{I}$ with
a VEV as,

\begin{equation}
\Psi^{I}\rightarrow\Psi^{I}+V^{I},
\end{equation}
where

\begin{equation}
V^{I}\sim\left(\begin{array}{cc}
V & 0\\
0 & v
\end{array}\right),\label{eq:168}
\end{equation}
to represent $U\left(N_{1}+N_{2}\right)\rightarrow U\left(N_{1}\right)\times U\left(N_{2}\right)$.
Thus plugging (\ref{eq:165}) - (\ref{eq:168}) into (\ref{eq:164}),
one obtains a mass term in the action as 
\begin{equation}
\mathrm{Tr}\left(V^{2}\Phi_{a}^{\dagger}\Phi_{a}+v^{2}\Phi_{a}\Phi_{a}^{\dagger}\right),
\end{equation}
and $\Phi_{a}$ can be interpreted as the heavy-light field acquiring
a mass through the VEV of the transverse mode $\Psi^{I}$.

With this Higgs mechanism in string theory, let us employ it in the
D4/D8 model by considering (b) of Figure \ref{fig:7}. In this configuration,
there is one pair of $\mathrm{D8/\overline{D8}}$-branes separated
from $N_{f}$ $\mathrm{D8/\overline{D8}}$-branes which is identified
as heavy flavor brane with an open string (the heavy-light string)
connecting them. We note that the configuration in (b) of Figure \ref{fig:7}
is generalized version of (a) of Figure \ref{fig:7} in the curved
spacetime. Then we can write down the D8-brane action with heavy flavor
by imposing the following replacement,

\begin{equation}
A_{a}\rightarrow\mathbf{A}_{a}=\left(\begin{array}{cc}
A_{a} & \Phi_{a}\\
\Phi_{a}^{\dagger} & 0
\end{array}\right),F_{ab}\rightarrow\boldsymbol{\mathrm{F}}_{ab}=\left(\begin{array}{cc}
F_{ab}+i\alpha_{ab} & f_{ab}\\
f_{ab}^{\dagger} & i\beta_{ab}
\end{array}\right),\label{eq:170}
\end{equation}
to (\ref{eq:33}), where $A_{a},F_{ab}$ are $N_{f}\times N_{f}$
matrix-valued fields as we have specified Section 2. $\Phi_{a}$ is
an $N_{f}\times1$ matrix-valued multiplet created by the heavy-light
string which is interpreted as the heavy-light meson field and\footnote{The index in the square brackets is ranked as $T_{\left[ab\right]}=\frac{1}{2!}\left(T_{ab}-T_{ba}\right)$
and we have chosen the gauge field as Hermitian field $\boldsymbol{\mathrm{A}}_{a}^{\dagger}=\boldsymbol{\mathrm{A}}_{a}$.}

\begin{align}
\alpha_{ab}= & 2\Phi_{[a}\Phi_{b]}^{\dagger},\ \beta_{ab}=2\Phi_{[a}^{\dagger}\Phi_{b]},\nonumber \\
f_{ab}= & 2\partial_{[a}\Phi_{b]}+2iA_{[a}\Phi_{b]}\equiv2D_{[a}\Phi_{b]},
\end{align}
we obtain the action (\ref{eq:164}) for $\Phi_{a}$ as 

\begin{align}
S\left[\Phi_{\mu,z}\right]= & -2\kappa\mathrm{Tr}\int dzdx^{4}f\left(z\right)\left(\partial_{\mu}\Phi_{\nu}^{\dagger}-\partial_{\nu}\Phi_{\mu}^{\dagger}\right)\left(\partial^{\mu}\Phi^{\nu}-\partial^{\nu}\Phi^{\mu}\right)\nonumber \\
 & -\kappa\mathrm{Tr}\int dzdx^{4}g\left(z\right)\left(\partial_{\mu}\Phi_{z}^{\dagger}-\partial_{z}\Phi_{\mu}^{\dagger}\right)\left(\partial^{\mu}\Phi_{z}-\partial_{z}\Phi^{\mu}\right)\nonumber \\
 & -\tilde{v}\mathrm{Tr}\int dzdx^{4}a\left(z\right)\left[2f\left(z\right)\Phi^{\dagger\mu}\Phi_{\mu}+g\left(z\right)\Phi_{z}^{\dagger}\Phi_{z}\right],
\end{align}
where $z$ is the Cartesian coordinates given in (\ref{eq:34}) and
the VEV of T-dualitied $\Psi^{4}=2\pi\alpha^{\prime}x^{4}$ is chosen
as \cite{key-100},

\begin{equation}
\Psi^{4}=\left(\begin{array}{cc}
-\frac{v}{N_{f}-1}\mathbf{1}_{N_{f}-1} & 0\\
0 & v
\end{array}\right),
\end{equation}
with 

\begin{equation}
\tilde{v}=v\kappa/\left(2\pi\alpha^{\prime}\right)^{2},f\left(z\right)=\frac{R^{3}}{4U\left(z\right)},g\left(z\right)=\frac{9}{8}\frac{U\left(z\right)^{3}}{U_{KK}},a\left(z\right)=\left[\frac{U\left(z\right)}{R}\right]^{3/2}.
\end{equation}
Note that $\Psi^{4}$ is the only transverse mode of D8-brane. The
heavy-light meson tower can be obtained by expanding $\Phi_{\mu,z}$
as what we have specified in Section 2.4. For example, the transverse
modes of heavy-light meson field is suggested as \cite{key-37,key-38},

\begin{equation}
\Phi_{\mu}=\sum_{n}\phi_{\left(n\right)}^{H}\left(z\right)B_{\mu}^{\left(n\right)}\left(x\right),\Phi_{z}=0,\label{eq:175}
\end{equation}
which leads to 

\begin{equation}
S\left[\Phi_{\mu,z}\right]=\sum_{n}\int d^{4}x\left[\frac{1}{2}\partial_{[\mu}B_{\nu]}^{\left(n\right)\dagger}\partial^{[\mu}B^{\left(n\right)\nu]}+m_{n}^{2}B_{\mu}^{\left(n\right)\dagger}B^{\left(n\right)\mu}\right],
\end{equation}
with the normalization

\begin{equation}
4\kappa\int dzf\left(z\right)\phi_{\left(n\right)}^{H}\phi_{\left(m\right)}^{H}=\delta_{mn},
\end{equation}
and eigenvalue equation,

\begin{equation}
-\frac{d}{dz}\left(g\frac{d\phi_{\left(n\right)}^{H}}{dz}\right)+2f\left(z\right)\left[-m_{n}^{2}+\tilde{v}^{2}a\left(z\right)\right]\phi_{n}^{H}=0.
\end{equation}
For the transverse modes longitudinal modes, the expansion is suggested
as,

\begin{align}
\Phi_{\mu}= & -\sum_{n}\frac{1}{2a\left(z\right)f\left(z\right)m_{n}^{2}}\frac{d}{dz}\left[a\left(z\right)g\left(z\right)\epsilon_{n}\left(z\right)\right]\partial_{\mu}D_{\nu}\left(x\right),\nonumber \\
\Phi_{z}= & \sum_{n}\epsilon_{n}\left(z\right)D_{n}\left(x\right),\label{eq:179}
\end{align}
leading to

\begin{equation}
S\left[\Phi_{\mu,z}\right]=\sum_{n}\int d^{4}x\left(\partial_{\mu}D_{n}^{\dagger}\partial^{\mu}D_{n}+m_{n}^{2}D_{n}^{\dagger}D_{n}\right),
\end{equation}
with the normalization

\begin{equation}
2\kappa=\int dza\left(z\right)g\left(z\right)\epsilon_{n}\left(z\right)\epsilon_{m}\left(z\right)=\frac{m_{n}^{2}}{\tilde{v}^{2}}\delta_{mn}.
\end{equation}
We note that, with the replacement (\ref{eq:170}), the Chern-Simons
action (\ref{eq:57}) for the D8-branes reduces to additional terms
as,

\begin{align}
\mathcal{L}_{\mathrm{CS}}\left[\Phi_{\mu,z}\right]= & -\frac{N_{c}}{24\pi^{2}}\left(d\Phi^{\dagger}Ad\Phi+d\Phi^{\dagger}dA\Phi+\Phi^{\dagger}dAd\Phi\right)\nonumber \\
 & +\frac{iN_{c}}{16\pi^{2}}\left(d\Phi^{\dagger}A^{2}\Phi+\Phi^{\dagger}A^{2}d\Phi+\Phi^{\dagger}AdA\Phi+\Phi^{\dagger}dAA\Phi\right)\nonumber \\
 & +\frac{5N_{c}}{48\pi^{2}}\Phi^{\dagger}A^{3}\Phi+\mathcal{O}\left(\Phi^{4},A\right).\label{eq:182}
\end{align}
Using the expansion (\ref{eq:175}) (\ref{eq:179}) and (\ref{eq:46}),
the DBI and CS term includes the interaction between light and heavy-light
mesons. 

It is also possible to obtain the baryon spectrum with heavy flavor
by considering baryon vertex as the instantons on the flavor brane
\cite{key-101,key-102,key-103}. Follow the steps in Section 2.5,
the equations of motion for the heavy-light meson field is derived
as,

\begin{align}
D_{M}D_{M}\Phi_{N}-D_{N}D_{M}\Phi_{M}+2iF_{MN}\Phi_{M}+\mathcal{O}\left(\lambda^{-1}\right) & =0,\nonumber \\
D_{M}\left(D_{0}\Phi_{M}-D_{M}\Phi_{0}\right)-iF^{0M}\Phi_{M}\nonumber \\
-\frac{1}{64\pi^{2}a}\epsilon_{MNPQ}\mathcal{K}_{MNPQ}+\mathcal{O}\left(\lambda^{-1}\right) & =0,
\end{align}
with $M,N=1,2,3,z$ and

\begin{equation}
\mathcal{K}_{MNPQ}=i\partial_{M}A_{N}\partial_{P}\Phi_{Q}-A_{M}A_{N}\partial_{P}\Phi_{Q}-\partial_{M}A_{N}A_{P}\Phi_{Q}-\frac{5i}{6}A_{M}A_{N}A_{P}\Phi_{Q},
\end{equation}
where $F=dA+\frac{i}{2}\left[A,A\right]$ is computed by the BPST
instanton solution given in (\ref{eq:82}) - (\ref{eq:85}). Thus
$\Phi_{\mu,z}$ can be solved as $\Phi_{\mu,z}=e^{\pm im_{H}t}\phi_{\mu,z}\left(x\right)$
by

\begin{align}
\phi_{0} & =-\frac{1}{1024a\pi^{2}}\left[\frac{25\rho}{2\left(x^{2}+\rho^{2}\right)^{5/2}}+\frac{7}{\rho\left(x^{2}+\rho^{2}\right)^{3/2}}\right]\chi,\nonumber \\
\phi_{M} & =\frac{\rho}{\left(x^{2}+\rho^{2}\right)^{3/2}}\sigma_{M}\chi,\label{eq:185}
\end{align}
where $\sigma_{M}$ is the embedded Pauli matrices as $\sigma_{M}/2=\left(t_{i},-\mathbf{1}_{N_{f}}\right)$
and $\chi$ is the $SU\left(N_{f}\right)$ spinor independent on $x,z$.
The soliton mass as the baryonic potential can be evaluated by inserting
(\ref{eq:185}) and the BPST solution (\ref{eq:82}) - (\ref{eq:85})
to the full action for the D8-branes. Afterwards one reaches to an
effective Hamiltonian for baryon state by following Section 2.5 as
it is given in \cite{key-102,key-103}. We note that \cite{key-39}
also gives another generalization with heavy flavor into black D4-brane
background.

\subsection{Interaction of hadron and glueball}

The interaction of hadrons relates to many significant topics in QCD
and nuclear physics, and its holographic description by the D4/D8
model has been reviewed briefly in Section 2 and \cite{key-23,key-24}.
In this section, we will take a look at the interaction in hadron
physics involving glueballs since the D4/D8 model provides explicit
definition of meson, baryon and glueball. 

The main idea to include the interaction of meson and glueball is
to consider the D8-brane action with a gravitational fluctuation.
Recall the discussion in Section 2.4 and 2.6, since meson is identified
as the gauge field on the D8-branes (created by $8-8$ string) and
glueball is identified as the gravitational polarization (close string),
the interaction of meson and glueball is nothing but the interaction
of open and close string which can be therefore included into the
D8-brane action when the metric fluctuation is picked up. For example,
when we put the gravitational polarization (\ref{eq:97}) into the
D8-brane action (\ref{eq:32}) with the meson tower given in (\ref{eq:37}),
by integrating out the $z$ dependence (the holographic coordinate)
it reduces to interaction action involving $\pi,\rho$ meson and exotic
glueball $G_{E}$ after some straightforward but messy calculations
as,

\begin{align}
S_{G_{E}}^{\pi-\rho}= & -\mathrm{Tr}\int d^{4}x\bigg\{ c_{1}\left[\frac{1}{2}\partial_{\mu}\pi\partial_{\nu}\pi\frac{\partial^{\mu}\partial^{\nu}}{M_{E}^{2}}G_{E}+\frac{1}{4}\left(\partial_{\mu}\pi\right)^{2}\left(1-\frac{\partial^{2}}{M_{E}^{2}}\right)G_{E}\right]\nonumber \\
 & +c_{2}M_{KK}^{2}\left[\frac{1}{2}\rho_{\mu}\rho_{\nu}\frac{\partial^{\mu}\partial^{\nu}}{M_{E}^{2}}G_{E}+\frac{1}{4}\left(\rho_{\mu}\right)^{2}\left(1-\frac{\partial^{2}}{M_{E}^{2}}\right)G_{E}\right]\nonumber \\
 & +c_{3}\left[\frac{1}{2}\eta^{\sigma\lambda}\partial_{[\mu}\rho_{\sigma]}\partial_{[\nu}\rho_{\lambda]}\frac{\partial^{\mu}\partial^{\nu}}{M_{E}^{2}}G_{E}-\frac{1}{8}\partial_{[\mu}\rho_{\nu]}\partial^{[\mu}\rho^{\nu]}\left(1+\frac{\partial^{2}}{M_{E}^{2}}\right)G_{E}\right]+c_{4}\frac{3}{2M_{E}^{2}}\rho_{\mu}\partial^{[\mu}\rho^{\nu]}\partial_{\nu}G_{E}\nonumber \\
 & +c_{5}\left[\partial_{\mu}\pi\left[\pi,\rho_{\nu}\right]\frac{\partial^{\mu}\partial^{\nu}}{M_{E}^{2}}G_{E}+\frac{1}{2}\partial_{\mu}\pi\left[\pi,\rho^{\mu}\right]\left(1-\frac{\partial^{2}}{M_{E}^{2}}\right)G_{E}\right]\nonumber \\
 & +\left[\frac{1}{2}\tilde{c}_{1}\left(\partial_{\mu}\pi\right)^{2}+\frac{1}{2}\tilde{c}_{2}M_{KK}^{2}\left(\rho_{\mu}\right)^{2}+\frac{1}{4}\tilde{c}_{3}\partial_{[\mu}\rho_{\nu]}\partial^{[\mu}\rho^{\nu]}+\tilde{c}_{5}\partial_{\mu}\pi\left[\pi,\rho^{\mu}\right]\right]G_{E}\bigg\},\label{eq:186}
\end{align}
where the coefficients $c_{i}$'s and $\tilde{c}_{i}$'s are coupling
constants and numerically computed as (in the unit of $\lambda^{-1/2}N_{c}^{-1}M_{KK}^{-1}$),

\begin{align}
c_{1} & =\int dZ\frac{\bar{H}_{E}}{\pi K}=62.6,\ \ \ \ c_{2}=2\kappa\int dZK\left(\psi_{1}^{\prime}\right)\bar{H}_{E}=7.1,\nonumber \\
c_{3} & =2\kappa\int dZK^{-1/3}\left(\psi_{1}\right)^{2}\bar{H}_{E}=69.7,\nonumber \\
c_{4} & =2\kappa M_{KK}^{2}\int dZ\frac{20ZK}{\left(5K-2\right)^{2}}\psi_{1}\psi_{1}^{\prime}H_{E}=10.6M_{KK}^{2},\nonumber \\
c_{5} & =\int dZ\frac{\psi_{1}\bar{H}_{E}}{\pi K}=2019.6N_{c}^{-1/2},\nonumber \\
\tilde{c}_{1} & =\int dZ\frac{H_{E}}{4\pi K}=16.4,\ \ \ \ \tilde{c}_{2}=\frac{1}{2}\kappa\int dZK\left(\psi_{1}^{\prime}\right)^{2}H_{E}=3.0,\nonumber \\
\tilde{c}_{3} & =\frac{1}{2}\kappa\int dZK^{-1/3}\left(\psi_{1}\right)^{2}H_{E}=18.1,\ \ \ \ \tilde{c}_{5}=\int dZ\frac{\psi_{1}H_{E}}{4\pi K}=508.2N_{c}^{-1/2}.
\end{align}
Then the associated amplitude of glueball decay can be further evaluated
by using the effective action $S_{G_{E}}^{\pi-\rho}$ with the coupling
constants. And one can also compute the effective action of meson
involving other types of glueball by changing the formulas of the
bulk gravitational polarization as it is discussed in \cite{key-28,key-29,key-30,key-31}.

The current setup to obtain an effective action of meson and glueball
interaction can also be generalized by including heavy flavor \cite{key-105}
which is to take into account the configuration (b) in Figure \ref{fig:7}
and the heavy-light meson field. The main idea is to pick up the gravitational
polarization in bulk metric when we write down the D8-brane action
with heavy flavor brane (i.e. with the replacement given in (\ref{eq:170})).
For example, by considering the gravitational polarization for the
exotic glueball $G_{E}$ in (\ref{eq:97}), the effective action of
heavy-light meson and glueball provides terms as,

\begin{align}
\partial^{2}G_{E}\partial_{\mu}Q_{\nu}^{\dagger}\partial^{\mu}Q^{\nu}, & \partial^{2}G_{E}Q_{\mu}^{\dagger}Q^{\mu},\partial^{2}G_{E}\partial_{\mu}Q^{\dagger}\partial^{\mu}Q,\partial^{2}G_{E}Q^{\dagger}Q,\nonumber \\
\partial^{\mu}\partial^{\rho}G_{E}\partial_{\mu}Q_{\nu}^{\dagger}\partial_{\rho}Q^{\nu}, & \partial^{\mu}\partial^{\nu}G_{E}\partial_{\mu}Q^{\dagger}\partial_{\nu}Q,\partial^{\mu}\partial^{\nu}G_{E}Q_{\mu,}^{\dagger}Q_{\nu,}\nonumber \\
G_{E}\partial_{\mu}Q_{\nu}^{\dagger}\partial^{\mu}Q^{\nu}, & G_{E}Q_{\mu}^{\dagger}Q^{\nu},G_{E}Q^{\dagger}Q,\nonumber \\
\partial_{\sigma}G_{E}\partial^{\rho}Q^{\sigma\dagger}\partial_{\rho}Q, & \partial_{\sigma}G_{E}\partial^{\rho}Q^{\sigma\dagger}Q_{\rho},\partial^{\mu}G_{E}Q_{\mu}^{\dagger}Q,
\end{align}
which are the same types as the interaction given in (\ref{eq:186}).
Here we have used $Q_{\mu},Q$ to denote the vector and scalar heavy-light
meson field, and the lowest heavy-light meson is identified to be
D-meson with a charm quark. Accordingly, the effective action with
heavy-flavor and glueball may be useful to study the oscillation of
D-meson pairs ($D-\bar{D}$) or B-meson pairs ($B-\bar{B}$) \cite{key-106,key-107}.
Note that since the heavy-light multiplet is created by the heavy-light
string, even if the heavy flavor is taken into account, the interaction
of heavy-light meson and glueball remains to be the open/close string
interaction through holography. Besides, in the presence of the heavy-light
meson and glueball, the effective action also mixes the interaction
terms of glueball, light and heavy-light meson which may describe
the various interaction in hadron physics. 

\begin{figure}
\begin{centering}
\includegraphics[scale=0.3]{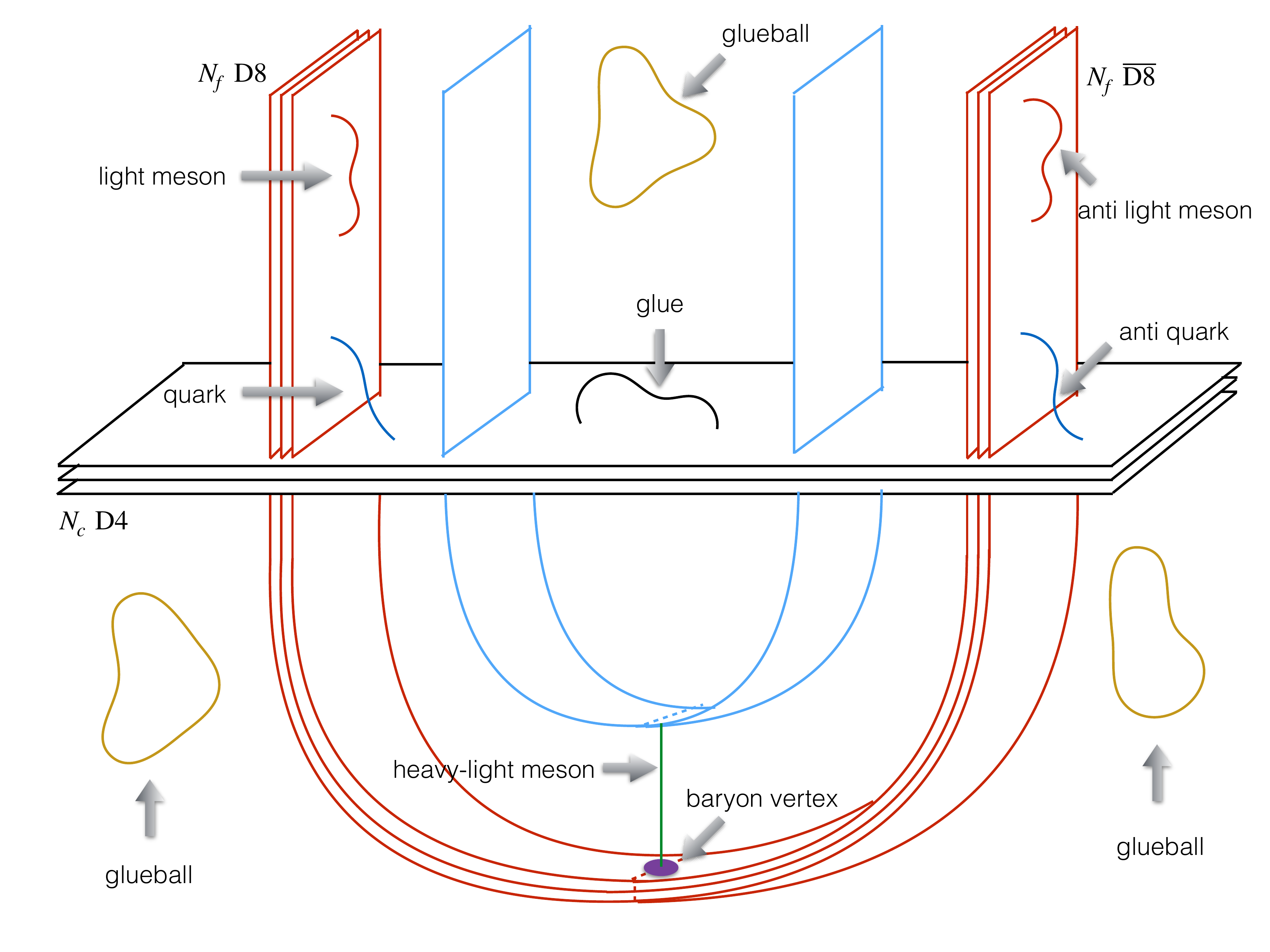}
\par\end{centering}
\caption{\label{fig:8} Strings as various hadrons in the D4/D8 model. }

\end{figure}

It is also possible to include the interaction of baryon (or baryonic
meson) and glueball in a parallel way, that is to consider the interaction
of baryonic $\mathrm{D4}^{\prime}$-branes and bulk close string \cite{key-108}.
Specifically, one can derive the Yang-Mills action presented in (\ref{eq:32})
with the gravitational polarization (\ref{eq:97}), then insert the
BPST instanton configuration (\ref{eq:82}) - (\ref{eq:85}) as baryon
under the large $\lambda$ rescaling (\ref{eq:77}). Afterwards, by
following the discussion in Section 2.5, we can obtain additional
terms to the collective Hamiltonian (\ref{eq:96}) as,

\begin{equation}
\Delta H\left(t\right)\simeq\cos\left(\omega t\right)\left\{ -\frac{27}{2}\pi^{2}+\left[-\frac{59049\pi^{4}}{20\rho^{2}}+\frac{27M_{E}^{2}+468}{16}\pi^{2}\left(2Z^{2}+\rho^{2}\right)\right]\lambda^{-1}+\mathcal{O}\left(\lambda^{-2}\right)\right\} .
\end{equation}
Using the standard technique in quantum mechanics for the time-dependent
perturbed Hamiltonian, it is possible to work out the decay rate of
baryon involving glueball and its associated select rule. We note
that, when the heavy flavor is included as in Section 2.3, moreover
the decay rate of heavy-light baryon or baryonic meson involving glueball
is able to be achieved. For example consider the exotic gravitational
polarization (\ref{eq:97}), we can reach to the time-dependent perturbed
Hamiltonian as \cite{key-109}, 

\begin{equation}
\Delta H\left(t\right)\simeq\lambda^{-1/2}M_{KK}^{-1}\left(\frac{5}{216\pi}m_{H}^{2}+\frac{15m_{H}}{32\rho^{2}}\right)G_{E}\chi^{\dagger}\chi,\label{eq:190}
\end{equation}
where we have taken the limit $m_{H}\rightarrow\infty$ followed $\lambda\rightarrow\infty$
to simplify the formula and $\chi^{\dagger}\chi=N_{Q}$ refers to
the number of heavy-flavor quark in the heavy-light meson. Then the
decay of heavy-light baryonic matter involving glueball can be evaluated
by using (\ref{eq:190}) to the quantum mechanical system (\ref{eq:96})
with heavy flavors. To close this section, we summarize the strings
as various hadrons in the D4/D8 model in Figure \ref{fig:8}, and
we can see the various interactions of hadrons are interaction of
strings through gauge-gravity duality in this model. 

\subsection{Theta dependence in QCD}

In Yang-Mills theory, there could be an topological term proportional
to the $\theta$ angle \cite{key-110}. In the large $N_{c}$ limit,
the full Lagrangian takes the form as,

\begin{equation}
S=\frac{N_{c}}{2\lambda}\mathrm{Tr}\int\ ^{\star}F\wedge F-i\frac{\lambda}{8\pi^{2}}\frac{\theta}{N_{c}}\mathrm{Tr}\int F\wedge F.
\end{equation}
While the value of $\theta$ angle may be experimentally small, it
leads to many interesting effects e.g. glueball spectrum \cite{key-111},
deconfinement transition \cite{key-112,key-113}, chiral magnetic
effect \cite{key-114,key-115}, especially its large $N_{c}$ limit
\cite{key-116}. Since the D4/D8 model is a holographic version of
QCD, it is possible to introduce the $\theta$ term to the dual theory
through the gauge-gravity duality.

The main idea to include an Yang-Mills $\theta$ term in holography
is to introduce coincident $N_{0}$ D0-branes acting to the $N_{c}$
D4-brane background (\ref{eq:14}). In this sense, we have to require
$N_{0}/N_{c}$ is fixed when $N_{c}\rightarrow\infty$. And for the
SUGRA approach, the dynamics of the Ramond-Ramond 1-form $C_{1}$
must be picked up into the IIA SUGRA action (\ref{eq:15}) in order
to include the charge of the D0-branes as,

\begin{equation}
S_{\mathrm{IIA}}^{10\mathrm{d}}=\frac{1}{2\kappa_{10}^{2}}\int d^{10}x\sqrt{-G}e^{-2\phi}\left[\mathcal{R}^{(10)}+4\partial_{\mu}\phi\partial\phi\right]-\frac{1}{4\kappa_{10}^{2}}\int d^{10}x\sqrt{-G}\left(\left|F_{4}\right|^{2}+\left|F_{2}\right|^{2}\right),\label{eq:192}
\end{equation}
where $F_{2}=dC_{1}$. To obtain an analytical solution, we assume
the D0-branes are smeared homogeneous along $x^{4}$, hence the associated
equations of motion to (\ref{eq:192}) can be solved as,

\begin{align}
ds^{2} & =H_{4}^{-1/2}\left[-H_{0}^{-1/2}f_{T}\left(U\right)\left(dx^{0}\right)^{2}+H_{0}^{1/2}dx_{i}dx^{i}\right]+H_{4}^{1/2}H_{0}^{1/2}\left[\frac{dU^{2}}{f_{T}\left(U\right)}+U^{2}d\Omega_{4}^{2}\right],\nonumber \\
H_{4} & =1+\frac{R^{3}}{U^{3}},H_{0}=1-\frac{U_{H}^{3}}{U^{3}}\frac{\Theta^{2}}{1+\Theta^{2}},e^{\phi}=H_{4}^{-1/4}H_{0}^{3/4},C_{1}=\frac{\Theta}{g_{s}}\frac{f_{T}}{H_{0}}dx^{4},F_{4}=3R^{3}g_{s}^{-1},
\end{align}
where $\Theta$ is a constant parameter. Take the near horizon limit
so that $H_{4}\rightarrow R^{3}/U^{3}$ and impose the double Wick
rotation as it is discussed in Section 2.1, we can obtain a D0-D4
bubble background associated to (\ref{eq:16}) as,

\begin{align}
ds^{2} & =H_{4}^{-1/2}\left[H_{0}^{1/2}\eta_{\mu\nu}dx^{\mu}dx^{\nu}+H_{0}^{-1/2}f\left(U\right)\left(dx^{4}\right)^{2}\right]+H_{4}^{1/2}H_{0}^{1/2}\left[\frac{dU^{2}}{f_{T}\left(U\right)}+U^{2}d\Omega_{4}^{2}\right],\nonumber \\
H_{4} & =\frac{R^{3}}{U^{3}},H_{0}=1-\frac{U_{KK}^{3}}{U^{3}}\frac{\Theta^{2}}{1+\Theta^{2}},e^{\phi}=H_{4}^{-1/4}H_{0}^{3/4},C_{1}=-i\frac{\Theta}{g_{s}}\frac{f}{H_{0}}dx^{4},F_{4}=3R^{3}g_{s}^{-1}.\label{eq:194}
\end{align}
Due to the presence of the D0-branes, we can see the dual theory to
(\ref{eq:194}) is pure Yang-Mills theory with a $\theta$ term if
a probe D4-brane located at the holographic boundary is taken into
account as,

\begin{align}
S_{\mathrm{D4}}= & \big[-T_{\mathrm{D4}}\mathrm{STr}\int d^{4}xd\tau e^{-\phi}\sqrt{-\det\left(G+2\pi\alpha^{\prime}F\right)}+g_{s}T_{\mathrm{D4}}\int C_{5}\nonumber \\
 & +\frac{1}{2}\left(2\pi\alpha^{\prime}\right)^{2}g_{s}T_{\mathrm{D4}}\int C_{1}\wedge F\wedge F\big]|_{U\rightarrow\infty}\nonumber \\
\simeq & -\frac{1}{2g_{\mathrm{YM}}^{2}}\mathrm{Tr}\int{}^{*}F\wedge F+i\frac{g_{\mathrm{YM}}^{2}}{8\pi^{2}}\theta\mathrm{Tr}\int F\wedge F+\mathcal{O}\left(F^{4}\right),
\end{align}
which further implies the bare $\theta$ angle relates to the $\Theta$
parameter in the solution (\ref{eq:194}) by

\begin{equation}
\Theta=\frac{\lambda}{8\pi^{2}}\left(\frac{\theta+2k\pi}{N_{c}}\right),k\in\mathbb{Z}
\end{equation}
as a fixed constant in the large $N_{c}$ limit.

With the geometry background (\ref{eq:194}), it is possible to evaluate
several properties of Yang-Mills theory with a $\theta$ term by following
the discussions in previous sections and let us take a brief look
at them for examples. First we focus on the the ground state energy
which can be evaluated by using (\ref{eq:104}) - (\ref{eq:108})
as,

\begin{equation}
F\left(\Theta\right)=-\frac{2N_{c}^{2}\lambda}{3^{7}\pi^{2}}\frac{M_{KK}^{4}}{\left(1+\Theta^{2}\right)^{3}}.\label{eq:197}
\end{equation}
In the expansion with respect to small $\Theta$, (\ref{eq:197})
reduces to minimized free energy difference,

\begin{equation}
\min_{k}\left[F\left(\Theta\right)-F\left(0\right)\right]\simeq\frac{1}{2}\chi_{g}\theta^{2}\left[1+\bar{b}_{2}\frac{\theta^{2}}{N_{c}^{2}}+\bar{b}_{4}\frac{\theta^{4}}{N_{c}^{4}}+\mathcal{O}\left(\theta^{6}\right)\right],
\end{equation}
as the energy of the $\theta$ vacuum. And the topological susceptibility
reads,

\begin{equation}
\chi_{g}=\frac{\lambda^{3}M_{KK}^{4}}{32\left(3\pi\right)^{6}},
\end{equation}
with

\begin{equation}
\bar{b}_{2}=-\frac{\lambda^{2}}{32\pi^{4}},\bar{b}_{4}=-\frac{5\lambda^{4}}{3\times2^{11}\pi^{8}}.
\end{equation}
Moreover, one can consider a constant $C_{1}$ in the black D4-brane
background (\ref{eq:14}) so that $F_{2}=dC_{1}=0$. Hence the ground
state energy of deconfined Yang-Mills theory with a constant $\theta$
term can be identified to $F_{\mathrm{deconf}.}$ presented in (\ref{eq:109}).
Then the QCD deconfinement phase transition can be obtained by comparing
the free energy (\ref{eq:197}) with $F_{\mathrm{deconf}.}$ in (\ref{eq:109})
which leads to the critical temperature as,

\begin{equation}
T_{c}\left(\Theta\right)=\frac{M_{KK}}{2\pi}\frac{1}{\sqrt{1+\Theta^{2}}}\simeq\frac{M_{KK}}{2\pi}\left[1-\frac{\lambda^{2}}{128\pi^{4}}\frac{\theta^{2}}{N_{c}^{2}}+\frac{3\lambda^{4}}{2^{15}\pi^{8}}\frac{\theta^{4}}{N_{c}^{4}}+\mathcal{O}\left(\frac{\theta^{6}}{N_{c}^{6}}\right)\right].
\end{equation}

Second, the QCD string tension also takes a correction due to the
presence of $\theta$ term. Consider an open string stretched in the
background (\ref{eq:194}) ending on a probe D4-brane at boundary.
Use the AdS/CFT dictionary, the Wilson loop in the dual theory relates
to the classical Nambu-Goto (NG) action $S_{NG}$ of the open string
corresponding to the tension $T$ with quark potential $V$ as,

\begin{equation}
\left\langle W\left(\mathcal{C}\right)\right\rangle \sim e^{-S_{NG}}\sim e^{-TV}.
\end{equation}
In the static gauge, the relevant string embedding can be chosen as
\begin{equation}
\tau=x^{0}\in\left[0,T\right],\sigma=x\in\left[-l/2,l/2\right],U=U\left(x\right),
\end{equation}
then the NG action is given by

\begin{equation}
S_{NG}=-\frac{1}{2\pi\alpha^{\prime}}\int d\tau d\sigma\sqrt{-g_{\tau\tau}g_{\sigma\sigma}}=-\frac{1}{2\pi\alpha^{\prime}}\int d\tau d\sigma\sqrt{-g_{00}\left[g_{xx}+g_{UU}U^{\prime}\left(x\right)^{2}\right]},
\end{equation}
To quickly evaluate the QCD tension, let us consider the limit $l\rightarrow\infty$.
In this limit, the open string must minimize its energy as possible
as it can, it forces the factor $\sqrt{-g_{00}g_{xx}}$ to become
minimal to take the value at $U=U_{KK}$ since the size of $x^{4}$
shrinks at $U=U_{KK}$. Therefore the QCD tension $T_{s}$ is obtained
from

\begin{equation}
S_{NG}\simeq-\frac{1}{2\pi\alpha^{\prime}}Tl\sqrt{-g_{00}g_{xx}}|_{U=U_{KK}}=-T_{s}Tl,
\end{equation}
as

\begin{equation}
T_{s}=\frac{1}{2\pi\alpha^{\prime}}\sqrt{-g_{00}g_{xx}}|_{U=U_{KK}}=\frac{\lambda}{27\pi}M_{KK}^{2}\frac{1}{\left(1+\Theta^{2}\right)^{2}}.
\end{equation}

Next, let us investigate the glueball mass with the background (\ref{eq:194}).
As the glueball corresponds to the gravitational fluctuation, by adding
a perturbation to metric presented in (\ref{eq:194}) as $g_{MN}\rightarrow g_{MN}+h_{MN}$,
it reduces to equation of motion for $h_{MN}$ as

\begin{equation}
\frac{1}{2}\nabla_{M}\nabla_{N}h_{\ K}^{K}+\frac{1}{2}\nabla^{2}h_{MN}-\left(\nabla^{K}\nabla_{M}h_{NK}+\nabla^{K}\nabla_{N}h_{MK}\right)-\frac{3}{2}h_{MN}=0.
\end{equation}
Here since IIA SUGRA can be obtained by the dimension reduction from
11d M-theory, $h_{MN}$ refers to the fluctuation on $\mathrm{AdS}_{7}$
which means $M,N$ runs over 0 - 6. Setting $h_{MN}=H_{MN}\left(U\right)e^{-ik\cdot x}$
with the ansatz 

\begin{equation}
H_{MN}\left(U\right)=\frac{U}{R}H\left(U\right)\mathrm{diag}\left(0,1,1,1,0,-\frac{3}{1+\Theta^{2}},0\right),
\end{equation}
it gives the eigen equation for $H\left(U\right)$ as,

\begin{equation}
H^{\prime\prime}\left(U\right)+\frac{4U^{3}-U_{KK}^{3}}{U\left(U^{3}-U_{KK}^{3}\right)}H^{\prime}\left(U\right)-\frac{M^{2}R^{2}}{U^{3}-U_{KK}^{3}}H\left(U\right)=0,
\end{equation}
which implies the mass spectrum $M$ with a correction due to $\Theta$,

\begin{equation}
M\simeq\frac{M\left(0\right)}{\sqrt{1+\Theta^{2}}}.
\end{equation}

The presence of $\theta$ also decreases the baryon mass $m_{B}$
as,

\begin{equation}
m_{B}=\frac{\lambda}{27\pi}N_{c}M_{KK}\frac{1}{\left(1+\Theta^{2}\right)^{3/2}},
\end{equation}
by imposing the metric presented in (\ref{eq:194}) into (\ref{eq:76})
which implies the evidence of metastable particles in QCD. By further
analyze the entanglement entropy on (\ref{eq:194}), it agree consistently
with the property of the possible metastable states in this model.
Besides, when we follow the discussion in Section 2.2, it is possible
to introduce flavored meson in the D0-D4 background. And the meson
mass also acquires the correction by the $\theta$ angle as \cite{key-40,key-41}.
Further follow the instantonic description for baryon in Section 2.5,
one can see the metastable baryonic spectrum in this model as \cite{key-117}.
In this sense, the Witten-Sakai-Sugimoto model in the D0-D4 brane
background is recognized as a holographic version of QCD with a $\theta$
term.

\section{Summary and outlook}

In this review, we look back to the fundamental properties of the
D4/D8 model which includes the D4-brane background, the embedding
of the $\mathrm{D8/\overline{D8}}$-branes and how to identify meson,
baryon, glueball in this model. Besides, we revisit some interesting
topics about QCD by using this model which relates to the deconfinement
transition, chiral phase, heavy flavor, various interaction of hadrons
and the $\theta$ term in QCD. This review illustrates that string
theory can provide a powerful method for studying the strongly coupled
regime of QCD, which is out of reach for the traditional methods of
perturbative QFT. We particularly note here there are additional interesting
approaches based on this model absent in the main text of this review,
they relates to the holographic Schwinger effect \cite{key-118,key-119,key-120,key-121},
the fluid/gravity correspondence \cite{key-122,key-123,key-124,key-125,key-126},
corrections to the instanton as baryon \cite{key-127,key-128}, the
approaches to the D3/D7 model \cite{key-129,key-130} and applications
to study neutron stars \cite{key-1+4,key-1+5}. With all of these
achievements, it may be possible to work out an exactly holographic
version of QCD based on the D4/D8 model in the future work, to reinterpret
the fundamental element of strong interaction according to gauge-gravity
duality.

\section*{Acknowledgements}

This work is supported by the National Natural Science Foundation
of China (NSFC) under Grant No. 12005033 and the Fundamental Research
Funds for the Central Universities under Grant No. 3132023198.

\section{Appendix A: The type II supergravity solution}

In this appendix, let us collect the $\mathrm{D}_{p}$-brane solution
in the type II SUGRA. We note that all the discussion in this appendix
is valid to the gravity solution presented in the main text if we
set $p=4$. In the string frame, the action for type II SUGRA sourced
by a stack of $N_{p}$ coincident $\mathrm{D}_{p}$-branes can be
written as, 

\begin{equation}
S_{\mathrm{II}}=\frac{1}{2\kappa_{10}^{2}}\int d^{10}x\sqrt{-g}\left[e^{-2\phi}\left(\mathcal{R}+4\partial_{M}\phi\partial^{M}\phi\right)-\frac{g_{s}^{2}}{2}\left|F_{p+2}\right|^{2}\right],\tag{A-1}\label{eq:213}
\end{equation}
where $2\kappa_{10}^{2}=16\pi G_{10}=\left(2\pi\right)^{7}l_{s}^{8}g_{s}^{2}$
is the 10d gravity coupling constant, $\mathcal{R},\phi,C_{p+1}$
is respectively the 10d curvature, dilaton and Ramond-Ramond $p+1$-form
field with $F_{p+2}=dC_{p+1}$. Note that, in string theory the dilaton
field may also be defined as $\Phi$ by

\begin{equation}
\Phi-\Phi_{0}=\phi,e^{\Phi_{0}}=g_{s}.\tag{A-2}
\end{equation}
Since a $\mathrm{D}_{p}$-brane for $p=4,5,6$ is magnetically dual
to $\mathrm{D}_{p}$-brane for $p=2,1,0$ and D3-brane is self dual,
we only consider the case for $p<7$ in (\ref{eq:213}). Vary the
action (\ref{eq:213}) with respect to $g_{MN},\phi,C_{p+1}$, the
associated equations of motion are collected as,

\begin{align}
0= & \mathcal{R}+4\nabla^{2}\phi-4\left(\nabla\phi\right)^{2},\nonumber \\
0= & \partial_{N}\left(\sqrt{-g}F^{NM_{1}...M_{p+1}}\right),\nonumber \\
0= & \mathcal{R}_{MN}-\frac{1}{2}g_{MN}\mathcal{R}+2\nabla_{M}\nabla_{N}\phi+2g_{MN}\left(\nabla\phi\right)^{2}-2g_{MN}\nabla^{2}\phi\nonumber \\
= & \frac{g_{s}^{2}}{2\left(p+1\right)!}e^{2\phi}\left[F_{M}^{\ K_{1}...K_{p+1}}F_{NK_{1}...K_{p+1}}-\frac{\left(p+1\right)!}{2}g_{MN}\left|F_{p+2}\right|^{2}\right].\tag{A-3}\label{eq:215}
\end{align}
The solution for (\ref{eq:215}) can be obtained by using the simply
homogeneous ansatz,

\begin{align}
ds^{2} & =H_{p}^{-\frac{1}{2}}\eta_{ab}dx^{a}dx^{b}+H_{p}^{\frac{1}{2}}\left(dr^{2}+r^{2}d\Omega_{8-p}^{2}\right),\ a,b=0,1...p\nonumber \\
e^{\phi} & =H_{p}^{-\frac{p-3}{4}},C_{01...p}=g_{s}^{-1}H_{p}^{-1},F_{r01...p}=\frac{\left(7-p\right)g_{s}^{-1}h_{p}^{7-p}}{r^{8-p}H_{p}^{2}},\tag{A-4}\label{eq:216}
\end{align}
where the harmonic function $H_{p}$ is solved through (\ref{eq:215})
as,

\begin{equation}
H_{p}\left(r\right)=1+\frac{h_{p}^{7-p}}{r^{7-p}}.\tag{A-5}
\end{equation}
Here $r$ refers to the radial coordinate vertical to the $\mathrm{D}_{p}$-brane,
$\Omega_{8-p}$ is the associated angle coordinate in the transverse
space. The constant $h_{p}$ relates to the charge of the $\mathrm{D}_{p}$-brane
computed as,

\begin{equation}
h_{p}^{7-p}=\left(2\sqrt{\pi}\right)^{5-p}g_{s}N_{p}\Gamma\left(\frac{7-p}{2}\right)l_{s}^{7-p}.\tag{A-6}
\end{equation}
The solution (\ref{eq:216}) representing extremal black $\mathrm{D}_{p}$-branes
reduces to the BPS condition as,

\begin{equation}
2\kappa_{10}^{2}g_{s}T_{\mathrm{D}_{p}}N_{p}=\int_{S^{8-p}}\ ^{\star}F_{p+2},\tag{A-7}
\end{equation}
due to the action for the Ramond-Ramond (R-R) field $C_{p+1}$ with
a source of $N_{p}$ coincident $\mathrm{D}_{p}$-branes,

\begin{equation}
S_{\mathrm{R-R}}=-\frac{1}{4\kappa_{10}^{2}}\int F_{p+2}\wedge\ ^{\star}F_{p+2}+g_{s}T_{\mathrm{D}_{p}}\int_{p+1}C_{p+1}.\tag{A-8}
\end{equation}
The equations of motion (\ref{eq:215}) also allow the near-extremal
solution as,

\begin{align}
ds^{2} & =H_{p}^{-\frac{1}{2}}\left[f\left(r\right)dt^{2}+\delta_{ij}dx^{i}dx^{j}\right]+H_{p}^{\frac{1}{2}}\left[\frac{dr^{2}}{f\left(r\right)}+r^{2}d\Omega_{8-p}^{2}\right],\nonumber \\
e^{\phi} & =H_{p}^{-\frac{p-3}{4}},C_{01...p}=g_{s}^{-1}H_{p}^{-1},F_{r01...p}=\frac{\left(7-p\right)g_{s}^{-1}h_{p}^{7-p}}{r^{8-p}H_{p}^{2}},\tag{A-9}\label{eq:221}
\end{align}
where $i,j$ run over the spacial index of the $\mathrm{D}_{p}$-branes.
The functions $f\left(r\right),H_{p}\left(r\right)$ are solved respectively
as,

\begin{equation}
f\left(r\right)=1-\frac{r_{H}^{7-p}}{r^{7-p}},H_{p}\left(r\right)=1+\frac{r_{p}^{7-p}}{r^{7-p}},\tag{A-9}
\end{equation}
where $r_{H}$ refers to the horizon of the $\mathrm{D}_{p}$-branes.
Notice the equation of motion (\ref{eq:215}) reduce to a constraint,

\begin{equation}
4\nabla_{M}\phi\nabla^{M}\phi-2\nabla^{2}\phi=\frac{p-3}{2}g_{s}^{2}e^{2\phi}\left|F_{p+2}\right|^{2},\tag{A-10}
\end{equation}
which implies,

\begin{equation}
r_{p}^{7-p}=\sqrt{h_{p}^{2\left(7-p\right)}+\left(\frac{r_{H}^{7-p}}{2}\right)^{2}}-\frac{r_{H}^{7-p}}{2}.\tag{A-11}
\end{equation}
So we have $r_{p}\rightarrow h_{p}$ if $r_{H}\rightarrow0$, thus
the near-extremal solution will return to the extremal solution in
this limit.

\section{Appendix B: Dimensional reduction for\textcolor{blue}{{} }spinors}

In this section, let us collect the dimensional reduction for spinor
and one can see various boundary conditions determine the associated
mass of fermion in lower dimension. Consider a complex massless spinor
$\Psi$ in $\mathbb{R}^{d+1}$ satisfying Dirac equation,

\begin{equation}
\gamma^{M}\partial_{M}\Psi=0,\tag{B-1}\label{eq:225}
\end{equation}
where $M$ runs over $\mathbb{R}^{d+1}$. When one of the spatial
direction is compactified on a circle $S^{1}$, $\mathbb{R}^{d+1}$
becomes to $\mathbb{R}^{d}\times S^{1}$. Let us denote the coordinates
on $\mathbb{R}^{d},S^{1}$ as $x^{\mu},y$ respectively. Then Fourier
series of $\Psi$ can be written as the summary of its modes on $S^{1}$
as,

\begin{equation}
\Psi\left(x^{\mu},y\right)=\sum_{k}e^{\frac{iky}{L}}\psi_{k}\left(x^{\mu}\right),\tag{B-2}\label{eq:226}
\end{equation}
where $L$ refers to the radius of $S^{1}$ and $k$ is integer or
half integer. Thus the boundary of the spinor $\Psi$ can be periodic
or anti-periodic as,

\begin{equation}
\Psi\left(x^{\mu},y\right)=\pm\Psi\left(x^{\mu},y+2\pi L\right),\tag{B-3}
\end{equation}
for $k$ is integer and half integer respectively. Mostly, anti-periodic
boundary condition for fermion is permitted since observables are
usually the combination of even power of spinors. Inserting (\ref{eq:226})
into (\ref{eq:225}), it leads to,

\begin{equation}
\left(\gamma^{\mu}\partial_{\mu}-i\gamma_{*}\frac{k}{L}\right)\psi_{k}\left(x^{\mu}\right)=0,\tag{B-4}
\end{equation}
where $\gamma_{*}=\gamma^{y}$. So we can see $\psi_{k}\left(x^{\mu}\right)$
is massive spinor in $\mathbb{R}^{d}$ with an effective mass $i\gamma_{*}\frac{k}{L}$
unless $k=0$. This implies under the dimension reduction, the spinor
in lower dimension is always massive if the anti-periodic boundary
condition is imposed. Note that in the low-energy theory, only the
mode with minimal $k$ is the concern, thus it means fermion is massless/massive
with periodic and anti-periodic boundary condition respectively in
the low-energy theory.

\section{Appendix C: Supersymmetric meson on the flavor brane}

While the D4/D8 model achieves great success, it contains issues.
The most important issue is that due to the remaining supersymmetry
on the D8-branes, the D4/D8 model contains supersymmetrically fermionic
meson (mesino) on the flavor $N_{f}$ $\mathrm{D8/\overline{D8}}$-branes
which should not be presented in QCD \cite{key-131}. As we have specified
in Section 2.1 that the supersymmetry on $N_{c}$ D4-branes breaks
down due to its compactified direction $x^{4}$, however there is
not any mechanism to break down the supersymmetry on the flavor branes
since the $N_{f}$ $\mathrm{D8/\overline{D8}}$-branes is perpendicular
to the compactified direction $x^{4}$. Therefore in principle, there
is no reason to neglect the supersymmetric fermions in this model.
So let us pick up the fermionic action for the D8-branes additional
to its bosonic action (\ref{eq:22}). Up to quadratic order, the fermionic
action for the D8-brane reads \cite{key-132,key-133,key-134},

\begin{equation}
S_{\mathrm{D8}}^{\left(f\right)}=i\frac{T_{\mathrm{D8}}}{2}\int d^{9}xe^{-\phi}\mathrm{STr}\sqrt{-\det\left[g_{ab}+\left(2\pi\alpha^{\prime}\right)\mathcal{F}_{ab}\right]}\bar{\Psi}\left(1-\Gamma_{\mathrm{D8}}\right)\left(\Gamma^{c}\hat{D}_{c}-\Delta+\hat{\mathrm{L}}_{D8}\right)\Psi,\tag{C-1}\label{eq:228}
\end{equation}
where $\Psi$ refers to 32-component Majorana spinor in 10d spacetime
and,

\begin{align}
\Gamma_{\mathrm{D8}} & =\frac{\sqrt{-\det\left[g_{ab}\right]}}{\sqrt{-\det\left[g_{ab}+\left(2\pi\alpha^{\prime}\right)\mathcal{F}_{ab}\right]}}\Gamma_{\mathrm{D8}}^{\left(0\right)}\Gamma^{11}\sum_{q}\frac{\left(-\Gamma^{11}\right)^{q}}{q!2^{q}}\Gamma^{a_{1}...a_{2q}}\mathcal{F}_{a_{1}a_{2}}...\mathcal{F}_{a_{2q-1}a_{2q}},\nonumber \\
\hat{\mathrm{L}}_{D8} & =-\frac{\sqrt{-\det\left[g_{ab}\right]}}{\sqrt{-\det\left[g_{ab}+\left(2\pi\alpha^{\prime}\right)\mathcal{F}_{ab}\right]}}\Gamma_{\mathrm{D8}}^{\left(0\right)}\sum_{q\geq1}\frac{\left(-\Gamma^{11}\right)^{q-1}}{\left(q-1\right)!2^{q-1}}\Gamma^{a_{1}...a_{2q-1}}\mathcal{F}_{a_{1}a_{2}}...\mathcal{F}_{a_{2q-1}c}g^{bc}\hat{D}_{b},\nonumber \\
\Gamma_{\mathrm{D8}}^{\left(0\right)} & =\frac{\epsilon^{a_{1}...a_{9}}}{9!\sqrt{-\det\left[g_{ab}\right]}}\Gamma_{a_{1}...a_{9}}=-\Gamma^{11}\Gamma^{\underline{5}},\nonumber \\
\Delta & =\frac{1}{2}\Gamma^{M}\partial_{M}\phi-\frac{1}{8}\frac{1}{4!}g_{s}e^{\phi}F_{MNPQ}\Gamma^{MNPQ},\nonumber \\
\hat{D}_{M} & =\nabla_{M}-\frac{1}{8}\frac{1}{4!}g_{s}e^{\phi}F_{MNPQ}\Gamma^{MNPQ}\Gamma_{M},\nonumber \\
\nabla_{M} & =\partial_{M}+\frac{1}{4}\omega_{M}^{\ \ \underline{N}\underline{P}}\Gamma_{\underline{N}\underline{P}}.\tag{C-2}\label{eq:229}
\end{align}
The action (\ref{eq:228}) is the fermionic action for D8-brane obtained
under T-duality. The notation in (\ref{eq:228}) and (\ref{eq:229})
is given as follows. The index labeled by capital letters $M,N,P,Q$
runs over 10d spacetime and labeled by lowercase letters runs over
D8-brane. The index with underline corresponds to index in the flat
tangent space used by elfbein e.g. $g_{MN}=e_{M}^{\underline{M}}\eta_{\underline{M}\underline{N}}e_{N}^{\underline{N}}$,
so we have e.g. $\Gamma^{a}=g^{ab}\partial_{b}X^{M}e_{M}^{\underline{N}}\Gamma_{\underline{N}}$.
$\Gamma^{M}$ refers to the Dirac matrix satisfying $\left\{ \Gamma^{M},\Gamma^{N}\right\} =2g^{MN}$
and $\omega_{M}^{\ \ \underline{N}\underline{P}}$ refers to the spin
connection. $F_{MNPQ}$ is the components of the Ramond-Ramond $F_{4}$
and $\phi$ is the dilaton field which are all given in Section 2.
The gamma matrix $\Gamma^{MNPQ}$ is given by $\Gamma^{MNPQ}=\Gamma^{[M}\Gamma^{N}\Gamma^{P}\Gamma^{Q]}$.
Here $\mathcal{F}_{ab}=F_{ab}+B_{ab}$ where $F_{ab}$ is the gauge
field strength on the flavor brane and $B_{ab}$ is the antisymmetric
tensor $B_{MN}$ induced on the flavor brane which can be set to zero.

Imposing the bubble solution given in (\ref{eq:16}) and supergravity
solutions for the dilaton $\phi$ and Ramond-Ramond $F_{4}$ to(\ref{eq:228}),
after some calculations it becomes,

\begin{equation}
S_{\mathrm{D8}}^{\left(f\right)}=\frac{iT}{\left(2\pi\alpha^{\prime}\right)^{2}\Omega_{4}}\int d^{4}xdZd\Omega_{4}\tilde{\Psi}P_{-}\left[\frac{2}{3}M_{KK}K^{-\frac{1}{2}}\Gamma^{\underline{m}}\nabla_{m}^{S^{4}}+K^{-2/3}\Gamma^{\underline{\mu}}\partial_{\mu}+M_{KK}\Gamma_{\underline{4}}\partial_{Z}\right]\Psi,\tag{C-3}\label{eq:230}
\end{equation}
where $\Gamma^{\underline{m}}\nabla_{m}^{S^{4}}$ is the Dirac operator
on $S^{4}$ i.e. the index $m$ runs over $S^{4}$ and,

\begin{equation}
\tilde{\Psi}=K^{-13/24}\Psi,K\left(Z\right)=1+Z^{2},T=\left(\frac{3}{2}\right)^{4}T_{\mathrm{D8}}\Omega_{4}\left(2\pi\alpha^{\prime}\right)^{2}\frac{\left(\frac{2}{3}M_{KK}R\right)^{21/2}}{M_{KK}^{5}},P_{-}=\frac{1}{2}\left(1-\Gamma_{\mathrm{D8}}\right).\tag{C-4}
\end{equation}
Since we are interesting in the fermionic part, the gauge field included
by $\mathcal{F}_{ab}$ has been turned off i.e. $\mathcal{F}_{ab}=0$.
Afterwards, in order to obtain a 5d effective action as the mesonic
action given in (\ref{eq:33}), we can decompose the spinor $\Psi$
into a 3+1 dimensional part $\psi\left(x,Z\right)$ as mesino, an
$S^{4}$ part $\chi$ and a remaining 2d part $\lambda$ as,

\begin{equation}
\Psi=\psi\otimes\chi\left(S^{4}\right)\otimes\lambda.\tag{C-5}\label{eq:232}
\end{equation}
And the associated gamma matrices can be chosen as,

\begin{align}
\Gamma^{\underline{\mu}} & =\sigma_{1}\otimes\gamma^{\underline{\mu}}\otimes\boldsymbol{1},\mu=0,1,2,3\nonumber \\
\Gamma^{\underline{4}} & =\sigma_{1}\otimes\gamma\otimes\boldsymbol{1},\nonumber \\
\Gamma^{\underline{5}} & =\sigma_{2}\otimes\boldsymbol{1}\otimes\tilde{\gamma},\nonumber \\
\Gamma^{\underline{m}} & =\sigma_{2}\otimes\boldsymbol{1}\otimes\tilde{\gamma}^{\underline{m}},m=6,7,8,9,\nonumber \\
\gamma & =i\gamma^{\underline{0}}\gamma^{\underline{1}}\gamma^{\underline{2}}\gamma^{\underline{3}},\nonumber \\
\tilde{\gamma} & =i\gamma^{\underline{6}}\gamma^{\underline{7}}\gamma^{\underline{8}}\gamma^{\underline{9}},\tag{C-6}\label{eq:233}
\end{align}
where $\sigma_{1,2,3}$ refer to the Pauli matrices. In this decomposition,
the 10d chirality matrix takes a very simple form as $\Gamma^{11}=\sigma_{3}\otimes\boldsymbol{1}\otimes\boldsymbol{1}$.
If we chose the $\sigma_{3}$ representation, $\lambda$ can be decomposed
by the eigenstates of $\sigma_{3}$ with

\begin{equation}
\sigma_{3}\lambda_{\pm}=\lambda_{\pm},\sigma_{1}\lambda_{\pm}=\lambda_{\mp},\sigma_{2}\lambda_{\pm}=\pm i\lambda_{\mp},\tag{C-7}
\end{equation}
where $\lambda_{\pm}$ refers to the two eigenstates of $\sigma_{3}$.
Since the kappa symmetry fixes the condition $\Gamma^{11}\Psi=\Psi$,
we have to chose $\lambda=\lambda_{+}$. Besides, as $\chi$ must
satisfy the Dirac equation on $S^{4}$, it can be decomposed by the
spherical harmonic function. So the eigenstates of $\Gamma^{\underline{m}}\nabla_{m}^{S^{4}}$
can be chosen as \cite{key-135},

\begin{equation}
\Gamma^{\underline{m}}\nabla_{m}^{S^{4}}\chi^{\pm l,s}=i\Lambda_{l}^{\pm}\chi^{\pm l,s};\Lambda_{l}^{\pm}=\pm\left(2+l\right),l=0,1...\tag{C-8}\label{eq:235}
\end{equation}
where $s,l$ are angular quantum numbers carried by spherical harmonic
function.

Put (\ref{eq:232}) into (\ref{eq:230}) with the decomposition (\ref{eq:233})
- (\ref{eq:235}) for $\chi$ and $\lambda$, we finally reach to
a 5d effective action for mesino field as,

\begin{equation}
S=iT\int d^{4}xdZ\bar{\psi}\left(-\frac{2}{3}M_{KK}\lambda K^{-1/2}+K^{-2/3}\gamma^{\underline{\mu}}\partial_{\mu}+M_{KK}\gamma\partial_{Z}\right)\psi.\tag{C-9}\label{eq:236}
\end{equation}
The 5d mesino $\psi$ can be further decomposed by working with

\begin{equation}
\psi\left(x,Z\right)=\left(\begin{array}{c}
\psi_{+}\\
\psi_{-}
\end{array}\right)=\sum_{n}\left[\begin{array}{c}
\psi_{+}^{\left(n\right)}\left(x\right)f_{+}^{\left(n\right)}\left(Z\right)\\
\psi_{-}^{\left(n\right)}\left(x\right)f_{-}^{\left(n\right)}\left(Z\right)
\end{array}\right],\tag{C-10}\label{eq:237}
\end{equation}
where $f_{\pm}^{\left(n\right)}\left(Z\right)$ are real eigenfunctions
of the coupled equations

\begin{align}
-\frac{2\lambda}{3}K^{-1/2}f_{+}^{\left(n\right)}+\partial_{Z}f_{+}^{\left(n\right)} & =\Lambda_{n}M_{KK}K^{-2/3}f_{-}^{\left(n\right)},\nonumber \\
-\frac{2\lambda}{3}K^{-1/2}f_{-}^{\left(n\right)}-\partial_{Z}f_{-}^{\left(n\right)} & =\Lambda_{n}M_{KK}K^{-2/3}f_{+}^{\left(n\right)},\tag{C-11}
\end{align}
with the normalizations

\begin{equation}
T\int dZK^{-2/3}f_{i}^{\left(n\right)}f_{j}^{\left(m\right)}=\delta^{mn}\delta_{ij},\ i,j=+,-,\tag{C-12}\label{eq:239}
\end{equation}
Plugging (\ref{eq:237}) - (\ref{eq:239}) into (\ref{eq:236}), the
action takes the canonical form as ($M_{n}=\Lambda_{n}M_{KK}$)

\begin{equation}
S=-\int d^{4}x\sum_{n}\left\{ \psi_{-}^{\left(n\right)\dagger}i\sigma^{\mu}\partial_{\mu}\psi_{-}^{\left(n\right)}+\psi_{+}^{\left(n\right)\dagger}i\bar{\sigma}^{\mu}\partial_{\mu}\psi_{+}^{\left(n\right)}+M_{n}\left[\psi_{-}^{\left(n\right)\dagger}\psi_{+}^{\left(n\right)}+\psi_{+}^{\left(n\right)\dagger}\psi_{-}^{\left(n\right)}\right]\right\} .\tag{C-13}\label{eq:240}
\end{equation}
Defining the Dirac spinor written in the Wely basis as,

\begin{equation}
\psi^{\left(n\right)}=\left(\begin{array}{c}
\psi_{+}^{\left(n\right)}\left(x\right)\\
\psi_{-}^{\left(n\right)}\left(x\right)
\end{array}\right),\tag{C-14}
\end{equation}
action (\ref{eq:240}) can be rewritten as, ($\left\{ \gamma^{\mu},\gamma^{\nu}\right\} =2\eta^{\mu\nu}$)

\begin{equation}
S=i\int d^{4}x\sum_{n}\left[\bar{\psi}^{\left(n\right)}\gamma^{\mu}\partial_{\mu}\psi^{\left(n\right)}+M_{n}\bar{\psi}^{\left(n\right)}\psi^{\left(n\right)}\right],\tag{C-15}\label{eq:242}
\end{equation}
leading to a standard action for fermion. As we can see, the fermionic
action illustrates the mesino mass takes the same order of meson mass
hence it should be not neglected in principle, and authors of \cite{key-136}
also confirm this conclusion which is consistent with the remaining
supersymmetry on D8-branes.

Moreover, when the bosonic gauge field is turned on, action (\ref{eq:228})
reduces to interaction terms of meson and mesino up to $\mathcal{O}\left(\mathcal{F},\Psi^{2}\right)$
as,

\begin{align}
S_{int}= & i\frac{T_{\mathrm{D8}}}{4}\int d^{9}x\sqrt{-g}e^{-\phi}\bar{\Psi}\Gamma_{\underline{5}}\Gamma^{11}\Gamma^{ab}\mathcal{F}_{ab}\left(\Gamma^{c}\hat{D}_{c}-\Delta\right)\Psi\nonumber \\
 & -i\frac{T_{\mathrm{D8}}}{2}\int d^{9}x\sqrt{-g}e^{-\phi}\bar{\Psi}\left(1-\Gamma_{\underline{5}}\right)\Gamma^{\underline{5}}\Gamma^{11}\Gamma^{a}\mathcal{F}_{a}^{\ b}\hat{D}_{b}\Psi.\tag{C-16}\label{eq:243}
\end{align}
Using the decomposition (\ref{eq:46}) for $A_{a}$ and (\ref{eq:233})
- (\ref{eq:235}) for $\Psi$, action (\ref{eq:243}) includes interaction
of $\pi$ meson and mesino as

\begin{align}
S_{int}= & \frac{M_{KK}}{f_{\pi}M_{int}^{2}}\sum_{n,p}\int d^{4}x\partial_{\mu}\pi\big[M_{KK}t_{\pi,n,p}\bar{\psi}^{\left(n\right)}i\gamma^{\mu}\gamma^{n+p+1}\psi^{\left(p\right)}\nonumber \\
 & +l_{\pi,n,p}\bar{\psi}^{\left(n\right)}\left(-\gamma\right)^{n+p+1}\partial^{\mu}\psi^{\left(p\right)},\tag{C-17}
\end{align}
where the coupling constant is evaluated numerically as,

\begin{align}
t_{\pi,n,p} & =\frac{1}{2}\int dZK^{-5/6}\left[f_{+}^{\left(n\right)}f_{+}^{\left(p\right)\prime}-f_{+}^{\left(p\right)}f_{+}^{\left(n\right)\prime}+iK^{-1/2}f_{+}^{\left(n\right)}f_{+}^{\left(p\right)}\right],\nonumber \\
l_{\pi,n,p} & =\int dZK^{-3/2}f_{+}^{\left(n\right)}f_{-}^{\left(p\right)}.\tag{C-18}
\end{align}
And one can further work out the interaction terms of $\rho$ meson
and mesino similarly. Since there is not any mechanism to suppress
the interaction of meson and mesino or break down the supersymmetry
on the D8-brane, we have to take into account these interactions in
this model in principle while they are absent in realistic QCD.

Although we do not attempt to figure out this issue completely in
this review, we give some comments which may be suggestive. The way
to break down the supersymmetry on D8-branes may follow the discussion
in \cite{key-22}, that is to compactify one of the directions of
D8-brane (which is vertical to the $N_{c}$ D4-branes) on another
circle then impose the periodic and anti-periodic boundary condition
to meson and mesino respectively. Afterwards the supersymmetry on
D8-branes would break down then the spectrum of meson and mesino is
separated by a energy scale $1/\beta_{s}$ where $\beta_{s}$ refers
to the size of the compactified direction of D8-brane. Another alternative
scheme is to consider that the bubble solution (\ref{eq:16}) has
a period $\beta_{T}$ with $\beta_{T}\gg1$, hence the dual theory
is non-supersymmetric above the size $\beta_{T}$ if we perform the
same dimension reduction as \cite{key-22}. Therefore it means the
supersymmetry gets to rise only at exactly zero temperature due to
$\beta_{T}=1/T$ which is ideal case, out of reach physically. So
the dual theory on the D8-brane would be non-supersymmetry at any
finite temperature.

\end{document}